\documentclass[journal,10pt,onecolumn,draftclsnofoot,]{IEEEtran}
\usepackage{subfigure}
\usepackage{graphicx}
\usepackage{textcomp}
\usepackage{subfigure}
\usepackage{amssymb}
\usepackage{amsfonts}
\usepackage{amsthm}
\usepackage[tbtags]{amsmath}
\usepackage[ruled,vlined]{algorithm2e}
\usepackage{csquotes}
\usepackage{textcomp}
\usepackage{color}
\usepackage{array}
\usepackage{bbm}
\usepackage{tabularx}
\newcolumntype{L}{>{\centering\arraybackslash}X}
\newcolumntype{M}[1]{>{\centering\arraybackslash}m{#1}}
\usepackage{caption}

\begin{document}

\title{A Survey on Machine and Deep Learning for Optical Communications}
\author{M. A. Amirabadi, S. A. Nezamalhosseini, M. H. Kahaei, and Lawrence R. Chen
\thanks{M. A. Amirabadi is with Department of Electrical Engineering, Ferdowsi University of Mashhad, Mashhad 9177948883, Iran, S. A. Nezamalhosseini, and M. H. Kahaei are with the School of Electrical Engineering, Iran University of Science and Technology (IUST), Tehran, 1684613114 Iran, Lawrence R. Chen is with Department of Electrical and Computer Engineering, McGill University, Montreal, Canada, e-mail: amirabadi@um.ac.ir, nezam@iust.ac.ir, kahaei@iust.ac.ir, lawrence.chen@mcgill.ca.}}

\maketitle

\begin{abstract}
The ever-growing complexity of optical communication systems and networks demands sophisticated methodologies to extract meaningful insights from vast amounts of heterogeneous data. 
Machine learning (ML) and deep learning (DL) have emerged as frontrunners in this domain, offering a transformative approach to data analysis and enabling automated self-configuration in optical communication systems.
The adoption of ML and DL in optical communication is driven by the exponential increase in system and link complexity, stemming from the introduction of numerous adjustable and interdependent parameters. 
This is particularly evident in areas like coherent transceivers, advanced digital signal processing, optical performance monitoring, cross-layer network optimizations, and nonlinearity compensation.
While the potential benefits of ML and DL are immense, the extent to which these methods can revolutionize optical communication remains largely unexplored. 
Additionally, many ML and DL algorithms have yet to be deployed in this field, highlighting the nascent nature of this research area.
To comprehensively address the application of ML and DL in optical communication, this paper presents a survey covering optical fiber communication (OFC), optical communication networking (OCN), and optical wireless communication (OWC).
This survey delves into unique challenges and complexities, showcasing how ML and DL effectively address them, and summarizing the key findings. 
Motivations for employing different ML and DL models in intelligent optical communications are analyzed from both the intrinsic characteristics of ML and DL and the external enabling techniques that facilitate their integration into optical systems.
The survey examines the utilization of ML and DL algorithms in a variety of optical communication applications, highlighting their impact on performance enhancement and system optimization. 
In addition to quantitative measures of the gains achieved by ML and DL approaches, the survey provides quantitative/qualitative comparisons to elucidate how different ML and DL algorithms can contribute to improving the performance of optical communication systems.
This survey acknowledges the challenges associated with implementing ML and DL in optical communication, and discussed potential solutions and future research directions.

\end{abstract}


\section{Introduction}

Artificial intelligence (AI) has emerged as a transformative force, enabling machines to perform tasks that traditionally required human intelligence. AI's capabilities extend to cognitive functions such as knowledge storage, problem-solving, and learning from experience.
Within the domain of optical communications, AI techniques like machine learning (ML) and deep learning (DL) are gaining traction, offering powerful solutions to address the computational complexities of existing analytical and numerical models.
ML, a branch of AI, focuses on algorithms that can learn patterns, models, and trends from data. These learned patterns can then be applied to make predictions or decisions on new data. DL, inspired by the human brain's neural network, has pushed the boundaries of ML, achieving even greater performance and accuracy.
The advancements in ML and DL techniques have spurred their integration into various domains, including optical communication. While AI has made significant progress in the learning aspects of ML and DL, its implementation in optical communication applications has gained momentum in recent years.
Researchers have successfully applied ML/DL to various optical fiber communication (OFC) media, including single-mode, few-mode, and multi-core fibers, and optical wireless communication (OWC) media, such as indoor, outdoor, and underwater environments. These applications span both physical layer and optical communication networking (OCN) domains.
In the physical layer, ML/DL algorithms are employed for tasks like:

\begin{itemize}

\item {Enhanced Modeling: ML/DL can capture the complex relationships between parameters in optical communication systems, leading to more accurate and reliable models for system design and performance evaluation.}

\item {Adaptive Signal Processing: ML/DL algorithm can be used for real-time signal processing tasks, such as noise reduction, error correction, and adaptive modulation, to optimize signal transmission and enhance quality of service.}

\item {Nonlinearity equalization: Compensation for signal distortions introduced by nonlinear optical effects in optical fibers.}

\item {Resource allocation: Efficient allocation of network resources such as bandwidth and power to optimize data transmission.}

\item {Constellation shaping: Optimization of optical signal constellation patterns to enhance transmission efficiency and error resilience.}

\end{itemize}

On the OCN front, ML/DL algorithms are used for:

\begin{itemize}

\item {Network Optimization: ML/DL can be employed to optimize network routing, resource allocation, and traffic management, leading to more efficient and resilient optical communication networks.}

\item {Fault Detection and Mitigation: ML/DL can detect and diagnose system faults, such as fiber cuts or optical signal impairments, enabling proactive network maintenance and improving overall network availability.}

\item {Resource Management: ML/DL can optimize resource allocation, such as power consumption and spectrum usage, to maximize the efficiency and sustainability of optical communication systems.}

\item {Optical performance monitoring (OPM): Real-time monitoring of network performance parameters to detect and diagnose potential issues.}

\item {Quality of transmission (QoT) estimation: Evaluation of network performance metrics to ensure data integrity and reliability.}

\end{itemize}

The ML algorithms applied in optical communication can be categorized into three main groups:

\begin{itemize}

\item {Supervised learning (SL): SL \cite{160} algorithms learn from labeled data, enabling them to make predictions or classifications. Examples include support vector machines (SVMs) \cite{161}, artificial neural networks (ANNs) \cite{162}, k-nearest neighbors (kNNs) \cite{163}, ensemble learning algorithms \cite{167}, and regression algorithms \cite{168}-\cite{170}.}

\item {Unsupervised learning (USL): USL \cite{171} algorithms extract patterns and groupings from unlabeled data. Examples include hierarchical clustering \cite{172}, K-means clustering \cite{173}, expectation maximization (EM) clustering \cite{174}, principal component analysis (PCA) \cite{175}, and independent component analysis (ICA) \cite{176}.}

\item {Reinforcement learning (RL): RL \cite{177} algorithms interact with an environment to learn optimal decision-making strategies. Examples include policy-based RL and value-based RL \cite{178}.}

\end{itemize}

DL \cite{179}, a subset of ML, utilizes ANNs with multiple layers to learn complex patterns from data. DL-based works in optical communication can be classified into four main categories:

\begin{itemize}

\item {Deep neural network (DNN): DNNs \cite{180}  are the most general form of neural networks, incorporating multiple layers to extract intricate patterns from data.}

\item {Recurrent neural network (RNN): RNNs \cite{181} are designed to handle sequential data, making them suitable for tasks such as optical signal processing and traffic prediction in optical networks.}

\item {Convolutional neural network (CNN): CNNs \cite{182} are specialized for processing spatial data, such as optical channel images or network topology maps.}

\item {Deep reinforcement learning (DRL): DRL \cite{183} combines RL with DL, enabling intelligent decision-making in complex and dynamic environments, such as optical networks.}

\end{itemize}

\subsection{Motivations}

The field of AI for optical communication is witnessing a remarkable pace of development, with new research advancements emerging at an accelerated rate. However, a comprehensive review of the literature reveals that many ML and DL algorithms remain unexplored, and numerous optical communication applications have yet to be fully addressed. This untapped potential underscores the nascent nature of this research area.
In light of this dynamic landscape, the need for comprehensive survey papers has become increasingly crucial. These surveys play a pivotal role in synthesizing the latest advancements and identifying the gaps in the field, enabling researchers to stay abreast of the cutting-edge developments and identify potential research directions.
By providing a comprehensive overview of the current state of AI for optical communication, survey papers serve as valuable resources for researchers seeking to navigate this rapidly evolving field. They facilitate the identification of promising research opportunities, enabling researchers to make informed decisions about their research endeavors and contribute to the advancement of this transformative technology.
As AI continues to permeate various aspects of optical communication, survey papers will undoubtedly play an even more indispensable role in fostering collaboration, accelerating progress, and shaping the future of this promising field.

\begin{figure*}[tp!]
\centering
\includegraphics[scale=0.4]{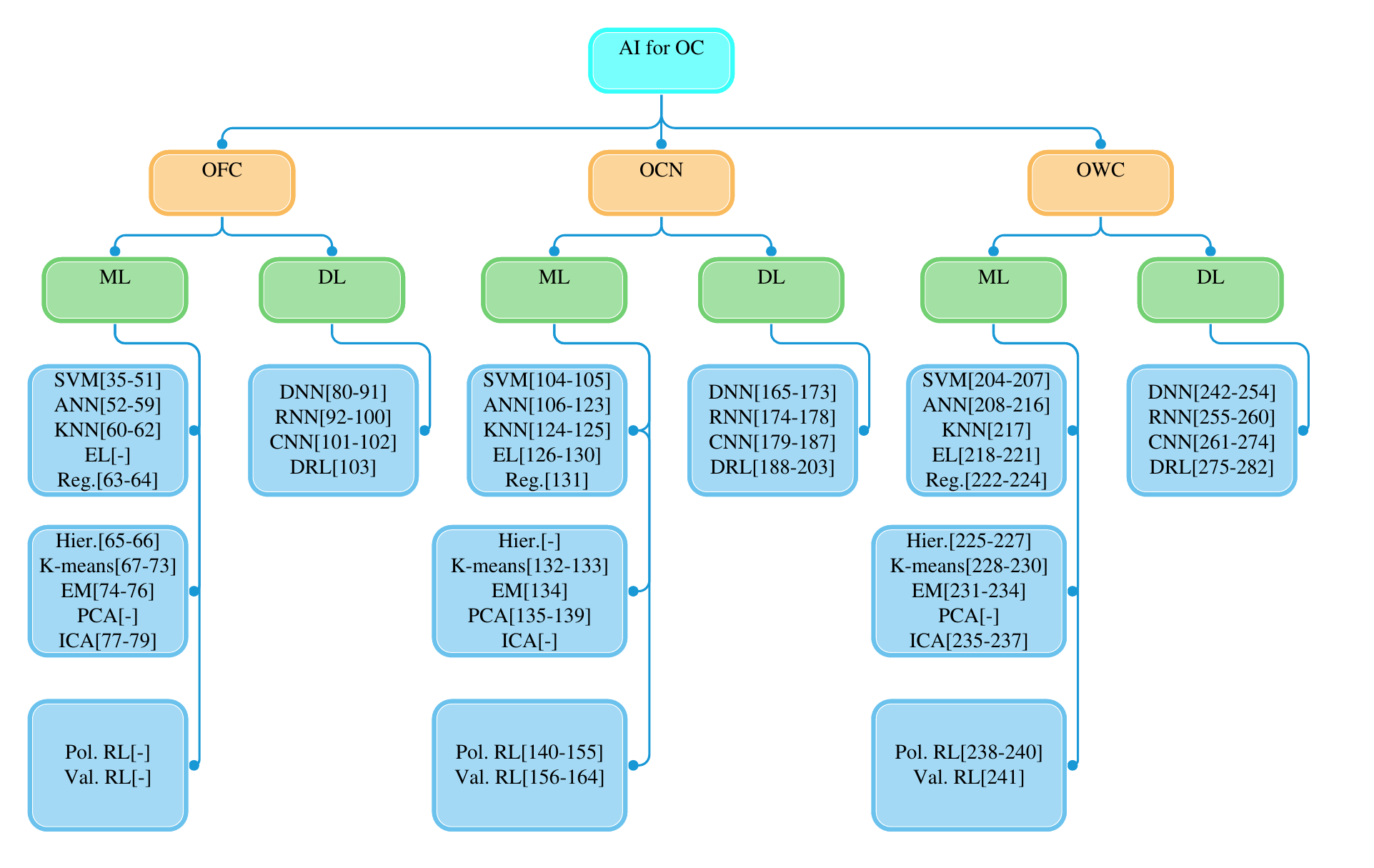}
\caption{Schematic diagram categorizing different AI methods applied in OFC, OCN, and OWC.}
\label{fig:20}
\end{figure*}

\begin{table}[tp!]
\tiny
\centering
\caption{List of Acronyms.}
\begin{tabular}{ |c|c|}
\hline
\textbf{Acronym} &	\textbf{Definition}	\\ \hline
Amplified Spontaneous Emission &ASE\\ \hline  
Artificial Intelligence &AI\\ \hline
Artificial Neural Network &ANN\\ \hline
Asynchronous Amplitude Histograms &AAH\\ \hline
Bit Error Rate &BER\\ \hline
Bit Rate-Modulation Format Identification &BR-MFI\\ \hline
Carrier Less Amplitude Phase &CAP\\ \hline
Channel State Information &CSI\\ \hline
Chromatic Dispersion &CD\\ \hline
Orthogonal Frequency Division Multiplexing &OFDM\\ \hline
Constant Module Algorithm &CMA\\ \hline
Convolutional Neural Network &CNN\\ \hline
Decision Tree &DT\\ \hline
Deep Learning &DL\\ \hline
Deep Neural Network &DNN\\ \hline
Deep Reinforcement Learning &DRL\\ \hline
Differential Group Delay &DGD\\ \hline
Digital Back-propagation &DBP\\ \hline
Elastic Optical Network &EON\\ \hline
Expectation Maximization &EM\\ \hline
Extended Gradient Boosting &XGB\\ \hline
Forward Error Correction &FEC\\ \hline
Free Space Optical &FSO\\ \hline
Gated Recurrent Units &GRU\\ \hline
Independent Component Analysis &ICA\\ \hline
Intensity Modulation/Direct Detection &IM/DD\\ \hline
Least Mean Square &LMS\\ \hline
Light Emitting Diode &LED\\ \hline
Long Short-Term Memory &LSTM\\ \hline
Machine Learning &ML\\ \hline
Maximum A Posteriori &MAP\\ \hline
Mean Absolute Error &MAE\\ \hline
Minimum Mean Square Error &MMSE\\ \hline
Mobile Ad-hoc Network &MANET\\ \hline
Multi-input Multi-output &MIMO\\ \hline
Multi-layer Perceptron &MLP\\ \hline
Non-return to Zero-on-off Keying &NRZ-OOK\\ \hline
Nonlinear Equalizer &NLE\\ \hline
Nonlinear Phase Noise &NLPN\\ \hline
Optical Burst Switching &OBS\\ \hline
Optical Communication Networking &OCN\\ \hline
Optical Fiber Communication &OFC\\ \hline
Optical Performance Monitoring &OPM\\ \hline
Optical Signal to Noise Ratio &OSNR\\ \hline
Optical Wireless Communication &OWC\\ \hline
Orbital Angular Momentum &OAM\\ \hline
Passive Optical Network &PON\\ \hline
Polarization Mode Dispersion &PMD\\ \hline
Principal Component &PC\\ \hline
Principal Component Analysis &PCA\\ \hline
Pulse Amplitude Modulation &PAM\\ \hline
Q-learning &QL\\ \hline
Quadrature Amplitude Modulation &QAM\\ \hline
Quadrature Phase Shift Keying &QPSK\\ \hline
Quality of Service &QoS\\ \hline
Quality of Transmission &QoT\\ \hline
Radial Basis Function Neural Network &RBFNN\\ \hline
Random Forest &RF\\ \hline
Received Signal Strength Indicator &RSSI\\ \hline
Recurrent Neural Network &RNN\\ \hline
Reinforcement Learning &RL\\ \hline
Routing and Resource Allocation &RRA\\ \hline
Routing and Spectrum Allocation &RSA\\ \hline
Routing and Wavelength Allocation &RWA\\ \hline
Routing, Core, and Resource Allocation &RCSA\\ \hline
Software Defined Network &SDN\\ \hline
Spatial Division Multiplexing &SDM\\ \hline
Supervised Learning &SL\\ \hline

\end{tabular}
\label{table:1}
\end{table}

\begin{table*}

\tiny
\centering
\caption{Summary of DRL applications in OFC, OCN, and OWC.}
\begin{tabular}{|M{.5cm}|M{0.75cm}|M{0.75cm}|M{0.75cm}|M{0.75cm}|M{12cm}|}

\hline

Ref	& Technique	& 	Media	& 	Year	& 	\# of Refs	& 	Highlights\\ \hline
\cite{153c}	& 	ML	& 	OCN	& 	2019	& 	73	& 	Authors reviewed the use of ML for failure management in optical networks\\ \hline
\cite{153d}		& ML	& 	OCN	& 	2021	& 	201	& 	Authors surveyed the use of ML techniques for routing optimization in SDN\\ \hline
\cite{153e}	& 	ML	& 	OCN	& 	2020	& 	110	& 	Authors provided an overview of routing and resource allocation based on ML in optical networks\\ \hline
\cite{153f}	& 	ML	& 	OCN	& 	2018	& 	43	& 	Authors described how ML can be used to improve the performance and efficiency of optical networks\\ \hline
\cite{153g}	& 	ML	& 	OCN	& 	2020	& 	132	& 	Authors surveyed the existing applications of ML for intelligent optical networks and classify the applications of ML in terms of their use cases\\ \hline
\cite{153h}	& 	ML	& 	OCN	& 	2018	& 	148	& 	Authors provided a review about the application of AI techniques for improving the performance of optical communication systems and networks\\ \hline
\cite{153i}	& 	ML	& 	OCN	& 	2018	& 	125	& 	Authors presented an overview of the application of ML to optical communications and networking\\ \hline
\cite{153j}	& 	ML	& 	OFC	& 	2019	& 	86	& 	Authors described the mathematical foundations of basic ML techniques from communication theory and signal processing perspectives, and to shed light on the types of problems in optical communications and networking that naturally warrant ML use\\ \hline
\cite{153k}	& 	ML	& 	OCN	& 	2018	& 	278	& 	Authors reviewed how ML algorithms are applied in the realm of SDN, from the perspective of traffic classification, routing optimization, QoS/ Quality of experience prediction, resource management, and security\\ \hline
\cite{153l}	& 	DL	& 	OCN	& 	2021	& 	42	& 	Authors highlighted the contributions of DL solutions to optical communications\\ \hline
\cite{153m}	& 	ML	& 	OCN	& 2018	& 	48	& 	Authors reviewed the existing ML approaches for coherent optical-OFDM in a common framework and reviewed the progress in this area with a focus on practical aspects and comparison with benchmark DSP solutions\\ \hline
\cite{153n}	& 	ML	& 	OCN	& 	2022	& 	108	& 	Authors overviewed the applications of ML technologies in QoT prediction for optical networks\\ \hline
\cite{153p}	& 	ML	& 	OCN	& 	2022	& 	79	& 	Authors reviewed OPM techniques where ML algorithms have been applied\\ \hline
[This work]	& 	ML, DL	& 	OFC, OCN, OWC	& 	2024	& 282	& Authors surveyed the applications of ML and DL in three domains: OFC, OCN, and OWC. By reviewing a wide range of works, the authors provide a comprehensive investigation of different methodologies applied in these applications\\ \hline
\end{tabular}
\label{table:2}
\end{table*}

\subsection{Related Works}

ML/DL has emerged as a transformative force in various fields, including optical communications. Several surveys have been conducted to comprehensively examine the applications and advancements of ML/DL in this domain (Table \ref{table:2}). These surveys provide valuable insights into the current state of research and offer promising directions for future exploration.
In \cite{153c}, the authors thoroughly investigate the application of ML for failure management in optical networks. Their study highlights the potential of ML to transform failure management practices and enhance network resilience.
ML techniques are also being explored for routing optimization in software-defined networks (SDNs). A recent survey \cite{153d} categorizes the use of ML techniques for routing optimization in SDNs into three main groups: SL, USL, and RL.
The authors of \cite{153e} provide a comprehensive overview of routing and resource allocation (RRA) in optical networks based on ML techniques. The authors first introduce the routing and wavelength allocation (RWA) problem in wavelength division multiplexing (WDM) optical networks, the routing and spectrum allocation (RSA) problem in elastic optical networks (EONs), and the routing, core, and resource allocation (RCSA) problem in spatial division multiplexing (SDM) optical networks.
The authors then delve into the challenges associated with QoT estimation, traffic estimation, and crosstalk prediction, which are crucial aspects of RRA in optical networks. They discuss how these challenges can be addressed using ML techniques.
In \cite{153f}, the authors explore the transformative potential of ML in optimizing the performance and efficiency of optical networks. They delve into ML concepts tailored specifically for the optical networking industry, encompassing algorithm selection, data and model management strategies, and integration with existing network control and management systems.
The authors meticulously detail four real-world networking case studies, showcasing the practical applications of ML in optical networks. These case studies illuminate how ML can revolutionize network operations, addressing challenges such as predictive maintenance, virtual network topology management, capacity optimization, and optical spectral analysis.
These case studies demonstrate the versatility and effectiveness of ML in tackling real-world challenges faced by optical network operators. 
The authors of \cite{153g} surveyed the existing applications of ML for intelligent optical networks. The authors classify the applications of ML in terms of their use cases, which are categorized into optical network control and resource management (RWA, bandwidth allocation, traffic management, network congestion control, self-healing networks, optical network monitoring and survivability), and optical network monitoring and survivability (fault detection and isolation, performance monitoring, network planning and optimization, security and intrusion detection).

A comprehensive review of recent research efforts is provided in \cite{153h}, highlighting the wide range of applications for AI in this domain. These applications encompass optical transmission, where AI algorithms can optimize signal processing and enhance the robustness of transmission systems, and optical network control and management, where AI can optimize resource allocation, improve network resilience, and facilitate self-configuration.
In \cite{153i}, the authors present an insightful overview of the application of ML to optical communications and networking. They delve into the motivations behind using ML in optical networks, exploring the challenges and opportunities that ML presents. The authors then comprehensively discuss the application of ML in optical networks, including optical transmission, where ML algorithms can optimize nonlinear equalization, constellation shaping, and adaptive modulation, and optical network control and management, where ML can optimize routing, resource allocation, and fault detection.
The authors of \cite{153j} delve into the mathematical underpinnings of fundamental ML techniques, including SL, USL, and RL, from the perspectives of communication theory and signal processing. They elucidate the theoretical foundations that enable these techniques to extract knowledge from data and make informed decisions.
Moreover, the authors provide an overview of ongoing ML research in optical communications and networking, highlighting the focus on physical layer issues. They discuss ML-based channel estimation techniques that can accurately model the nonlinearities of optical channels, ML-based signal detection methods that can overcome noise and interference, and ML-based equalization algorithms that can compensate for channel distortions.

In \cite{153k}, the authors thoroughly explore the application of ML algorithms in SDN, focusing on traffic classification, routing optimization, quality of service (QoS)/quality of experience prediction, resource management, and security. They demonstrate the ability of ML to enhance the efficiency, reliability, and intelligence of SDN networks.
The authors of \cite{153l} showcase the contributions of DL solutions to optical communications, highlighting the potential of DNN, RNN and DRL in various optical communication tasks. They discuss the challenges and opportunities associated with using DL in optical communications, emphasizing its transformative potential.
In \cite{153m}, the authors provide a comprehensive review of existing ML approaches for coherent optical-orthogonal frequency division multiplexing (OFDM), systematically analyzing their progress and comparing them to benchmark DSP solutions. They focus on the practical aspects and effectiveness of ML-based coherent optical-OFDM techniques, including linear prediction, volterra series, and neural networks.
The authors of \cite{153n} present an overview of ML-based QoT prediction techniques, examining recent studies that utilize random forests, neural networks, and SVMs for accurate QoT estimation.
In \cite{153p}, the authors review the application of ML algorithms in OPM, focusing on signal quality monitoring, impairment monitoring, network fault detection and localization, bit rate-modulation format identification (BR-MFI). They demonstrate the effectiveness of ML in enhancing OPM capabilities and enabling proactive network management.

\subsection{Novelties and Contributions}

While existing surveys have provided valuable insights into AI applications in optical communication, they have largely focused on ML methods while overlooking the advanced capabilities of DL algorithms.
Additionally, these surveys have neglected AI applications in OWC, an emerging field with immense potential for ML and DL applications. 
Moreover, they have primarily focused on categorizing research based on applications, overlooking the significance of analyzing algorithms. This approach limits the comprehensive understanding of ML and DL methodologies employed in optical communications.
By addressing the limitations of existing surveys and offering a comprehensive and algorithm-centric perspective, this paper aims to serve as a valuable resource for researchers, practitioners, and educators seeking to navigate the evolving landscape of AI in optical communication.
The novelties and contributions of this paper are summarized as follows:

\begin{itemize}

\item {Up-to-date survey of AI applications in OFC, OCN, and OWC: This paper provides a comprehensive overview of recent advancements in AI applications across various OFC, OCN, and OWC domains.}

\item {Comprehensive and general review of ML and DL methods: The paper presents a broad spectrum of ML and DL techniques applied in diverse optical communication applications, offering a comprehensive understanding of the capabilities and applicability of these methods.}

\item {Consideration of numerous research findings: The paper delves into numerous research findings related to the design, deployment, complexity and performance of various ML and DL methods, providing practical insights for practitioners.}

\item {Quantitative measures of ML and DL gains: The paper includes quantitative measures of the gains achieved using ML and DL approaches for addressing specific issues or challenges, facilitating a deeper understanding of their impact on optical communication systems.}

\item {Comparisons between ML/DL and conventional algorithms: The paper presents quantitative and qualitative comparisons between ML/DL and conventional algorithms, providing valuable insights into the advantages and limitations of these approaches.}

\item {Summarizing the reviwed papers: The paper provides tables summarizing the comparative/quantitative performance and complexity, deployed train and test data type, objective of the paper (output), the input data (features), the adopted metrics, and adopted algorithms.}

\item {Future Directions and Transformative Potential: The paper summarizes the key findings, discusses the open challenges and future research directions, and highlights the transformative potential of ML and DL for the advancement of optical communication.}

\end{itemize}

This survey comprehensively explores the multifaceted applications of ML and DL in the realm of OFC, OCN, and OWC, delving into the diverse ML and DL algorithms that underpin these advancements, providing a clear and general overview for readers to navigate this rapidly evolving field.
This comprehensive survey serves as a valuable resource for researchers and practitioners alike, providing a thorough understanding of the applications and methodologies involved in harnessing the power of ML and DL to revolutionize optical communication systems.
Moreover, in this survey, we summarize the key findings of ML and DL implementations on OFC, OCN, and OWC through Tables \ref{table:3}-\ref{table:8} which holds immense value for multiple reasons:

\begin{itemize}

\item {Reproducibility and Transparency:
Summarizing key findings allows others to understand the research and potentially reproduce the results. This is crucial for scientific progress and building trust in ML models.
Transparency regarding performance, complexity, data, and metrics helps evaluate the model's strengths and weaknesses objectively.}

\item {Knowledge Sharing and Comparison:
Summaries facilitate knowledge sharing within the field, allowing researchers to build upon existing work and avoid duplication of effort.
Comparisons between different approaches are made easier by readily available summaries of key findings. This helps determine the best method for a specific task or situation.}

\item {Informed Decision-Making:
Practitioners can make informed decisions about adopting or adapting ML models by understanding their performance, complexity, data requirements, and objectives.
Summarizing metrics and objective functions helps assess the model's alignment with business goals and identify potential biases.}

\item{Efficiency and Collaboration:
Having clear summaries eliminates the need to read through entire research papers to grasp the core findings, saving time and effort.
Summarizing key findings encourages collaboration by providing a common ground for researchers and practitioners from different backgrounds.}
\end{itemize}

Here's a breakdown of the value for each aspect we mentioned in summaries:

\begin{itemize}

\item {Performance: Understanding how well the model performs on actual data helps gauge its real-world effectiveness.}
\item {Complexity: Knowing the model's complexity (number of parameters, training time) informs decisions about deployment and computational resources.}
\item {Train/Test Data Type: Understanding the type of data used for training and testing helps assess the model's generalizability and potential biases.}
\item {Objective: Knowing the model's objective function clarifies its purpose and helps evaluate its success in achieving that goal.}
\item {Input Data: Understanding the format and characteristics of the input data helps determine if the model is suitable for a specific task or needs adaptation.}
\item {Metrics: Knowing the metrics used to evaluate the model provides insights into its strengths and weaknesses, allowing for informed comparisons.}
\item {Adopted Algorithms: Summarizing the chosen algorithms helps understand the approaches taken and facilitates further research in that direction.}
\end{itemize}

Overall, summarizing key findings in ML is crucial for transparency, knowledge sharing, informed decision-making, efficiency, and collaboration. This practice ultimately accelerates progress and promotes responsible development within the field. 
In summary, the survey could help not only who already has some working knowledge of using ML or DL in OFC, OCW, and OCN, but also who is just trying to get started in the field.

\subsection{Paper Organization}

This review dives deep into the exciting world of ML and DL applications in optical communication. To navigate this vast landscape, we've created a clear four-tiered structure illustrated in Fig. \ref{fig:20}.
At the core, the paper dissects three key applications: OFC, OCN, and OWC. Each gets its own dedicated section (Sections II, III, and IV).
Zooming in further, we explore both ML and DL implementations within each application. Think of it like nested subsections (A and B within each section). This distinction lets us compare and contrast these approaches effectively.
Within the ML realm, we delve into three main branches: SL, USL, and RL. This breakdown helps us systematically examine how various ML techniques tackle different optical communication challenges.
Similarly, the DL world unfolds with its four main architectures: DNNs, RNNs, CNNs, and DRL. Each gets its own spotlight, highlighting its unique strengths and applications within optical communication.
For each subcategory, we begin with a visual representation of the ML or DL paradigm, followed by a comprehensive review of its existing implementations in OFC/OCN/OWC. This approach paints a clear picture of how diverse ML and DL applications empower different domains of optical communication.
Finally, Section V delves into insightful discussions, identifies key challenges, and charts future research directions. Section VI concludes the review, leaving you with a comprehensive understanding of this dynamic field.

\section{Optical fiber communication}
\subsection{Machine Learning}

\begingroup
\tabcolsep = 1.0pt
\def\arraystretch{1}
\begin{table*}
\tiny
\centering
\caption{Summary of ML applications in OFC.}
\begin{tabular}{|M{.5cm}|M{3cm}|M{3cm}|M{1cm}|M{3cm}|M{2.5cm}|M{2cm}|M{2.5cm}|}
\hline
Ref	& Performance	& Complexity	& Train/Test 	& Objective	& Input data & Metrics & Adopted algorithms\\ \hline

\cite{10} & Improves the sensitivities by 1.2 dB for 16-QAM, 1.3 dB for 64-QAM, 1.8 dB for 16-APSK and 1.3 dB for 32-APSK at BER of  1e-3& It has lower computational complexity than conventional linear detection schemes& Sim/EXP&Learn the relationship between received and transmitted symbols & Received signal&BER &k-means, Bit-based SVM \\ \hline

\cite{12} &Achieves BER of 1e-6 at SNR of 18 dB, which is a significant improvement over the BER of 1e-3 achievable with hard decision detector & It has 10 times lower complexity than conventional detector & SIM/SIM& Learn the relationship between received and transmitted symbols&Received signal & Average identification accuracy& HD detector, SVM detector\\ \hline

\cite{14} &Reduces the BER by up to a factor of 10 compared to conventional equalizers & Reduces complexity by 50\% over conventional equalizers& EXP/EXP&Compensate for the nonlinearity of the VCSEL link and improve the BER of the transmitted signal &Received signal & Sensitivity, Specificity, Accuracy, F1-score, BER&SVM-RBF, SVM-Linear, SVM-Poly \\ \hline

\cite{16} & Extends the optimum launched power by 2 dB compared to the Volterra&N-SVM has 6 times lower computational load compared to SVM& SIM/SIM& Compensate for the fiber nonlinearities& Received signal &BER, Computational complexity &Volterra, Fast Newton-based SVM NLE \\ \hline

\cite{39}  & Enhances Q-factor by 2 dB compared to Volterra& It has lower complexity than Volterra& SIM/SIM& Compensate for the fiber nonlinearities& Received signal& BER, Spectral efficiency&Conventional adaptive filter, ANN-based equalizer\\ \hline

\cite{64} & Provides up to 4 dB improvement in Q-factor compared to previously reported equalizers&The complexity of the RBFNN-NLE is independent of the number of nonlinear terms, while for Volterra it grows exponentially with the order of the nonlinear terms &SIM/SIM & Equalize the received signal by compensating for the impairments caused by nonlinearities&Received signal &BER &RBF-NN, volterra \\ \hline

\cite{43}  &Offers 1.5 dB improvement in OSNR compared to LMS & It is more complex than LMS equalizer&SIM/SIM &Classify the received signal points into different bit classes &Received signal &BER, Q-factor &ANN, volterra, Linear equalizer \\ \hline

\cite{44}  &Reveals Q-factor improvement of 3 dB and 1 dB with respect to the linear equalizer and Volterra, respectively &Reduces complexity by 50\% compared to the conventional linear equalizer &SIM/SIM &Classify the received signal points into different bit classes &Received signal &BER, SER, MSE, NMSE, EVM & ANN, volterra, DFE\\ \hline

\cite{57}   &Reduces the noise figure of EDFAs by up to 2 dB, improve the gain flatness of EDFAs by up to 3 dB, and reduce the power consumption of EDFAs by up to 10\% &Computes the gain adjustments in real time, even for large networks, while traditional methods may require significant time to compute EDFA gain adjustments &EXP/EXP &Predict the optimal gain and noise figure of the EDFA &Input power, Output power, Gain, Noise figure, Gain flatness &MSE, PAPR, BER &Proportional-integral-derivative control, RL \\ \hline

\cite{101}  & Achieves up to 0.7 dB improvement in Q-factor, compared to conventional detectors& It has lower computational cost than conventional approach and can be implemented in real time& EXP/EXP& Mitigate the impairments caused by nonlinearities in the system link& Received signal& BER, Q-factor, Eye diagram& Volterra equalizer NFT-based transmission scheme, Statistical OSNR estimation technique, kNN-based detector\\ \hline

\cite{102}  &Achieves 0.5 dB BER improvement in the 800-km SMF & It has low-complexity and zero-redundancy &EXP/EXP &Compensate for both linear and nonlinear impairments &Received signal &BER, Q-factor, Eye diagram &NDA-kNN, linear equalizer, volterra, DFE \\ \hline
 
\cite{106}  &Reduces the nonlinearity-induced signal distortion by up to 3 dB & It has low-complexity and can be easily implemented in real-time& SIM/SIM&Compensating inter/intra-channel Kerr nonlinearity in SDM based transmission &Received signal &BER, Q-factor, Computational complexity &Sparse Identification for Nonlinear Optics, Orthogonal Matching Pursuit, Simultaneous Perturbation Stochastic Approximation, Two-Step Iterative Soft-Thresholding \\ \hline

\cite{107a}  &Obtains 0.06 meters MSE and 0.08 meters MAE &Linear regression model predicts fault locations within minutes, compared to the hours or days using conventional methods &EXP/EXP & Predict the actual fault location for new fault events& optical time-domain reflectometry measurement, Cable type, Soil type, Deployment depth, Environmental factors &MAE, RMSE, R-squared & Linear regression, Single-layer perceptron neural network\\ \hline

\cite{20}  & Achieves a Q-factor improvement of up to 2.5 dB compared to Volterra& Fuzzy-Logic C-means has complexity of $O(NK)$, where N is the number of symbols in the received signal and K is the number of clusters, which is lower than other NLE techniques ($O(N^2)$)& SIM/SIM&Group the received signal points into different clusters based on their similarity and then map to the transmitted symbols & Received signal& BER, SNR, NMSE&Hierarchical clustering, Fuzzy-Logic C-means clustering, K-means, Fast-Newton SVM, ANN, volterra \\ \hline

\end{tabular}
\label{table:3}
\end{table*}
\endgroup

\begingroup
\tabcolsep = 1.0pt
\def\arraystretch{1}
\begin{table*}
\ContinuedFloat
\tiny
\centering
\caption{Summary of ML applications in OFC (cont.).}
\begin{tabular}{|M{.5cm}|M{3cm}|M{3cm}|M{1cm}|M{3cm}|M{2.5cm}|M{2cm}|M{2.5cm}|}
\hline
Ref	& Performance	& Complexity	& Train/Test 	& Objective	& Input data & Metrics & Adopted algorithms\\ \hline

\cite{117}  & Reduces the OSNR requirement up to 0.8 dB at BER of 1e-2 compared to the conventional QAM demodulation technique, for signals affected by inter-channel interference& It is a computationally efficient and well-suited for real-time applications& EXP/EXP& Identify the cluster centroids in the constellation of the received symbols in short time windows& Received signal& MSE, BER, Q-factor& K-means, LSTM\\ \hline

\cite{118}  & Increases the signal Q-factor by up to 2.158 dB compared to linear equalization at 500 km& It has complexity of $O(N^2)$, where N is the number of data points, while K-means has complexity of $O(KI)$, where K is the number of clusters and I is number of iterations &SIM/SIM & Group the received signal points into clusters based on their density& Received signal&BER, Q-factor &Linear equalizer, Fuzzy-logic C-means, Hierarchical, K-means \\ \hline

\cite{119}  &Achieves up to 0.4 dB Q factor improvement & It is relatively simple to implement and requires minimal computational resources& SIM/SIM& Classify the received signal points into different bit classes& Received signal& BER, Q-factor, Eye diagram, NMSE& RF, SVM, ANN\\ \hline

\cite{121}  &Achieves SNR improvement of up to 9 dB compared to conventional PCM-based D-radio over fiber systems & It is relatively simple, computationally efficient, and suitable for real-time implementation in D-radio over fiber systems &EXP/EXP &Groups the data points into a predefined number of clusters based on their similarity &Digitized RF signal &PAR, BER, Spectral efficiency, Complexity & PCM and uniform vector quantization, K-means-based multi-dimensional quantization\\ \hline

\cite{127}  & Reduces the MSE of phase noise estimation by up to 20 dB compared to time-domain approaches& It is up to 10 times faster than conventional methods&SIM/EXP &Estimate the amplitude and phase noise parameters of the laser &Time-domain signal of the laser under test & MSE, RMSE, Correlation coefficient& RF, SVM, XGB\\ \hline

\cite{132a}  & It is more robust to noise and non-linear distortions than CMA-based equalizer& It is less complex than CMA-based equalizer &SIM/SIM & Decompose the received signal into its constituent independent components& Received signal&BER, Q-factor, Eye diagram &Multi-tap ICA, CMA \\ \hline

\cite{132b}  & Achieves the same BER as conventional methods with 3 dB less SNR& It has higher complexity and memory requirement than CMA&EXP/EXP & Separate the received signal into two independent components, each of which corresponds to one of the transmitted polarization states& Received signal&BER, Q-factor, Complexity of medium &CMA, Decision-directed ICA \\ \hline

\cite{132c}  &Up to 3 dB reduction in BER & It is less complex than conventional TSs-based CEs &SIM/SIM & Separate the transmitted signal from the noise and impairments without using any training symbols& Received signal& BER, MSE& Maximization of negentropy, Maximum likelihood, Minimization of mutual information, Fast ICA\\ \hline

\end{tabular}
\label{table:3b}
\end{table*}
\endgroup

\begin{figure}
\centering
\includegraphics[scale=0.6]{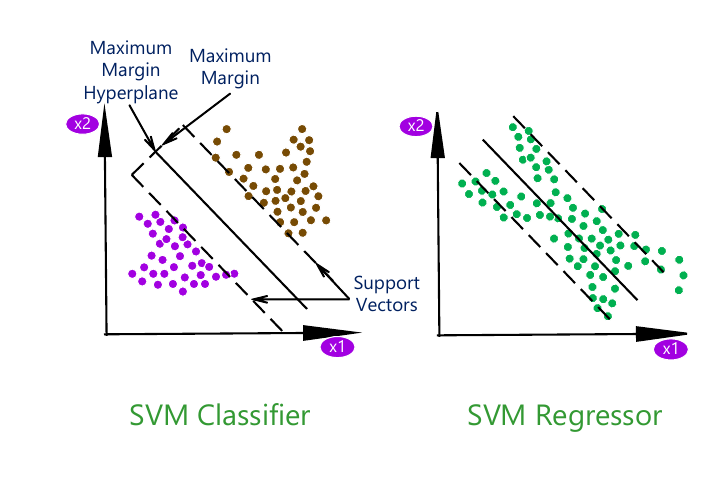}
\caption{Descriptions of SVM classifier and regressor.}
\label{fig:25}
\end{figure}

\textbf{Support Vector Machine Algorithm:} 
SVMs (Fig.  \ref{fig:25}) have emerged as one of the most widely used ML algorithms in OFC. Their versatility and effectiveness in addressing various challenges make them a valuable tool for enhancing the performance and reliability of optical communication systems.
SVMs effectively mitigate the impact of nonlinear phase noise (NLPN), ensuring accurate signal detection and transmission at high bit rates \cite{2, 3, 4}.
SVMs can compensate for laser phase noise, minimizing signal distortions and enhancing system stability \cite{5}.
SVMs can compensate for phase skew, modulator imperfections between in-phase and quadrature (IQ), improving signal integrity and reducing error rates \cite{5}.
SVMs can effectively handle modulator nonlinearity, preventing signal distortion and ensuring reliable data transmission \cite{6}.
SVMs can mitigate the effects of the fiber Kerr effect, which causes nonlinearities that limit transmission capacity.
SVMs can suppress amplified spontaneous emission (ASE) noise, improving signal quality and reducing error rates.
SVMs can effectively detect signals in both linear and nonlinear channels, ensuring reliable transmission over extended distances \cite{7}.

SVMs have also demonstrated remarkable performance in fault diagnosis, particularly in identifying and locating faults in optical systems. By transforming the classification task into a nonlinear programming problem \cite{8}, SVMs can accurately identify and localize faults, enabling timely intervention and restoration of service.
In the realm of signal detection, SVMs have been successfully applied to M-ary signal detection, which involves classifying the received bit sequences. This is typically achieved by employing $log_2(M)$ hyperplanes \cite{10} or transforming the problem into a multi-layer binary classification problem \cite{11}.
Experimental results have shown that SVM-based detectors outperform conventional detectors in various modulation schemes, such as 16-quadrature amplitude modulation (QAM), 64-QAM, 16-APSK, and 32-APSK \cite{10}. These improvements in signal detection translate into significant gains in system performance and reliability.
SVMs have also been employed in conjunction with constant module algorithm (CMA) equalizers \cite{6, 13} to enhance the performance of high-density carrier less amplitude phase (CAP) modulation. This approach demaps the rotated constellations directly without any correction, significantly reducing bit error rate (BER) compared to traditional hard decision methods \cite{12}.
Nonlinear equalizers (NLEs) is a crucial technique for compensating for nonlinearities in optical fiber channels, which can distort and degrade transmitted signals.
SVMs have also found widespread application in equalizing fiber linear and nonlinear effects, particularly in higher-order modulations where conventional equalizers struggle to meet performance requirements \cite{14, 15}. Experimental results have shown that SVM-based NLEs achieve at least 3.5 dB performance improvement at 60 Gbps and error-free transmission at 40 Gbps transmission rate, compared to Volterra-based NLE \cite{14}. Additionally, SVM-NLEs offer controllable computational complexity, making them suitable for optical link bandwidth design.
SVM-NLEs have been successfully applied to single/multi-channel WDM and OFDM systems in long-range communication \cite{16}-\cite{19}. They effectively tackle fiber nonlinearity, the interaction between nonlinear effects and stochastic noises, and inter-channel nonlinear crosstalk effects. SVM-NLEs outperform digital back-propagation (DBP) for multi-channel quadrature phase shift keying (QPSK) as DBP cannot handle these effects \cite{19}. In particular, SVM-NLEs improve the optimum launched power by 2 dB in 40 Gbps 16-QAM coherent optical-OFDM at 2000 km transmission, compared to Volterra-based NLE \cite{16}.

In conclusion, SVMs have emerged as a powerful tool for addressing various challenges in OFC, including signal detection, fault diagnosis, and equalization. Their versatility, effectiveness, and ability to handle complex scenarios make them an invaluable asset in the development of next-generation optical communication systems. As research continues to advance, SVMs are poised to play an even more significant role in shaping the future of OFC.

\begin{figure}
\centering
\includegraphics[scale=0.9]{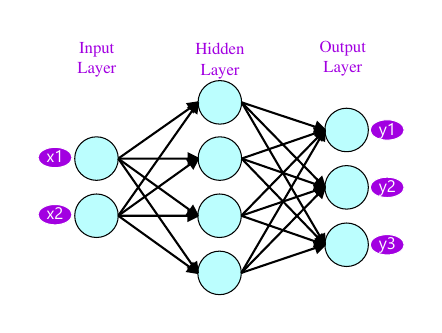}
\caption{Schematic of an ANN.}
\label{fig:35}
\end{figure}

\textbf{Artificial Neural Network Algorithm:} 
ANNs (Fig. \ref{fig:35}) have emerged as powerful tools for enhancing the performance of OFC systems. Their ability to learn complex relationships between input and output data makes them well-suited for tasks such as NLE, gain adjustment, and signal constellation shaping.
However, effectively utilizing ANNs in OFC systems requires careful consideration of their hyperparameters, which control the network's structure and learning process. These hyperparameters can be manually or automatically tuned to achieve optimal performance.
In \cite{39,64}, the authors emphasize the importance of hyperparameter tuning for ANN applications in OFC. They highlight the need for precise adjustments to ensure accurate fitting of the network to the training data. The layer type plays a crucial role in the ANN's ability to handle different data types. Radial basis function neural networks (RBFNNs) are suitable for linear data \cite{38}, while multi-layer perceptrons (MLPs) are better suited for nonlinear data \cite{40}.
A comprehensive study on ANN hyperparameter tuning for OFC is presented in \cite{81o}. This work provides valuable insights into optimizing ANN performance in OFC systems.
ANNs have demonstrated their effectiveness in performing NLE, achieving impressive performance improvements.
In \cite{39}, the authors demonstrate a 2 dB Q-factor enhancement for ANN-based NLE compared to Volterra-based NLE for a 40 Gbps 16-QAM coherent optical-OFDM system at 2000 km.
In \cite{38}, a RBFNN-based NLE is employed for a 16-QAM coherent optical-OFDM system. This approach achieves an 8.7 dB Q-factor at an optimum launched power of -3 dBm and a transmission distance of 1200 km, demonstrating the potential of RBFNN-based NLE for increasing data rate and reach.
The ANN-based NLE can effectively mitigate inter/intra-band cross/self-phase modulation interference, enhancing the nonlinearity tolerance of coherent optical-OFDM systems. In \cite{43}, ANN-based NLE is shown to double the transmission distance (up to 320 km) for a coherent optical-OFDM system compared to the least mean square (LMS) algorithm, with a 0.7 dB improvement in optical signal to noise ratio (OSNR).
Furthermore, ANN-based NLE has demonstrated significant performance improvements over linear equalization and Volterra-based NLE for 80 Gbps 16-QAM coherent optical-OFDM systems. In \cite{44}, ANN-based NLE achieves Q-factor improvements of 3 dB and 1 dB, respectively, compared to linear equalization and Volterra-based NLE, after transmission distances of 1000 km.
Amplification gain adjustment is another crucial aspect of OFC systems, particularly in the context of power constraints. In \cite{57}, the authors propose an ANN-based approach for optimizing the gain of cascaded amplifiers to achieve predefined input and output powers for the entire link. This method minimizes noise figure and gain flatness, improving overall system performance.
The proposed ANN-based gain adjustment scheme results in a noise figure of 30.06 dB and a gain flatness of 5.26 dB for a 6-amplifier system, while maintaining input and output powers around 3 dBm with less than 0.1 dB error.

These examples illustrate the remarkable potential of ANNs for enhancing the performance of OFC systems. As research continues, ANN-based techniques are expected to play an increasingly prominent role in optimizing signal transmission, improving spectral efficiency, and enabling more robust and reliable OFC.

\begin{figure*}
\centering
\includegraphics[scale=0.5]{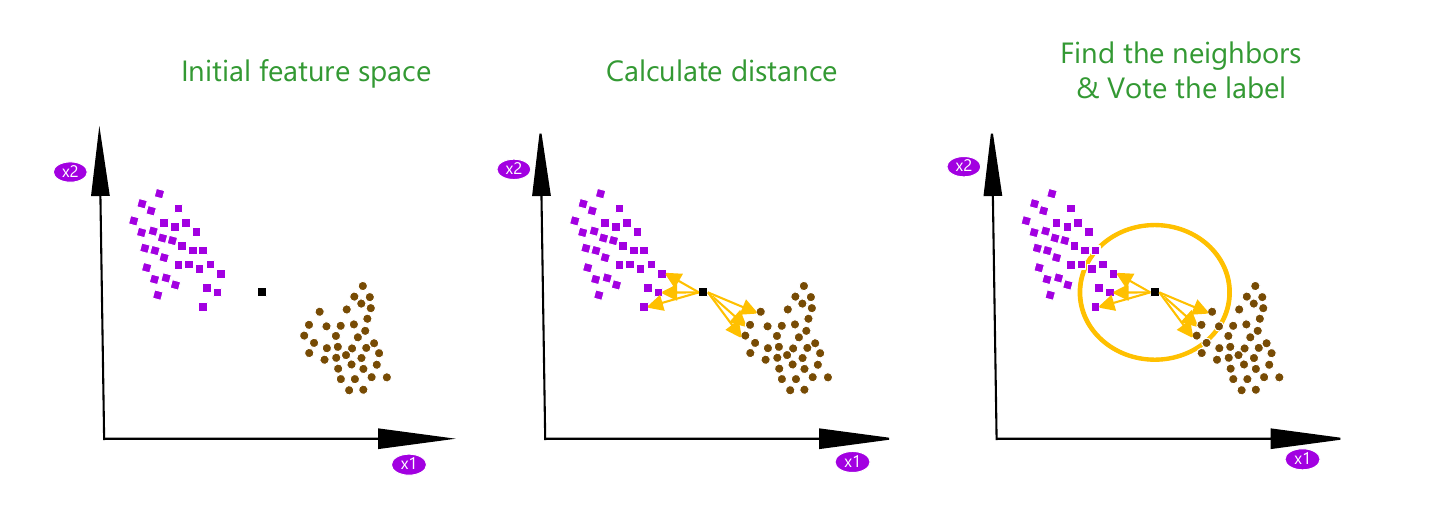}
\caption{Visualization of kNN classification procedure.}
\label{fig:26}
\end{figure*}

\textbf{k-Nearest Neighbors Algorithm:} kNN algorithm (Fig. \ref{fig:26}) is effectively applied in \cite{100} as a detector to deal with circularly symmetric non-Gaussian impairments such as NLPN (in dispersion unmanaged links), I/Q imbalance, and ASE noise in a 112 Gbps 16-QAM optical phase conjugation system with 800-km dispersion-managed and 1280-km dispersion-shifted-fiber links. 
If training points are distributed non-uniformly, the input point might be located to one point of the nearest class and several points of a farther class and finally allocated to the farther class. 
To deal with this issue, authors of \cite{101} assigned distance-related weights to each of the surrounding points and then applied kNN. 
Compared with the maximum likelihood post-compensation approach, in the presence of fiber impairments, the kNN detector enhances the linewidth tolerance by 180 kHz and improves the nonlinear tolerance for the dispersion neglected, managed, and unmanaged links by 1.7, 1.0, and 0.4 dBm, respectively. 
The kNN dependency on the amount of input data and the $k$ value leads to high complexity. As an alternative, the density parameter of the test data can be used to rapidly extract the center noiseless data and label them as the classification references, then apply the kNN to classify the remaining test data. Authors of \cite{102} applied this approach and achieved efficient nonlinearity mitigation with low complexity and zero data redundancy. 
Experimental results show that this method achieves 0.5 and 2 dB BER improvement in 800-km SMF 16-QAM and 80-km SMF 64-QAM transmission systems, respectively, and is robust to noise and has fast convergence.

\begin{figure}
\centering
\includegraphics[scale=0.5]{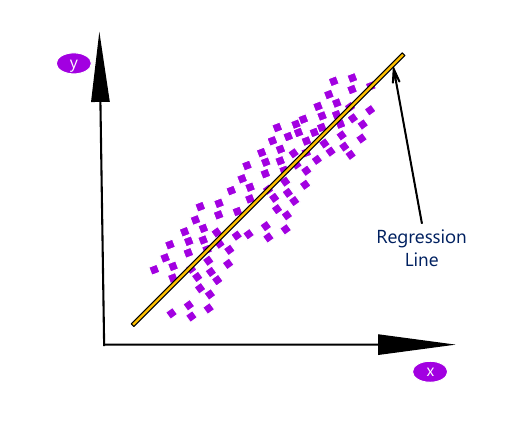}
\caption{An example of linear regression algorithm.}
\label{fig:27}
\end{figure}

\textbf{Regression Algorithms:}
Regression algorithms (Fig. \ref{fig:27}), such as linear regression, least absolute shrinkage selection operator (Lasso) regression, and ridge regression, have been widely applied in various optical communication applications.
Efficient compensation of inter/intra-channel Kerr nonlinearity is crucial for achieving high-performance SDM-based transmission. One approach to address this challenge is to utilize sparsity-promoting techniques, such as Lasso regression.
Lasso regression, a type of linear regression that incorporates a penalty term to encourage fewer coefficients to be nonzero, has demonstrated its effectiveness in identifying the interaction between symbols in different time slots and selecting the minimum number of relevant perturbation terms for compensating inter/intra-channel Kerr nonlinearity.
Lasso regression's low complexity makes it suitable for real-time implementation in SDM transmission systems. Additionally, its sparsity-inducing property ensures that only the most significant perturbation terms are employed, reducing computational overhead and improving overall system efficiency.
The application of Lasso regression for compensating inter/intra-channel Kerr nonlinearity in SDM-based transmission has been demonstrated in several studies \cite{106, 107}. These studies have shown that Lasso regression can effectively mitigate the detrimental effects of nonlinearity, enabling high-capacity and high-quality SDM transmission over long distances.

As SDM technology continues to evolve, Lasso regression is expected to play an increasingly important role in compensating inter/intra-channel Kerr nonlinearity and enabling the realization of next-generation high-speed optical communication systems.

\begin{figure}
\centering
\includegraphics[scale=0.35]{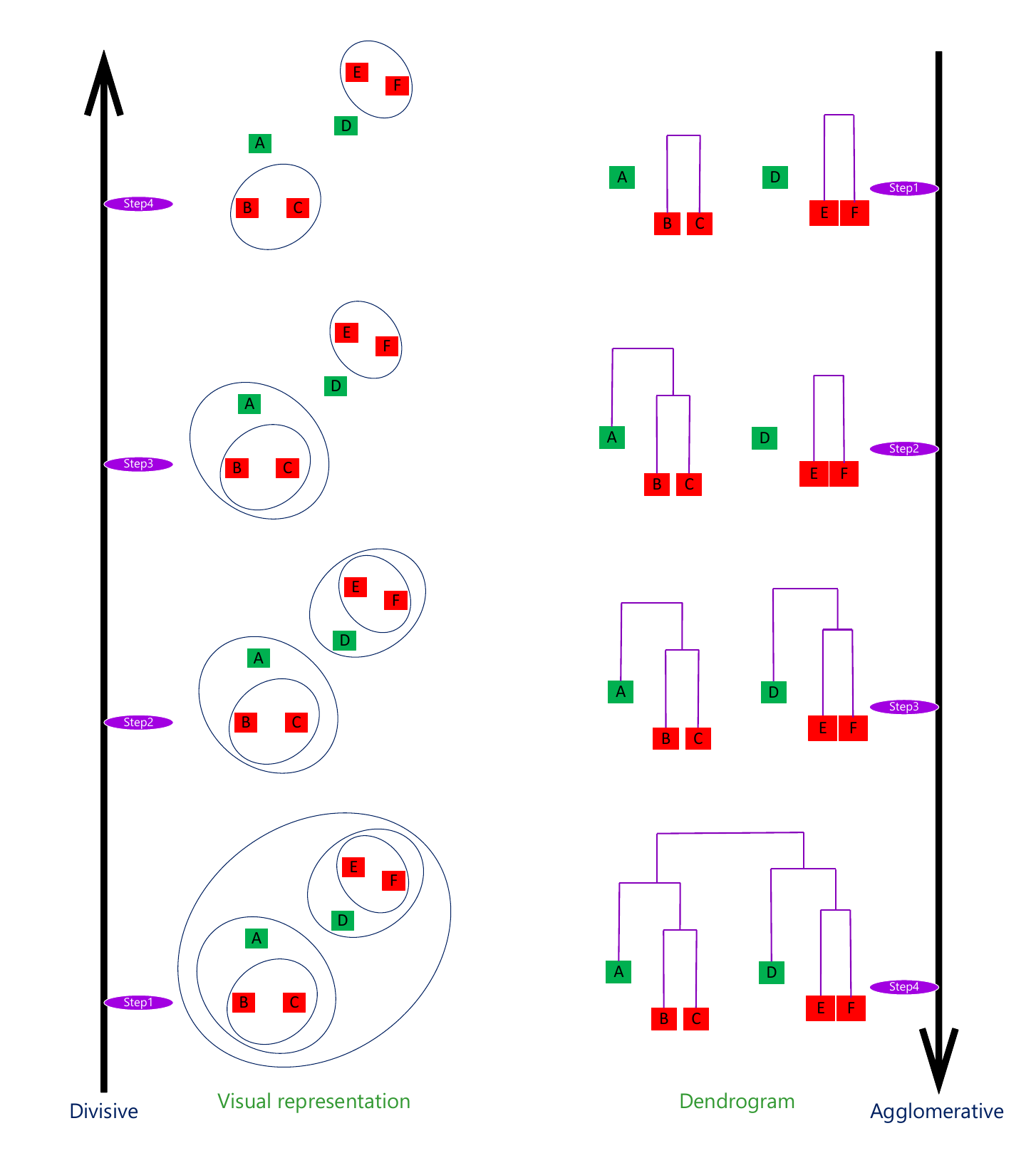}
\caption{Demonstration of hierarchical clustering.}
\label{fig:28}
\end{figure}

\textbf{Hierarchical Clustering Algorithm:} 
Hierarchical clustering (Fig. \ref{fig:28}),is a powerful USL algorithm that groups similar data points into a hierarchical structure. In \cite{116}, the authors explored the application of hierarchical and Fuzzy-Logic C-means clustering for NLE. A recent study \cite{20} provides compelling experimental evidence that Fuzzy-Logic C-means outperforms hierarchical clustering in terms of NLE performance.
For low-level modulation formats, Fuzzy-Logic C-means exhibits superior performance at the optimum launched power, while hierarchical clustering performs more effectively at high launched powers. This finding suggests that Fuzzy-Logic C-means is more robust to variations in launched power, ensuring consistent performance across a broader range of operating conditions.
Furthermore, Fuzzy-Logic C-means demonstrates significant performance gains over other NLE algorithms, including Kmeans, fast-Newton SVM, ANN, and Volterra-based NLE. These gains are particularly evident for BPSK and QPSK modulation formats, with Q-factor improvements of 2.5 dB and 0.4 dB, respectively.

The superior performance of Fuzzy-Logic C-means clustering for NLE stems from its ability to capture complex nonlinearities in optical channels and adapt to dynamic channel conditions. This enhanced performance makes Fuzzy-Logic C-means a promising candidate for real-world NLE applications in optical communication systems.

\begin{figure*}
\centering
\includegraphics[scale=0.5]{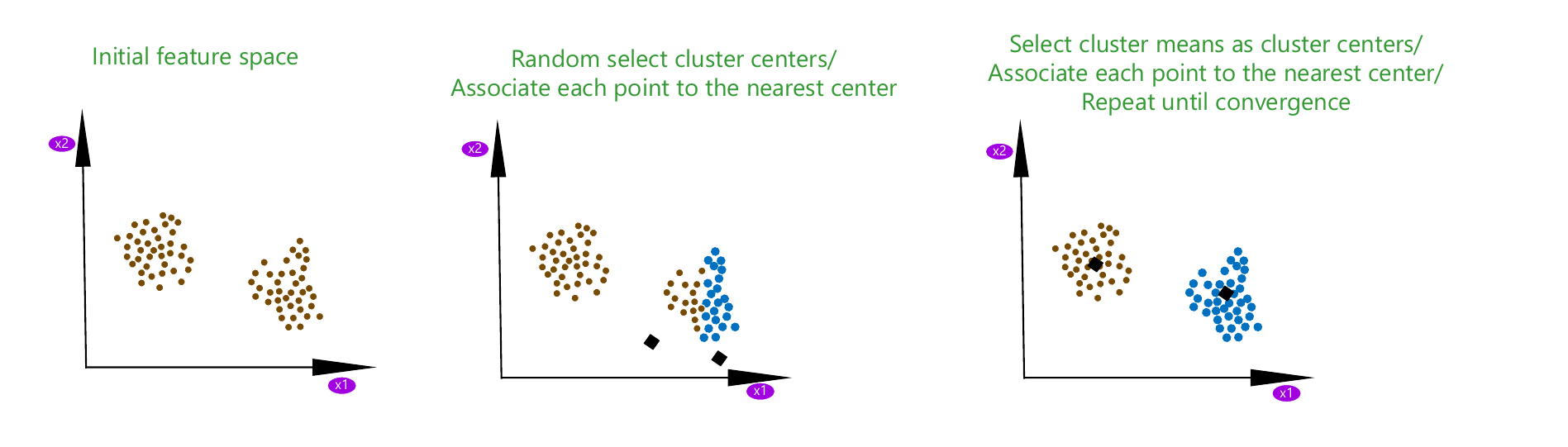}
\caption{Example of K-means clustering.}
\label{fig:29}
\end{figure*}

\textbf{K-Means Clustering Algorithm:} 
K-means clustering (Fig. \ref{fig:29}), a versatile ML algorithm, has emerged as a powerful tool for addressing signal distortion and enhancing performance in various optical communication applications. Its ability to identify patterns and group data points based on their similarities makes it well-suited for mitigating fiber nonlinearity effects, optimizing transmission strategies, and enabling efficient resource management.
In \cite{117}, the authors employed time windowing within the K-means algorithm to dynamically track signal variations in multi-carrier 16-QAM Nyquist-WDM systems. This approach effectively compensated for time-varying distortions caused by I/Q imbalance, NLPN, and inter-channel interference, resulting in reduced OSNR requirements and improved BER performance.
In \cite{118}, authors introduced a modified version of the density-based clustering algorithm algorithm, which incorporates K-means clustering on the noisy “un-clustered” symbols. This approach demonstrated superior Q-factor performance compared to linear equalization for 24.72 Gbps DQPSK coherent optical-OFDM transmission over 500 km fiber links.
To mitigate fiber non-Gaussian nonlinear effects in dispersion-managed and unmanaged fibers, \cite{119} employed a Parzen window-based detector to design decision boundaries. This method achieved Q-factor improvements up to 0.4 dB, highlighting the effectiveness of K-means clustering in addressing nonlinearities.
The K-means algorithm has also been successfully applied for spectrally efficient digitized radio over fiber transmission \cite{120,121}. By grouping highly correlated neighboring data points into multi-dimensional vectors and adopting K-means clustering for quantization, this approach achieved significant SNR improvements compared to pulse coding modulation, enabling transmission at 30 Gbps with EVMs of 8\% and 1\% for 4 and 7 quantization bits, respectively.
In optical phase-modulated radio over fiber systems, \cite{122} and \cite{123} demonstrated the efficacy of K-means clustering for RF phase recovery. Authors of \cite{122} successfully demodulated three-channel, 8-PSK subcarrier multiplexed signals at 5 GHz and 40 km transmission. Furthermore, \cite{123} experimentally demonstrated efficient signal demodulation of three-channel WDM signals with 6-GHz RF carrier frequency, at 2.5 Gbps per channel and 78.8 km transmission.

These examples highlight the versatility and effectiveness of K-means clustering in addressing various challenges in optical communication systems. Its ability to group data points based on their similarities makes it a valuable tool for mitigating signal distortions, optimizing transmission strategies, and enabling efficient resource management. As optical communication technology continues to evolve, K-means clustering is likely to play an increasingly prominent role in enhancing system performance and reliability.

\begin{figure*}
\centering
\includegraphics[scale=0.5]{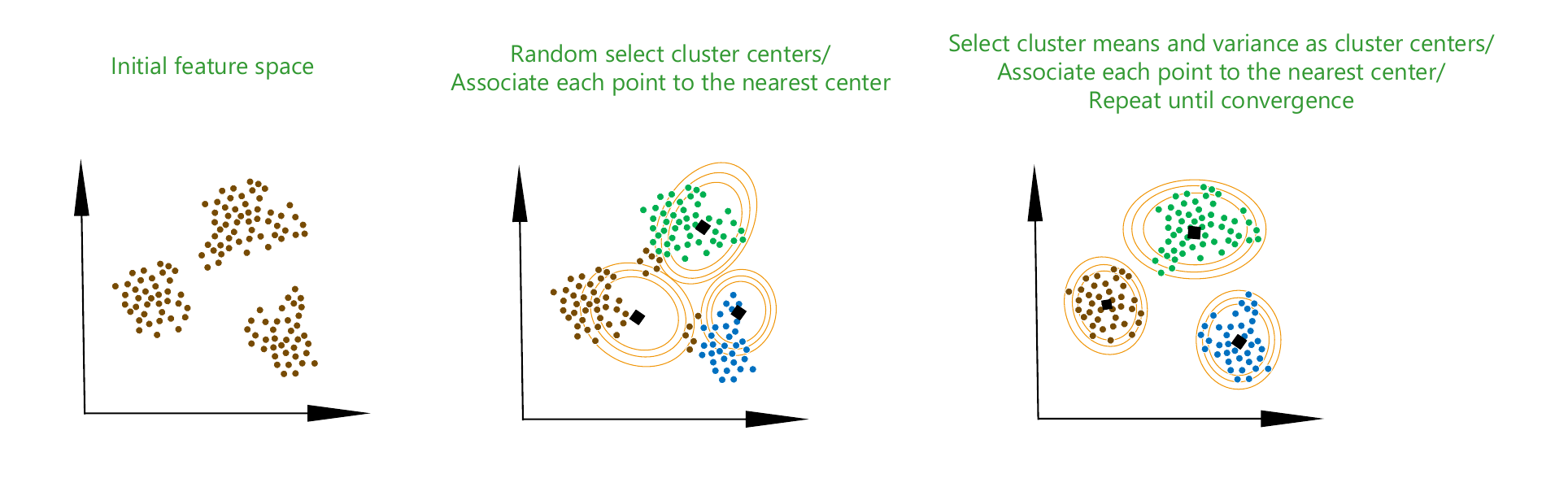}
\caption{Example of EM clustering.}
\label{fig:30}
\end{figure*}

\textbf{Expectation Maximization Clustering Algorithm:} 
EM clustering (Fig. \ref{fig:30}) has emerged as a promising technique for mitigating laser amplitude and phase noise in optical communication systems. This method leverages Bayesian analysis to evaluate the content of an incoming optical signal and determine the probability that it represents amplitude or phase noise.
In \cite{127,128}, simulations and experiments demonstrated the effectiveness of EM clustering for noise mitigation. The approach was shown to improve the launch power tolerance by 1.5 dB and reduce the OSNR requirement by 0.3 dB, enabling more robust and efficient optical transmission.
Moreover, \cite{128a} explored the application of EM clustering in conjunction with a soft decision error-control code. This combination enabled iterative mitigation of NLPN, further enhancing the performance of optical systems.

The integration of EM clustering into optical communication systems offers a valuable tool for combating laser noise and achieving improved signal quality. As research continues, this technique is expected to play an increasingly significant role in optimizing optical transmission and enabling the next generation of high-speed communication networks. In summary, EM clustering presents a promising approach for mitigating laser amplitude and phase noise. Its ability to identify and counteract these noise sources holds the potential to significantly enhance the performance and reliability of optical communication systems. As research progresses, EM clustering is poised to become an essential tool in the optimization of optical transmission and the development of future optical communication networks.

\begin{figure}
\centering
\includegraphics[scale=0.6]{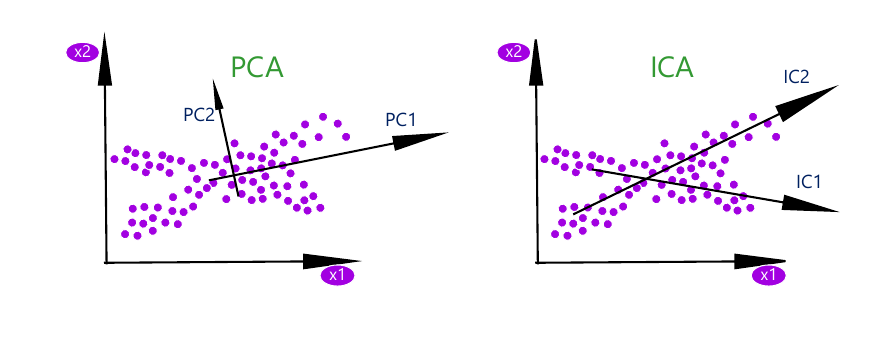}
\caption{Example of data fitted using PCA and ICA.}
\label{fig:22}
\end{figure}

\textbf{Independent Component Analysis Algorithm:} 
ICA (Fig. \ref{fig:22}) has emerged as a promising technique for blind equalization and phase recovery in coherent transmission. By exploiting the statistical independence of the channel and signal components, ICA can effectively mitigate the effects of PMD and PDL in coherent optical communication systems.
In \cite{132a}, ICA is employed for blind equalization and phase recovery in coherent transmission. Simulation results show that ICA achieves similar performance to conventional CMA equalization in QPSK and outperforms CMA in 16-QAM, demonstrating its effectiveness in mitigating PMD and PDL, especially in high-order modulation formats.
Authors of \cite{132b} applied ICA for polarization demultiplexing, a crucial task in coherent optical systems. This method addresses the singular convergence problem of the CMA algorithm while maintaining comparable polarization tracking performance. ICA's ability to separate and decouple polarization components proves to be effective in robust polarization demultiplexing.
ICA is further extended in \cite{132c} for channel equalization in DDO-OFDM and PM coherent optical-OFDM systems. Without the need for conventional training symbols, the ICA-based equalizer achieves performance comparable to training symbol-based channel estimation methods. This remarkable capability enables blind equalization in these systems, enhancing their flexibility and adaptability.

The use of ICA for blind equalization and phase recovery, polarization demultiplexing, and channel equalization in diverse optical communication systems demonstrates its versatility and effectiveness in addressing challenges related to signal impairments and complexity. As ICA research continues to advance, its role in optical communication is expected to grow even more prominent, paving the way for more robust, efficient, and flexible optical communication systems.

\subsection{Deap Learning}
\begin{figure}
\centering
\includegraphics[scale=0.9]{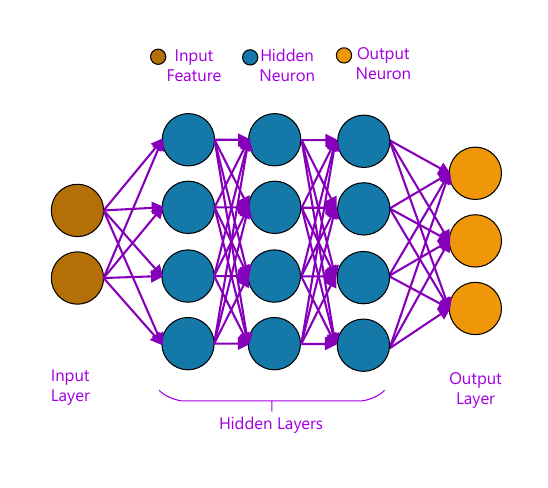}
\caption{Schematic of DNN.}
\label{fig:11a}
\end{figure}
\begin{figure}
\centering
\includegraphics[scale=0.9]{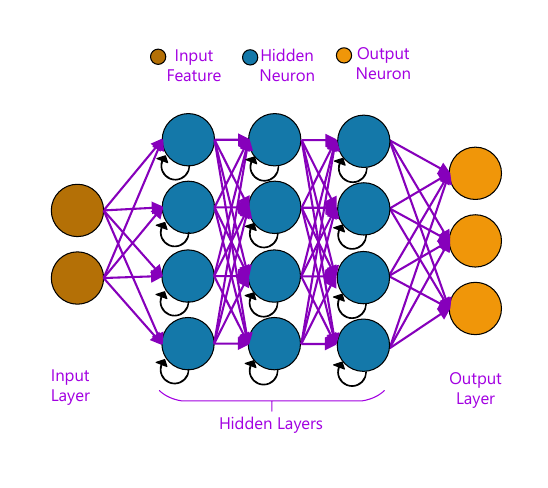}
\caption{Schematic of RNN.}
\label{fig:11c}
\end{figure}
\begin{figure*}
\centering
\includegraphics[scale=0.7]{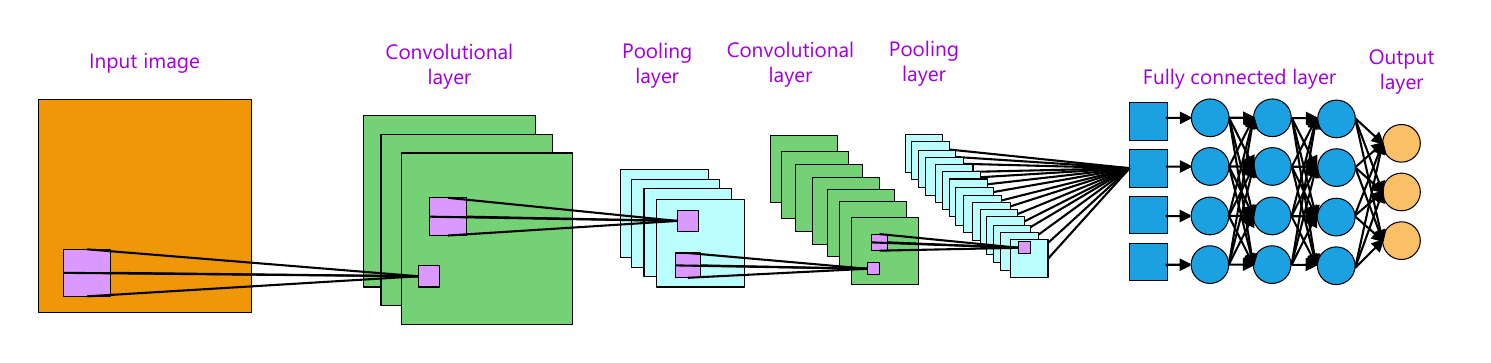}
\caption{Structure of CNN algorithm.}
\label{fig:11d}
\end{figure*}

\begingroup
\tabcolsep = 1.0pt
\def\arraystretch{1}
\begin{table*}[tp!]
\tiny
\centering
\caption{Summary of DL applications in OFC.}
\begin{tabular}{|M{.5cm}|M{3cm}|M{3cm}|M{1cm}|M{3cm}|M{2.5cm}|M{2cm}|M{2.5cm}|}
\hline
Ref	& Performance	& Complexity	& Train/Test 	& Objective	& Input data & Metrics & Adopted algorithms\\ \hline
 
\cite{76}  &Reduces SNR penalty by 2 dB &It is 1000 times faster than DBP at 3200 km transmission & SIM/EXP& Learn the inverse of the nonlinear Schrödinger equation, which is the equation that describes the propagation of light in a fiber-optic link& Received signal&BER, SER, Q-factor, Complexity &DNN, DBP, FFNN, CNN, RNN \\ \hline

\cite{77}  &Outperforms the Semi-Definite Relaxation Row-by-Row detector by up to 0.5 dB in BER & It has a computational complexity of $O(N^2)$, where N is the number of modes in the fiber which is significantly lower than the computational complexity of the zero forcing and SDR-RBR detectors ($O(N^3)$ and $O(N^4)$)&SIM/EXP & Detect the transmitted signal from the received signal in a new MDM optical transmission system& Received signal&BER, Q-factor, MAE, RMSE, MSE & Zero forcing, Semi-definite relaxation row-by-DNN\\ \hline

\cite{81}  &Achieves BERs below the 6.7\% HD-FEC threshold at distances beyond 40 km, which is a significant improvement over conventional IM/DD systems, which typically require error correction at much shorter distances &It is less complex than traditional methods &SIM/EXP &Recover the original data &A set of optical signals that are generated at the transmitter &BER, SER, NMSE, Capacity, Spectral efficiency, Energy efficiency, Complexity &MLP, CNN, RNN \\ \hline

\cite{88}  &Predicts the Raman gain profile with an accuracy of over 99\%, which is significantly higher than the accuracy of traditional design methods (80\%) & It is significantly less complex than traditional methods, which require a deep understanding of the physics of Raman scattering&SIM/EXP &Learn the mapping between the Raman gain profile and the pump wavelength and power &Raman gain profile, Pump wavelength, Pump power, Fiber parameters &MSE, Latency, Complexity &DNN, SVM, DTs \\ \hline

\cite{88i}  &Mitigates the nonlinearity by gains up to 0.13 bit/4D &Reduces computational complexity of conventional constellation shaping techniques &SIM/EXP &Generate constellation shapes that optimize the transmission performance over the FMF channel &A large set of training symbols is generated, taking into account the FMF channel characteristics &BER &Uniform distributed QAM, Maxwell-Boltzmann distributed QAM, DNN-based GCS \\ \hline

\cite{88a}&Reduces the BER by 20\% compared to a conventional equalizer &It is 50\% less computationally complex than the conventional equalizer &EXP/EXP &Learn the temporal correlations in the received signal, which allows it to reduce the number of input symbols required for equalization &Received signal &BER, Q--factor & LSTM, GRU, Bi-LSTM\\ \hline

\cite{88f}  &Reduces the number of FLOPs by 34\% compared to the pure RNN approach & It has low computational complexity, making it suitable for real-time applications& SIM/SIM&Learn the relationship between the channel parameters, the transmitted symbols, and the received symbols &Received signal &MSE, MAE, BER &FNN, RNN, CNN, CNN-RNN \\ \hline

\cite{88h}  &Reduces the BER 100 times & Reduces the complexity by up to 60\% compared with traditional non-cascade FNN/RNN-based equalizers& SIM/EXP&Compensate for the linear and nonlinear impairments &Received signal &BER, NMSE, EVM &No equalization FFE, Traditional non-cascade FNN/RNN, Cascade FNN/RNN, Combination of FFE and traditional non-cascade FNN/RNN \\ \hline

\cite{88m}  &Reduces the number of required epochs and training symbols to train an equalizer by up to 90\% and 62.5\%, respectively & It has relatively low complexity, making it a practical solution for real-time applications& SIM/EXP& Equalize the received symbols from an optical link& A set of training symbols that are transmitted over an optical link &MSE, BER, NMSE, Q-factor, EVM &FNN, RNN \\ \hline

\cite{88d}  & Reduces the BER of an optical fiber communication system by up to 10 dB&Reduces the complexity of an optical transceiver by up to 50\% &SIM/EXP & Predict the received signal from the transmitted signal and channel model& Transmitted signal, Channel model& BER, Spectral efficiency&FFNN, bidirectional RNN \\ \hline

\cite{92a}  &Achieves a similarity of 97.0\% and 85.6\% for the intensity and phase patterns, respectively, while the IGP algorithm achieves a similarity of 92.0\% and 75.0\%, and the SVM method achieves a similarity of 94.0\% and 80.0\% &Decomposes a 10-mode fiber in 0.02 seconds, while the IGP algorithm takes 100 seconds and the SVM method takes 10 seconds &SIM/SIM &Reconstruct the modal coefficients of the optical fiber output &The near-field intensity pattern of the optical fiber output, The phase information of the optical fiber output &MSE, Correlation coefficient &Gerchberg-Saxton, CNN, Genetic \\ \hline

\cite{92c}  &Achieves BER reduction of 100 times for two-mode groups & It has complexity of $O(N^2)$, where N is the number of coefficients, which is much lower than the complexity of the Volterra equalizer $O(N^3)$ &SIM/EXP &Compensate for the nonlinear impairments that occur in the OAM-MDM IM/DD transmission system & Received signal&BER, Q-factor, Spectral efficiency, Complexity &CNN, volterra, Linear equalizer \\ \hline

\cite{150a}  & Reduces BER by up to 50\% compared to greedy search&Reduces equalizer complexity by up to 80\% compared to greedy search & SIM/SIM& Automatically optimize the structure of the Volterra equalizer for any given channel and signal parameters& Structure parameters of the Volterra equalizer, Channel characteristics, Signal parameters& BER, Equalizer's complexity&Auto-Volterra, Greedy search, volterra-Pruning \\ \hline

\end{tabular}
\label{table:4}
\end{table*}

\textbf{Deep Neural Network Algorithm:} 
DNNs, a type of ANN with multiple layers (Fig. \ref{fig:11a}), have emerged as powerful tools for modeling complex relationships between inputs and outputs. Grid search is the most widely used method for DNN hyperparameter tuning, but it is computationally expensive. To address this challenge, two suboptimal manual search techniques with high efficiency were proposed in \cite{81o}.
DNN algorithms are well-suited for equalization in optical communication systems due to their ability to learn complex relationships. In \cite{73}, DNNs were employed for compensation of fiber linear and nonlinear effects with low complexity. This approach effectively addressed the interplay of deterministic and stochastic nonlinearities \cite{75}.
The DNN method can effectively compensate for nonlinear inter-carrier cross-talk even in the presence of frequency stochastic variations. By applying the DNN to the split-step Fourier method, also known as DNN-based DBP, the complexity was significantly reduced compared to conventional DBP \cite{76}. This was demonstrated in a 32x100 km fiber-optic link, where the DNN-based DBP exhibited lower complexity compared to the conventional DBP.
In \cite{77}, DNNs were applied for implementing the inverse channel matrix in mode division multiplexing (MDM)-based transmission. Experiments revealed that DNN outperformed conventional MIMO detectors with zero forcing detector and Semi-Definite Relaxation Row-by-Row.
DNN-based modeling offers high accuracy, low latency, and low complexity, making it well-suited for modeling optical modulators \cite{81}. Experimental results showed that DNN could achieve BER below the 6.7\% hard-decision FEC threshold with a data rate of 42 Gbps at distances beyond 40 km. Moreover, DNN outperformed conventional intensity modulation/direct detection (IM/DD) pulse amplitude modulation (PAM)2/PAM4 with the feedforward equalizer.
DNNs can effectively learn the relationship between Raman gain profile and pump powers and wavelengths, accurately predicting the gain distribution in optical amplifiers. This capability has led to the development of DNN-based optical amplifier models that achieve high accuracy, with maximum error of 0.6 dB \cite{88}.
However, using DNNs solely at the receiver side can result in suboptimal solutions. To address this limitation, end-to-end DNNs have been proposed, which incorporate both the transmitter and receiver sides of the optical communication system. These DNNs optimize the transceiver configuration by minimizing a loss function, leading to significantly improved performance.
End-to-end DNNs have demonstrated their effectiveness in various optical communication tasks, including autoencoding \cite{88i}-\cite{83}, constellation shaping \cite{84, 85}, and power adjustment \cite{86}. Autoencoding algorithms have been developed for probabilistic, geometric, and joint geometric, probabilistic constellation shaping in MDM transmission, demonstrating improved performance over traditional methods.
Moreover, end-to-end DNN-based constellation shaping has been shown to effectively mitigate the nonlinearity in optical communication systems, achieving gains up to 0.13 bit/4D \cite{88i}. This highlights the potential of DNNs to revolutionize constellation shaping and enhance the overall performance of optical communication systems.

\textbf{Recurrent Neural Network Algorithm:} 
RNNs (Fig. \ref{fig:11c}) have emerged as powerful tools for nonlinear equalization in optical communications. Their ability to handle temporal dependencies makes them well-suited for compensating for the nonlinear effects that arise in optical channels.
In \cite{88a}-\cite{88c}, RNNs were employed as nonlinear equalizers in 56 Gbps 4-PAM MDM-based transmission systems. The authors experimentally demonstrated that RNNs could achieve the same BER performance as ANNs while offering a 70\% reduction in computational complexity.
In \cite{88f}, a hybrid CNN-RNN-based equalizer was developed for 100 Gbps 16-QAM 2000 km transmission systems. The hybrid approach demonstrated comparable BER performance to DBP equalization with significantly fewer floating-point operations compared to end-to-end MLP, CNN, RNN, and bidirectional RNN models.
Authors of \cite{88h} proposed cascade DNN-RNN-based equalizers to compensate for both linear and nonlinear effects in 100 Gbps 4-PAM systems at 15 km transmission distances. Experimental results demonstrated that the cascade approach outperformed the combination of feedforward and non-cascade DNN-RNN-based equalizers and achieved a BER lower than the 7\% HD-FEC threshold with received optical power greater than 5 dBm.
However, a significant limitation of RNN-based nonlinear equalizers is their reliance on a specific operating condition. This makes them unsuitable for dynamic optical networks, such as data center networks, where optical links may undergo frequent reconfiguration.
To address this challenge, researchers have proposed transfer learning-aided DNN/RNN-based algorithms that can be trained on a small amount of data from a similar condition and then adapt to a new operating condition with minimal retraining. In \cite{88m}, a transfer learning-aided DNN/RNN-based algorithm was demonstrated to achieve a reduction of 90\%/87.5\% in training epochs and 62.5\%/53.8\% in dataset size compared to training from scratch.
RNNs have also been applied for auto-encoding in optical communications. In \cite{88e}, an RNN-based autoencoder was developed for communication over dispersive nonlinear channels. The autoencoder was shown to improve the performance of the IM/DD system compared to DNN.
Meanwhile, authors of \cite{88d} presented a bidirectional RNN based autoencoding algorithm for communication over dispersive channels with IM/DD. This approach was verified to further improve the performance of the IM/DD system compared to DNN.
Finally, in \cite{88g}, researchers investigated the use of different RNN architectures, including LSTM, GRU, and Vanilla-RNN, for compensating fiber nonlinear effects. The results indicated that Vanilla-RNN had the lowest complexity among the tested architectures.

Overall, RNNs have shown promising performance for nonlinear equalization and auto-encoding in optical communications. Their ability to handle temporal dependencies makes them well-suited for addressing the challenges of nonlinear effects and dynamic operating conditions in optical networks. As RNN research continues, we can expect further advancements in their application to improve the performance and efficiency of optical communication systems.

\textbf{Convolutional Neural Network Algorithm:} 
CNNs (Fig. \ref{fig:11d}) have emerged as promising tools for addressing challenges in orbital angular momentum (OAM) communication systems. In \cite{92a}, CNNs are employed to predict the mode coefficients of an OAM signal by analyzing interference patterns extracted from distribution histograms. The authors developed simulations demonstrating high reconstruction accuracy, achieving 97.0\% and 85.6\% similarities between reconstructed and given intensity and phase patterns, respectively.
In \cite{92c}, CNNs are applied for nonlinearity equalization in an OAM system with IM/DD. Experiments show that the CNN-based equalization outperforms conventional Volterra-based NLE by 3 dB and 1.5 dB in two-mode groups, achieving significantly improved signal quality.

The application of CNNs in OAM communication systems holds immense potential for enhancing performance and robustness. These networks can be leveraged for various tasks, including mode classification, nonlinearity compensation, and error correction, enabling more efficient and reliable OAM signal transmission. As research in this area continues to progress, CNNs are expected to play an increasingly crucial role in shaping the future of OAM communication.

\begin{figure}
\centering
\includegraphics[scale=0.7]{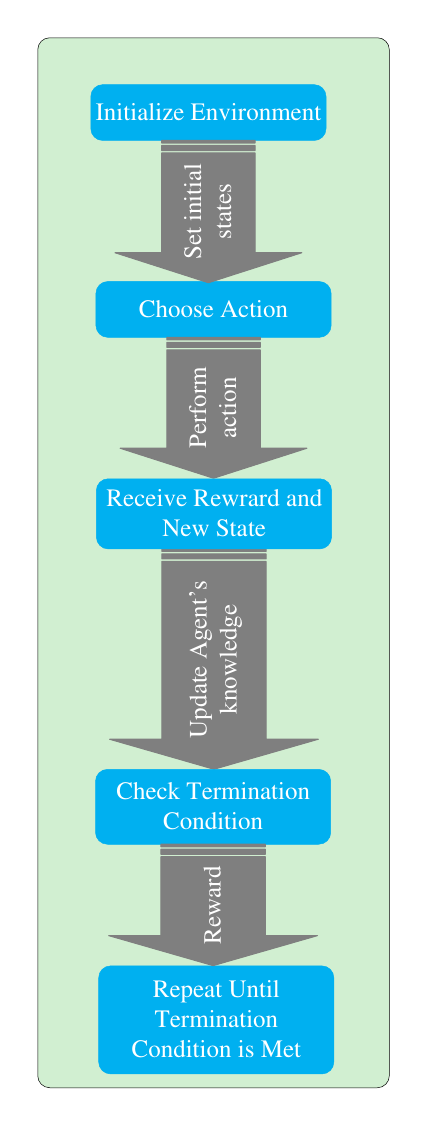}
\caption{Schematic of DRL algorithm.}
\label{fig:34}
\end{figure}

\textbf{Deep Reinforcement Learning Algorithms} 
DRL (Fig. \ref{fig:34}) combines RL with DL, enabling agents to make decisions from unstructured input data without manual engineering of the state space. Volterra-based NLE, a powerful equalization technique in optical communications, offers significant performance gains but comes with a high computational complexity. Traditional blind optimization methods for Volterra-based NLE are often inefficient and can result in multiple suboptimal structures.
In \cite{150a}, DRL is introduced as an efficient approach to optimize the structure of Volterra-based NLE. DRL learns the functional relationship between optimization objectives and the structure parameters, enabling it to search for the optimal structure directly without relying on exhaustive search or heuristics.
DRL and greedy search are experimentally applied to three Volterra-based equalization schemes: Volterra-FFE, volterra-DFE, and Volterra-Pruning, for a 50 Gbps PAM-4 IM/DD system. The results demonstrate that DRL outperforms greedy search in terms of complexity reduction and performance improvement. With little or no BER loss, DRL can significantly reduce the computational complexity of Volterra-based equalization compared to greedy search. Additionally, DRL-based pruning techniques enhance the efficiency of Volterra-based equalization by removing redundant or unnecessary taps from the Volterra filter, further reducing complexity without compromising performance.
The application of DRL for Volterra-based NLE optimization showcases the potential of AI techniques to enhance the performance and efficiency of optical communication systems. As DRL research continues to advance, its role in optimizing complex communication algorithms is expected to grow even more prominent in the future.

\section{Optical communication networks}
\subsection{Machine Learning}

\begingroup
\tabcolsep = 1.0pt
\def\arraystretch{1}
\begin{table*}
\tiny
\centering
\caption{Summary of ML applications in OCN.}
\begin{tabular}{|M{.5cm}|M{3cm}|M{3cm}|M{1cm}|M{3cm}|M{2.5cm}|M{2cm}|M{2.5cm}|}
\hline
Ref	& Performance	& Complexity	& Train/Test 	& Objective	& Input data & Metrics & Adopted algorithms\\ \hline
 
\cite{28} &Achieves 95-99\% accuracy, which is slightly higher than semi-analytical (90-95\%) and the cognitive (90-95\%) approaches &Reduces the computing time by up to 90\% compared to semi-analytical and 50\% compared to the cognitive methods&EXP/EXP & QoT estimation&Features that describe the lightpath & Accuracy, Precision
Recall, F1 score, Computing time
& Semi-analytical, Cognitive, SVM\\ \hline

\cite{30} & Achieves an average prediction accuracy of 95\%, which is higher than traditional methods (80\%)& It has linear computational complexity, while traditional risk-aware models have quadratic or even higher computational complexity& EXP/EXP& OPM and failure risk prediction& Time series data for various optical network indicators&Accuracy &RF, XGB, SVM \\ \hline

\cite{52}  &Reduces the RMSE of OSNR monitoring by up to 30\%, CD monitoring by up to 40\%, and PMD monitoring by up to 90\% compared to traditional methods &Does not require any clock or timing recovery, which significantly reduces the complexity of the implementation &SIM/SIM &Learn the relationship between the AHH and OSNR, CD, and PMD &AHH of received signal &RMSE & ANN, SVM, RF\\ \hline

\cite{45}  &Estimates both linear and nonlinear SNRs with 0.04 dB and 0.20 dB errors, respectively, which is better than traditional methods (errors of up to 1 dB) & It has less complexity than traditional methods & SIM/SIM&Estimate the linear and nonlinear noise components of the received signal & Received signal& MSE, PSNR, SSIM& SVR, RF, RNN\\ \hline

\cite{46}  &FFNN predicts QoT with MAE of 0.1 dB, which is better than that of traditional methods (MAE of 1 dB) &GBM is the most complex model, followed by SVM, RF, RBF, and BP&SIM/SIM &Predict the OSNR of a new WDM channel that is being established &OSNR of WDM channels on established connections, Randomly generated network states &MAE, RMSE, MSE, MPE, MBE, R-squared & RF, XGB, SVM \\ \hline

\cite{55}  & Classifies all modulation formats with 99.6\% accuracy& It has lower computational complexity than SVM &EXP/EXP &Classify the AAHs into different modulation format classes &AAH of received signal &Accuracy, Computational complexity, Robustness to link impairments &ANN, SVM, kNN \\ \hline

\cite{59}  &Reduces NLI variance estimation MSE by up to 80\% & It has computational complexity of $O(N)$, while split-step Fourier method has complexity of $O(N^3)$ & SIM/EXP&Predict the nonlinear distortions experienced by the optical signals as they propagate through the fiber link &Fiber link parameters, Optical signal parameters, Monitored signal parameters &BER, MSE, Computational complexity & SVM, ANN\\ \hline

\cite{60}  &Achieve 95\% accuracy in predicting the OSNR, which is better than traditional techniques (80\%) &Reduces complexity of QoT estimation by up to 90\% compared to traditional methods &SIM/EXP &QoT estimation in EON & Alien wavelength performance monitoring data, Lightpath provisioning data&Accuracy, Precision, Recall, F1-score &ANN, SVM, RF \\ \hline

\cite{61}  &Increases network capacity by up to 30\% &Reduces time required to plan a new optical network by 50\% &EXP/EXP &OSNR prediction in software-defined network &Network topology, Network traffic demand, Network monitoring data &Planning time, Planning accuracy, Network resource utilization, Network performance &Random search, Grid search, Bayesian optimization \\ \hline

\cite{62}  &Improves the capacity of the optical network by more than 17\% &Achieves up to a 100-fold reduction in complexity compared to traditional methods &SIM/SIM &Predict the optimal symbol rate and modulation format for the BVTs in the network &Optical signal parameters, ROADM filtering parameters, Network topology, Traffic demand &Average network capacity, Spectral efficiency, Latency, Packet loss &ANN \\ \hline

\cite{58}  &Detects 98\% of faults, compared to 85\% for traditional methods &Reduces the average fault detection time by 20\% &SIM/EXP & Detect and diagnose faults in the network in a proactive and intelligent manner& Network telemetry data, Historical fault data, Knowledge base&Detection accuracy, False alarm rate, MTTR, Proactive reaction time, Cost of ownership & Random search, Grid search, Bayesian optimization\\ \hline

\cite{102a}  & Obtains 100\% MFI accuracy, and MSE of 0.69 for OSNR monitoring, which is similar to that of ANN& It has low computational overhead, making it suitable for real-time implementation in IM/DD O-OFDM transceivers&EXP/EXP &Classify the received signal into different modulation formats and OSNR levels&Received signal &Accuracy, Computational complexity &kNN, SVM, RF \\ \hline

\cite{102d}  &Achieves 100\% accuracy in the presence of nonlinear effects & Achieves 4.6 times reduction in complexity compared with DL-based method& SIM/SIM& Learn a decision boundary that separates the received signal AHs corresponding to different modulation format classes& AH of received signal& Accuracy, Precision, Recall, F1-score& RF, SVM, kNN, DNN\\ \hline

\cite{102i}  &Achieve 100\% accuracy at OSNRs lower than that required for 20\% FEC limit, which is similar to DNN and SVM & It has at least an order of magnitude lower complexity than DNN and SVM&SIM/SIM &Learn the relationship between the feature vector and the modulation format and OSNR &AH of pre-equalized received signal & Accuracy, Precision, Recall, F1-score, MAE& SVM, RF, DNN\\ \hline

\cite{124}  &Obtains up to 50\% reduction in total deployment cost & Achieves up to 80\% reduction in planning time&SIM/SIM &Minimize the total deployment cost of green field passive optical networks&Location of ONUs, Maximum split ratio, Cost of trenching and laying fibers, Cost of optical splitters &Total deployment cost, Average deployment cost per area, Coverage, Average number of subscribers per PON &K-means, Genetic algorithms, Tabu search, MILP \\ \hline

\end{tabular}
\label{table:5}
\end{table*}
\endgroup

\begingroup
\tabcolsep = 1.0pt
\def\arraystretch{1}
\begin{table*}
\ContinuedFloat
\tiny
\centering
\caption{Summary of ML applications in OCN (cont.).}
\begin{tabular}{|M{.5cm}|M{3cm}|M{3cm}|M{1cm}|M{3cm}|M{2.5cm}|M{2cm}|M{2.5cm}|}
\hline
Ref	& Performance	& Complexity	& Train/Test 	& Objective	& Input data & Metrics & Adopted algorithms\\ \hline
 
\cite{124a}  & Detects 95\% of malicious nodes, while previous approaches can only detect 70\% of malicious nodes& It is a computationally efficient clustering algorithm& SIM/SIM & Classify the nodes into either behaving or not-behaving classes&Burst header packet features, Network traffic information & Accuracy, False positive rate, True positive rate& K-means, Gaussian mixture model, Self-training, Modified self-training\\ \hline

\cite{128b}  &Achieves a BER performance that is within 2 dB of the optimal detector for a wide range of SNRs &Requires $O(K^3)$ operations per iteration, which is much lower than multistage ($O(K^3)$) and decorrelating ($O(K^4)$) &SIM/SIM & Learn the relationships between these input data and the desired output& Signal parameters, Channel parameters, Noise parameters& BER& Matched filter, Decorrelating, EM-based\\ \hline

\cite{129}  &Achieves an accuracy of 99.9\% for BR-MFI& It has low complexity, and is suitable for real-time applications &EXP/EXP &Joint monitoring of OSNR, CD, and PMD, and for autonomous bit
rate and modulation format identification
 &Asynchronous delay-tap plots of the received optical signal &Accuracy &PCA, SVM, kNN \\ \hline
\cite{29}  &Achieves a false positive rate of 0\% and a true positive rate of 99\%, which is better than that of existing fiber optic perimeter detection systems (10-20\%) &Reduces the complexity by a factor of 10 &EXP/EXP & Identifying the intrusion from environmental events &Back scattered optical signal from the fiber optic cable &Sensitivity, Specificity, Accuracy, Precision, F1-score & PCA, SVM, kNN\\ \hline

\cite{133}  & The proposed self-healing framework reduces packet loss by up to 88\% compared to shortest path and multiple path routing& It has expensive computational process, especially for large networks&SIM/SIM &Learn a policy that allows the network to gracefully recover from failures and attacks &Network topology, Traffic demand matrix, Link failure probabilities &Average delay, Packet loss rate, Average throughput, Fairness index &QL, SARSA, SARSA(lambda) \\ \hline

\cite{134}  &A reduction in blocking probability of up to 90\% compared to deflection routing & It has relatively low complexity, making it suitable for real-time implementation in OBS networks & SIM/SIM& Reduce blocking probability in buffer-less OBS networks& State of the network & Blocking probability, Average delay, Packet loss rate& Random routing, Least loaded routing, Most loaded routing, RL-based routing\\ \hline

\cite{136}  &Reduces the loss probability of high-priority bursts by up to 50\% &It is computationally efficient and can be implemented in real time &SIM/SIM &Make decisions about how to route and resolve contentions for each burst &Network topology, Traffic demand, Current network state &Average delay, Packet loss rate, Average throughput &RL-based Alternative Routing, Integrated RL-based Routing and Contention Resolution, Static routing, Deflection routing \\ \hline

\cite{140}  & The proposed algorithm reduces the blocking probability by up to 20\% compared to the classical shortest path trees& It is able to reduce the number of messages required by a factor of 100& SIM/SIM&Determine a RWA solution that minimizes the blocking probability of multicast requests &Multicast requests &Average delay, Packet loss rate, Average throughput, Fairness index &RL, MOSP algorithm \\ \hline

\cite{141}  &Reduces average packet delay by up to 20\% compared to traditional LSSR algorithms &In general, the complexity of RL-based LSSR is higher than the complexity of traditional LSSR algorithms & SIM/SIM&Route traffic in the network in a way that minimizes the overall load cost & Network topology, Traffic demand, Routing policy, Load costs&Average delay, Packet loss rate, Average throughput, Fairness index &RL-SSR, Dijkstra's algorithm, Bellman-Ford algorithm, Link state routing \\ \hline

\cite{143}  &Improves the accuracy of SNR estimation by up to 10\% &Reduces the computational cost of the model by 20\% &SIM/EXP &Predict the values of network parameters with reduced uncertainty &Monitored physical parameters of the network &MSE, RMSE, MAE, MPE, R-squared, Accuracy, Precision, Recall, F1-score &Gaussian processes, Statistical active learning \\ \hline

\cite{145b}  &Reduces delay by up to 15\% and increases capacity by up to 25\% compared to device-to-device-only &It is more complex than the traditional algorithms for VLC and device-to-device heterogeneous network optimization, such as the greedy algorithm and the genetic algorithm &SIM/SIM &Find the optimal data transmission routes in the network, which minimize the total delay and maximize the total throughput &Number of VLC transmitters, Number of device-to-device users, Transmission range of VLC transmitters, Transmission range of device-to-device users, Channel bandwidth, Data rate &Average delay, Packet loss rate, Average throughput, Fairness index &Centralized RL, Distributed RL \\ \hline

\cite{145g}  &Achieves up to 30\% higher throughput compared to traditional methods &Reduces the average latency by 15\% compared to the static threshold-based approach and 10\% compared to the dynamic threshold-based approach &SIM/SIM &Learn the optimal allocation of radio and light resources &Network topology, Channel characteristics, User traffic, Environmental factors &Average delay, Packet loss rate, Average throughput, Fairness index &QL, SARSA \\ \hline

\cite{145h}  &Outperforms the average downlink data rate of QL and Sarsa by 13\% and 14\%, respectively &Reduces the handover delay by 30\% compared to the existing schemes &SIM/SIM &Learn a policy that maximizes the expected reward over time, resulting in an adaptive vertical handover scheme that improves the performance of hybrid VLC-IR networks in ship cabins &Location of the user device (UD), RSS from the VLC and IR access points (APs), Channel conditions between the UD and the APs &Handover success rate, Average handover delay, Average system throughput, Packet loss rate, Average delay &Sarsa-lambda, RL, QL, Sarsa \\ \hline

\end{tabular}
\label{table:5b}
\end{table*}
\endgroup

\begingroup
\tabcolsep = 1.0pt
\def\arraystretch{1}
\begin{table*}
\ContinuedFloat
\tiny
\centering
\caption{Summary of ML applications in OCN (cont.).}
\begin{tabular}{|M{.5cm}|M{3cm}|M{3cm}|M{1cm}|M{3cm}|M{2.5cm}|M{2cm}|M{2.5cm}|}
\hline
Ref	& Performance	& Complexity	& Train/Test 	& Objective	& Input data & Metrics & Adopted algorithms\\ \hline
 
\cite{145i}  &Achieves an average system throughput that is up to 20\% higher than the SSS and up to 10\% higher than the iterative optimization method &Achieves complexity improvements of up to 98\% compared to the exhaustive search method and up to 75\% compared to the iterative optimization method &SIM/SIM &Learn a policy that maximizes the expected reward over time, resulting in an optimized load balancing scheme for the hybrid LiFi-WiFi network &Number of users associated with each LiFi and WiFi AP, Channel conditions between the users and the Aps, Data rates of the users & Average delay, Packet loss rate, Average throughput, Fairness index&SSS, Exhaustive search, RL \\ \hline

\cite{146} &Achieves a BLP of 0.001, while SPF and DWNV achieve BLPs of 0.015 and 0.005, respectively, under a load of 0.8 Erlangs & It has a computational complexity of $O(n)$, where n is the number of paths, while SPF and DWNV have computational complexities of $O(n^2)$ and $O(n^3)$, respectively& SIM/SIM&Learn a policy that minimizes the BLP over time, resulting in an optimized OBS network &Status of the links and wavelengths, Number of bursts queued at each node, Estimated burst arrival rates &BLP, Average delay, Network throughput &Shortest path, Least loaded path, Random path selection, Random wavelength selection \\ \hline

\cite{137}  & Achieves a BLP reduction of up to 50\% compared to the best existing algorithm&The average delay is reduced by up to 30\% compared to the best existing algorithm& SIM/SIM& Learn policies that maximize their own expected rewards over time, resulting in an optimized path selection scheme for the OBS network& Number of bursts waiting to be transmitted at each node, Available bandwidth on each link, Estimated delay on each path&Average delay, Packet loss rate, Average throughput, Fairness index &MARL, Shortest path, Random path \\ \hline

\end{tabular}
\label{table:5c}
\end{table*}
\endgroup

\textbf{Support Vector Machine Algorithm:} 
In the dynamic and demanding landscape of optical communication networks, accurate and reliable OPM plays a critical role in ensuring network stability, efficiency, and reliability. 
SVMs have emerged as powerful tools for QoT estimation and OPM in OCNs. In \cite{28}, SVMs were successfully applied for QoT estimation, achieving a remarkable accuracy of 99.95\% in lightpath classification. This accuracy surpasses the performance of traditional analytical and cognitive Case-Based Reasoning methods, which typically suffer from lower accuracy and higher computational complexity.
The effectiveness of SVMs in QoT estimation is further demonstrated in \cite{30}, where they were employed for OPM and failure risk prediction. The SVM-based approach significantly enhanced OCN stability compared with conventional methods, providing more accurate predictions of optical equipment failure states with an accuracy of 95\%.

The superior performance of SVMs in QoT estimation and OPM stems from their ability to handle nonlinear and complex relationships between network parameters and QoT metrics. Their ability to learn from labeled data and generalize to unseen data enables them to accurately predict QoT degradation and potential failures, ensuring the reliable operation of OCNs. As SVMs continue to gain traction in optical communication, their applications in QoT estimation and OPM are expected to expand further. Moreover, their integration with other ML techniques, such as DL, is poised to unlock even more sophisticated and effective QoT estimation and OPM strategies.

\textbf{Artificial Neural Network Algorithm:} 
ML, particularly ANNs, has emerged as a powerful tool for enhancing OPM, enabling more sophisticated and intelligent network monitoring capabilities.
ANN-based OPM utilizes the strength of ANN algorithms to effectively estimate various network parameters, such as OSNR, chromatic dispersion (CD), differential group delay (DGD), and polarization mode dispersion (PMD), which are crucial for maintaining network quality and reliability.
The selection of appropriate features significantly impacts the performance of ANN-based OPM. A wide range of features have been explored in the literature, including the eye diagram, delay-tap plots, asynchronously sampled signal amplitudes \cite{52}, asynchronous amplitude histograms (AAH) \cite{53}, and asynchronous diagrams. AAH features, which do not require timing or clock recovery, offer the potential for low-cost monitoring solutions.
Extensive simulations and experimental results have demonstrated the superior accuracy of ANN-based OPM compared to traditional methods. For instance, ANN was able to achieve 0.04 and 0.20 dB error for the linear and nonlinear SNR estimation, respectively, over 19,800 sets of realistic fiber transmissions \cite{45}. In another study, ANN-based OSNR prediction with 0.25 dB root mean square error (RMSE) was obtained on a mesh network \cite{46}. 
The application of ANN in OPM extends beyond OSNR estimation to encompass the simultaneous identification of multiple impairments, such as CD, DGD, and PMD. ANN has been successfully employed for 40 Gbps O dB, non-return to zero-on-off keying (NRZ-OOK), RZ-OOK, RZ-differential QPSK (RZ-DQPSK), and DPSK systems \cite{47}-\cite{51}. Additionally, ANN can be used for MFI in heterogeneous OCNs \cite{54}, achieving 99\% accuracy for six widely-used modulation formats for 12 dB OSNR in the presence of CD and PMD \cite{55}.
In \cite{59}, ANNs were employed to calibrate the deviations of existing fiber nonlinearity models. By combining modeling and monitoring, ANNs were able to better estimate the variance of fiber nonlinear interference noise, improving the accuracy of OSNR predictions.
In \cite{60}, ANN-based QoT estimation was developed for lightpath-level prediction in EONs with alien wavelength. An ANN was trained using 1,200 data points, achieving a remarkable 95\% accuracy for OSNR prediction. This enabled real-time optimization of network performance and resource allocation.
Furthermore, ANNs were deployed for OSNR prediction in a SDN monitoring database \cite{61}. By analyzing network traffic and environmental conditions, ANNs were able to predict OSNR accurately, enabling the adjustment of spectral efficiency and configuration of probabilistic shaping-based bandwidth variable transmitters. This approach maximized network capacity and ensured optimal transmission performance.

Beyond its role in OPM, ANN has also been explored for various other applications in optical communication networks. ANN can be used to select the optimal symbol rate and modulation format in bandwidth variable transceivers \cite{62}, leading to higher average capacity compared with a fixed symbol rate with a standard QAM transceiver. Additionally, ANN can predict network traffic \cite{64} and allocate flexible bandwidth \cite{65}.
ANN offers a cost-effective approach for cognition-driven learning and fault management techniques, providing proactive fault detection and identification compared to traditional fixed threshold-triggered operations \cite{58}. ANN has also been investigated for attack detection in OCNs, demonstrating superior accuracy in detecting out-of-band power jamming attacks compared to other ML techniques, such as SVM, decision tree (DT), and Naive Bayesian \cite{63}.

In conclusion, ANN has emerged as a powerful tool for enhancing OPM and overall network performance in optical communication systems. Its ability to accurately estimate various network parameters, identify impairments, and optimize resource allocation makes it a valuable asset for improving the reliability, efficiency, and security of optical networks. As research continues, ANN is expected to play an increasingly crucial role in shaping the future of optical networking.

\textbf{k-Nearest Neighbors Algorithm:} 
The kNN algorithm has shown remarkable potential in joint MFI and OSNR monitoring for OFDM systems. In \cite{102a}, kNN was employed to achieve 100\% classification accuracy for estimating 4/16/32/64/128-QAM modulation formats with a launched power of -11 dBm. Simultaneously, a MSE of 0.69 was achieved for OSNR monitoring, comparable to that obtained using ANNs, while kNN demonstrated significantly lower computational resource consumption.
Furthermore, kNN has been utilized as a classifier for QoT estimation at the lightpath level in \cite{102b}, considering both linear and nonlinear impairments. This application demonstrates the ability of kNN to handle the complexities of OFDM systems and provide accurate QoT estimations.

These studies highlight the effectiveness of kNN for joint MFI and OSNR monitoring in OFDM systems, offering a computationally efficient and accurate alternative to ANNs. As research continues, kNN is expected to play an increasingly important role in enhancing the performance and reliability of OFDM-based optical communication systems.

\begin{figure}
\centering
\includegraphics[scale=0.4]{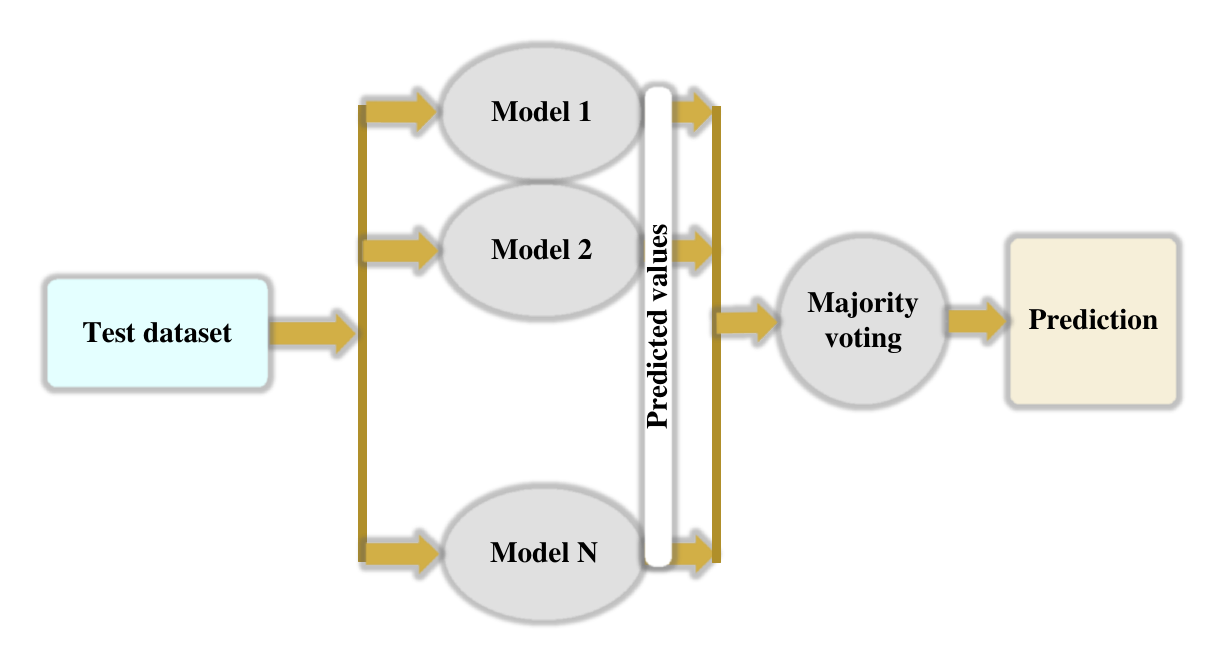}
\caption{Schematic of an ensemble learning algorithm.}
\label{fig:15}
\end{figure}

\textbf{Ensemble Learning Algorithms:} 
EL methods (Fig. \ref{fig:15}), such as random forest (RF), offer a compelling combination of low complexity and favorable performance, making them particularly attractive for optical communication applications. RF has emerged as the most widely used ensemble learning method in this field due to its ability to achieve high accuracy while maintaining computational efficiency.
A recent study \cite{102d} presents an MFI algorithm based on RF, leveraging distinct features extracted from the optical channel response in the presence of nonlinear effects. The proposed algorithm achieves 100\% MFI accuracy for PM 4/8/16/32/64-QAM modulations, outperforming kNN, SVM, and DNN. Notably, RF's computational complexity is significantly lower than DNN, making it a more practical choice for real-time implementation.
The performance of ensemble learning methods is primarily determined by the structure of the DTs and the overall forest configuration. RF has been successfully employed for QoT estimation in \cite{103, 104}. This application involves predicting whether the BER of an unestablished lightpath meets the required threshold based on various lightpath characteristics, such as total length, number of lightpaths, length of the longest lightpath, traffic volume, and modulation format.
In EONs, the use of CD uncompensated links is prevalent, resulting in varying CD values across intermediate nodes. RF has demonstrated its robustness for OSNR monitoring in a wide CD range \cite{102f}. Additionally, RF has been applied for joint MFI and OSNR estimation in \cite{102i}.
Simulation results indicate that RF, DNN, and SVM can all achieve 100\% MFI accuracy for 16 GBaud PM-4/8/16/32/64-QAM systems with OSNR levels as low as 5.12 dB, 7.45 dB, 10.74 dB, 15.15 dB, and 18.22 dB, respectively. These results demonstrate RF's ability to operate at significantly lower OSNR levels compared to the required OSNR at a BER of 2e-2. Moreover, RF outperforms DNN and SVM in terms of mean absolute error (MAE) for OSNR estimation, achieving values of 0.24 dB, 0.29 dB, 0.31 dB, 0.38 dB, and 0.66 dB for PM-4/8/16/32/64-QAM, respectively. Finally, RF's computational complexity is one order of magnitude lower than DNN and SVM, making it a more efficient choice for practical implementation.

In summary, RF has emerged as a powerful and versatile ensemble learning method for a wide range of optical communication applications. Its ability to achieve high accuracy while maintaining low complexity makes it a compelling choice for real-time implementation. As RF research continues to advance, its potential to further revolutionize optical communication is expected to grow even more prominent.

\textbf{Regression Algorithms:} 
The prevention or minimization of fiber faults is paramount in OCNs to ensure uninterrupted and reliable data transmission. Traditional methods, such as optical time-domain reflectometry, are employed for fault detection and localization. However, these traditional approaches often exhibit discrepancies between the measured and actual fault distances.
In \cite{107a}, the authors propose an alternative approach using linear regression to improve fault localization accuracy. By utilizing linear regression, they achieved a significantly reduced fault tracing delay, minimizing revenue loss and ensuring faster network restoration. This study highlights the potential of ML techniques in enhancing fault localization in OCNs.

The application of ML in fault detection and localization holds immense promise for revolutionizing network operations and enhancing the resilience of optical communication systems. As research continues to advance, ML techniques are expected to play an increasingly crucial role in shaping the future of fault management in OCNs.

\textbf{K-Means Clustering Algorithm:} 
K-means clustering has been successfully employed for various tasks in optical communication networks, including optimizing green field passive optical network (PON) deployment and preventing burst header packet flooding in optical burst switching (OBS) networks.
In \cite{124}, K-means is applied to minimize the total deployment cost of green field PONs. The authors demonstrate that this approach can significantly reduce the total deployment cost by up to 50\% compared to the benchmark random-cut sectoring solution. This cost reduction is achieved by efficiently grouping nodes into clusters based on their geographical proximity and optical characteristics.
Burst header packet flooding is a malicious attack in OBS networks where an edge node continuously sends burst header packets to reserve resources while not sending any data bursts. This behavior can disrupt network operation and lead to denial of service attacks.
To address burst header packet flooding, the authors of \cite{124a} propose a K-means-based classification scheme to distinguish between behaving and not-behaving nodes. Their experimental results show that K-means can achieve up to 90\% accuracy in classifying nodes with only 20\% of data. This accuracy increases to 65.15\% and 71.84\% for classifying nodes into three categories: behaving, not-behaving, and potentially not-behaving, with 20\% and 30\% of data, respectively.

The application of K-means in optical network deployment and burst header packet flooding prevention demonstrates its versatility and effectiveness in addressing critical challenges in the field of optical communication. As K-means research continues to advance, its potential to further enhance network performance and security is expected to grow even more prominent.

\textbf{Expectation Maximization Clustering Algorithm:} 
In \cite{128b}, the authors propose two novel blind and unblind multi-user detection schemes for optical code division multiple access systems, leveraging the power of EM. In this approach, EM iteratively solves an unconstrained likelihood function to estimate the interference, followed by a one-dimensional constrained Boolean likelihood function to detect the bit, utilizing the estimated interference.
Simulation results demonstrate the superior performance of the proposed detectors compared to well-known sub-optimum detectors such as multistage and decorrelating detectors. This enhanced performance stems from the ability of the EM-based detectors to effectively handle the multi-user interference and achieve higher BER.

The proposed blind and unblind multi-user detection schemes using EM offer a promising approach for enhancing the performance of optical CDMA systems. These schemes are particularly well-suited for scenarios where CSI is unavailable or limited, enabling robust detection in challenging environments. As research continues in this area, EM-based detectors are expected to play an increasingly crucial role in the development of next-generation optical code division multiple access systems.

\begin{figure}
\centering
\includegraphics[scale=0.5]{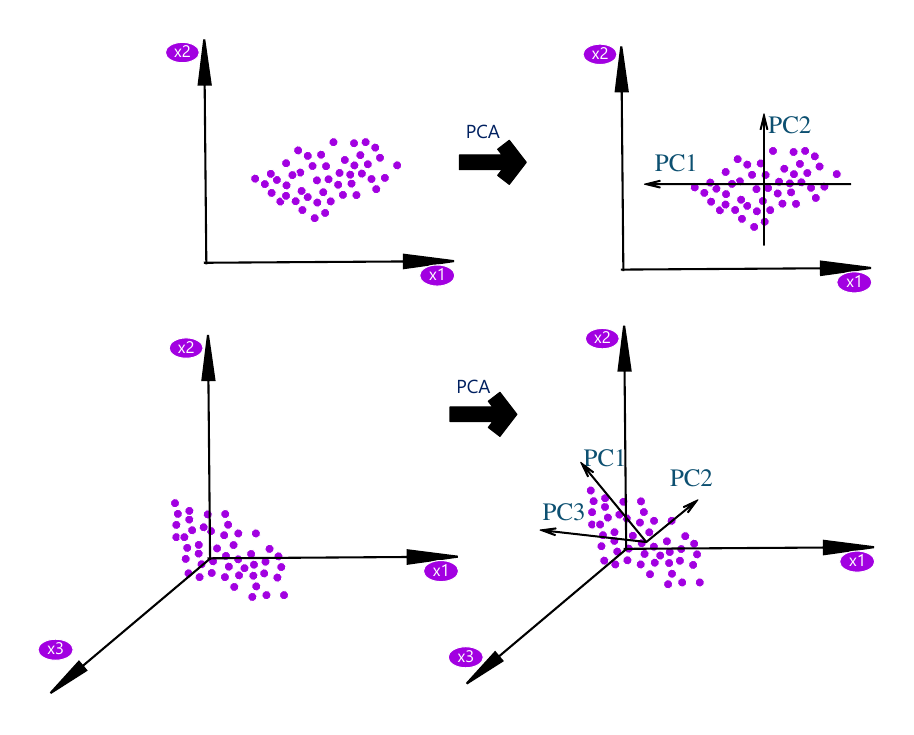}
\caption{PCA concept illustration.}
\label{fig:21}
\end{figure}

\textbf{Principle Component Analysis Algorithm:} 
PCA (Fig. \ref{fig:21}) has emerged as a powerful tool for analyzing and extracting meaningful information from complex optical communication data. 
In \cite{129}, PCA is employed for joint monitoring of OSNR, CD, and PMD. This enables real-time monitoring of these critical parameters, ensuring optimal optical transmission.
PCA can also be used for autonomous bit rate and modulation format identification in optical systems. This eliminates the need for manual configuration and enables seamless adaptation to changing network conditions.
PCA can be used to identify anomalies in network traffic or intrusion attempts, facilitating timely alarm generation or filtering. Studies \cite{130, 131} demonstrate the efficacy of PCA in identifying abnormal flows or attacks.
In fiber optic perimeter intrusion detection systems, PCA can be applied to distinguish intrusions from environmental events. In \cite{29}, the authors extract frequency-domain features using PCA and classify them using SVMs to effectively detect intrusions.
Ultra-high resolution optical spectrum distortion detection: The detection of signal distortions in ultra-high resolution optical spectra can be challenging due to the dynamic nature of optical channels and the difficulty of obtaining a reference spectrum. Studies \cite{132} show that PCA can effectively extract features from ultra-high resolution spectra and SVMs can be used for classification, enabling robust distortion detection.

These applications highlight the versatility and effectiveness of PCA in various domains of optical communication. As PCA research continues to advance, its role is expected to grow even more prominent, enabling more efficient, reliable, and intelligent optical communication systems.

\begin{figure}
\centering
\includegraphics[scale=0.7]{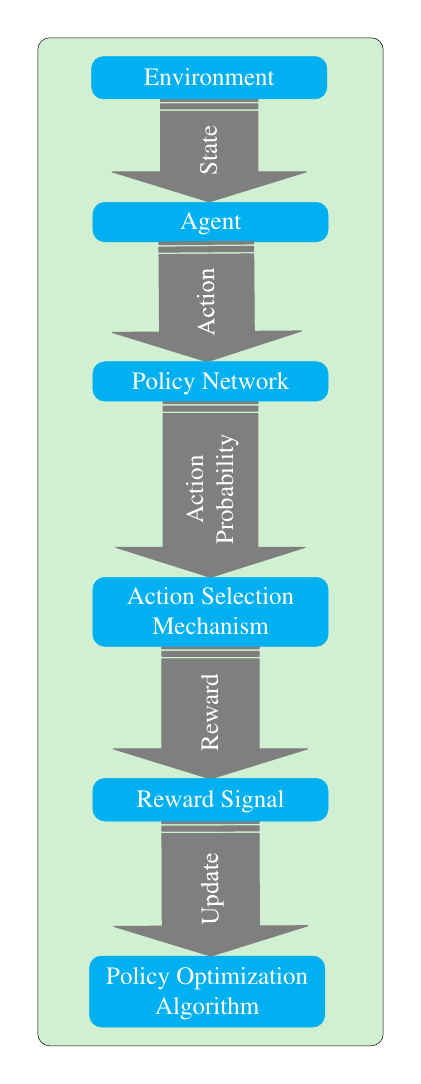}
\caption{Concept of policy-based RL.}
\label{fig:32}
\end{figure}

\textbf{Policy-based Reinforcement Learning Algorithm:} 
Policy-based RL (Fig. \ref{fig:32}), a subfield of RL, emerges as a powerful approach for learning optimal decision-making strategies in complex and dynamic environments like optical communication systems. Transparent optical networks have security issues and require self-healing to be protected from failures and attacks. Authors of \cite{133} applied RL to combine all the available resources in the network with an intelligent strategy to combat the attacks.
Results show that the self-organized routing outperforms the shortest path and multiple path routing, and reduces packet loss ranging from 8\% to 88\% for the case of one, two, three, and four failing wavelengths.
Authors of \cite{134, 135} employed RL to reduce blocking probability in buffer-less OBS networks due to the wavelength contention. This method applied deflection routing to fix wavelength contention and a dynamic Offset Time allocation to reduce losses due to Insufficient Offset Time.
Simulations indicate an effective reduction in blocking probability while maintaining a reasonable end-to-end delay for each burst.
Authors of \cite{136} employed an RL-based routing approach to reduce effectively burst loss probability compared to shortest path routing. Moreover, by combining this approach and the reactive deflection routing, an integrated RL-based routing and contention resolution approach is presented which effectively reduces the loss probability.
In \cite{140}, authors presented an RL-based algorithm for multicast routing and wavelength assignment in wavelength-routed networks. 
This algorithm adapts to topology and traffic changes and has no global knowledge of the network. 
Measurements showed less blocking probability compared to the classical shortest path trees built by protocols like MOSPF (a well-known multicast protocol).
In \cite{141}, authors applied RL for online routing.
The blocking probability of this method is comparable with other event-dependent routing methods such as the Success-to-the-top algorithm.
Authors of \cite{142} employed RL to present a provisioning strategy for connectivity services with different priorities to conserve QoS requirements and maximize provider benefits. 
In \cite{143}, RL is applied for QoT estimation in terms of SNR, power level, and noise figure, and reduced the design margins.
Considering a European backbone network (28 nodes, 41 links), this method reduced the QoT inaccuracy from 4.2 dB to 0.02 dB.

RL has emerged as a powerful tool for addressing complex decision-making problems in optical wireless networks. Several RL-based approaches have been proposed for various optical wireless networks applications, including spectrum sensing, network selection, route selection, vertical handover, and load balancing.
In \cite{144}, the authors developed an RL-based algorithm for distributed spectrum sensing in cognitive radio-based optical wireless networks. The algorithm enables autonomous secondary users to learn their optimal sensing policies, reducing collisions and improving channel utilization.
Authors of \cite{145} presented a RL-based context-aware indoor network selection algorithm that considers the asymmetric traffic characteristics and network performance. The algorithm exploits time-location-dependent load distribution and employs knowledge transfer to enhance its performance and efficiency.
The limited coverage of VLC necessitates its integration with device-to-device technology. In \cite{145b}, the authors utilized RL for route selection in an indoor VLC-device-to-device heterogeneous network. Simulation results demonstrated that RL-based route selection significantly improves data rate and reduces delay.
Hybrid indoor networks, comprising LTE, WLAN, and VLC, require efficient network selection to ensure high user quality of experience. In \cite{145g}, transfer-based RL is applied for context-aware network selection in such hybrid networks, addressing the challenges posed by dynamic environments and complex service requirements. Simulations showed that transfer-based RL can significantly enhance the performance and efficiency of network selection.
To address seamless vertical handover in hybrid VLC-IR networks, an RL-based approach is proposed in \cite{145h}. Based on Sarsa($\lambda$) algorithm, this scheme dynamically adjusts the vertical handover decision based on real-time network conditions. Simulation results demonstrated that the proposed algorithm outperforms conventional vertical handover approaches, achieving higher average doptical wireless networkslink data rates and better handoff performance.
For dynamic load balancing in hybrid LiFi-WiFi networks, RL has been employed in \cite{145i,145j,145k}. Simulations with five users showed that RL-based load balancing significantly improves network throughput compared with signal strength strategy and iterative algorithms. While exhaustive search achieves the highest throughput, its computational complexity grows exponentially with the number of users. RL, on the other hand, exhibits a quadratic relationship with the number of users, making it a more scalable approach. Moreover, RL achieves user satisfaction levels comparable to exhaustive search.

These examples highlight the potential of RL in enhancing the performance, efficiency, and flexibility of optical wireless networks. As RL research continues to advance, its role in optical wireless networks is expected to grow even more prominent, enabling the development of more intelligent, adaptive, and robust optical wireless networks architectures.

\begin{figure}
\centering
\includegraphics[scale=0.7]{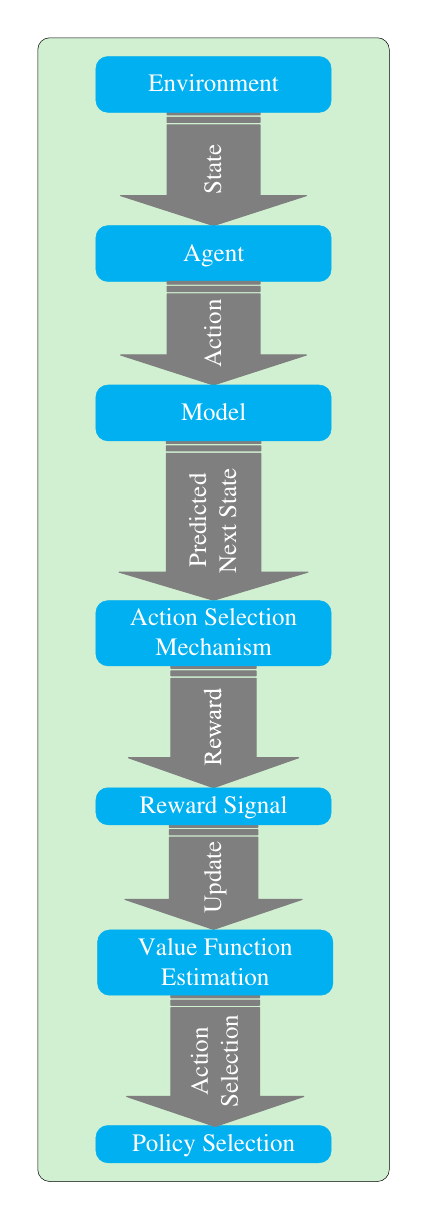}
\caption{Concept of value-based RL.}
\label{fig:33}
\end{figure}

\textbf{Value-based Reinforcement Learning Algorithm:} 
Value-based RL (Fig. \ref{fig:33}) is a branch of RL that focuses on learning the optimal value function, which represents the expected cumulative reward associated with each state. Q-learning (QL), a value-based RL algorithm, has emerged as a powerful tool for deflection routing in OBS networks. This algorithm has been successfully applied in several studies to address the challenges of OBS networks, including congestion, nonlinearity, and resource contention.
In \cite{146}, the authors investigated the performance of QL-based deflection routing compared with traditional routing strategies, such as shortest path selection and DWNV (which is known for its adaptability to dynamic network conditions). Their results demonstrated that QL consistently outperformed these traditional methods, achieving lower blocking probabilities and higher throughput, even under varying traffic conditions.
Furthermore, the simplicity of QL was highlighted due to its reliance on a look-up table representation, reducing computational complexity compared with other RL algorithms.
In \cite{148}, the authors extended the application of QL to path/wavelength selection in OBS networks. They calculated Q values for a set of pre-computed path/wavelength combinations at the output node and selected the one with the minimum burst loss probability.
However, this approach, known as single-agent path selection, relies solely on feedback received at the ingress node, overlooking the impact of path selections made by other network nodes. To address this limitation, the authors of \cite{137} formulated the path selection problem in OBS networks as a multi-agent RL problem. They developed a QL algorithm that utilizes agents at different source nodes to collaboratively learn the network dynamics and select the optimal path for each burst.
Simulation results demonstrated that the multi-agent RL approach outperformed both shortest path selection and single-agent RL, achieving significant reductions in blocking probability and increased throughput.

In conclusion, QL has emerged as a promising RL algorithm for deflection routing in OBS networks. Its ability to learn from experience and adapt to dynamic network conditions makes it a valuable tool for enhancing the performance and reliability of these networks. As research continues in this area, QL is expected to play an increasingly crucial role in shaping the future of OBS technology.

QL has emerged as a powerful tool for optimizing resource allocation and network performance in OWC. Research has demonstrated its effectiveness in various aspects of OWC systems, including access point selection, wavelength allocation, transmit power control, and vertical handover optimization.
In \cite{149}, QL was employed for resource allocation in LiFi-WiFi access networks supporting the Internet of Things service. The results showed that QL could effectively reduce the Internet of Things migration ratio, increase the internet provider revenue, and improve the virtual network acceptation ratio.
In \cite{145f}, a QL algorithm was developed for access point selection in a hybrid WiFi-VLC system. Simulations revealed that the proposed algorithm significantly improved the total throughput and enhanced fairness among connected users compared to traditional methods.
In \cite{149b}, QL was applied to allocate access points and wavelengths in a VLC network. While MILP could also be used for this task, QL provided a more efficient and scalable solution without requiring accurate network information.
Multi-agent QL was employed in \cite{149c} for access point transmit power allocation in an RF-VLC network, ensuring QoS constraints. The proposed algorithm outperformed conventional methods, demonstrating its effectiveness in handling multi-agent interactions and dynamic network conditions.
In heterogeneous RF-VLC networks, AP assignment typically involves either quasi-static network selection or dwell time-based vertical handover from VLC to RF. However, the first approach can lead to outdated decisions in rapid movements, while the second method may cause handover delays. As an alternative, QL was used in \cite{149d,149e} to optimize time-to-trigger values for vertical handover between LTE and VLC based on SNR measurements. Simulations showed that this approach significantly improved the average throughput of mobile devices compared to fixed time-to-trigger methods.

\subsection{Deep  Learning}

\begingroup
\tabcolsep = 1.0pt
\def\arraystretch{1}
\begin{table*}[tp!]
\tiny
\centering
\caption{Summary of DL applications in OCN.}
\begin{tabular}{|M{.5cm}|M{3cm}|M{3cm}|M{1cm}|M{3cm}|M{2.5cm}|M{2cm}|M{2.5cm}|}
\hline
Ref	& Performance	& Complexity	& Train/Test 	& Objective	& Input data & Metrics & Adopted algorithms\\ \hline
 
\cite{80}  &Achieves OSNR monitoring MSE of 1.2 dB, 0.4 dB, and 1 dB for 112 Gbps PM-QPSK, 112 Gbps PM 16-QAM, and 240 Gbps PM 64-QAM signals, respectively, which is better than those of conventional methods (2-3 dB) &Does not require any additional hardware, which further reduces the complexity of the system &SIM/EXP &Identify the modulation format and estimate the OSNR of the received signals &A set of AHs of received optical signals &Accuracy, Complexity &CNN, RNN, Hybrid CNN-RNN \\ \hline

\cite{81b}  &Reduces spectrum fragmentation by up to 20\% compared to traditional RSA algorithms &Reduces average latency by up to 15\% compared to traditional RSA algorithms &SIM/EXP &Predict the optimal RSA strategy for a new network topology and traffic demands &Network topology, Traffic demands, Routing and spectrum allocation strategies &Blocking probability, Spectrum fragmentation, Delay &DNN-based RSA, Traditional RSA \\ \hline
\cite{87}  & Achieves MAE of 0.18 dB, representing a 72\% reduction compared to the conventional method (0.63 dB)&Exhibites a response time of 0.2 ms, which is substantially faster than the conventional method (50 ms) & SIM/EXP& Predict the OSNR of a new optical signal waveform without the need for any manual feature extraction& OSNR of an optical signal transmitted over a fiber optic link& BER, Q-factor, Symbol rate, Spectral efficiency, Complexity&LSTM-NN, SVM, BPNN \\ \hline

\cite{88l}  &Achieves MAE of 0.04 dB, 0.04 dB, and 0.06 dB for QPSK, 16QAM, and 64QAM signals, respectively, which is significant over traditional OSNR estimation methods (1 dB) &The proposed method is computationally efficient and can be implemented in real time &SIM/SIM &Predicting OSNR and nonlinear noise power &Received signal &MAE, Maximum estimation error for nonlinear noise power, Complexity, Tolerance to fiber nonlinearity &LSTM, EVM, Correlation function and statistical moments, Fractional Fourier transform \\ \hline

\cite{89}  &Achieves routing accuracy of 99.2\%, compared to 99.5\% for the shortest path algorithm &It is 10 times faster than the shortest path algorithm &SIM/EXP &Route traffic through the network in a way that optimizes performance while respecting the privacy constraints of each domain &Network topology data, Traffic demand data, Domain privacy constraints &Routing accuracy, Convergence time &Heuristic-algorithm-based routing, AI-based routing \\ \hline

\cite{89c}  & Reduces MAE by 15\% over traditional methods& Reduces prediction time by 20\% over traditional methods& SIM/EXP&Predict future network traffic by providing it with the current network state and other relevant data &Traffic volume, Packet size, Flow type, Source and destination IP addresses, Time of day, Day of the week, Special events &MAE, RMSE, MPE &LSTM, GRU, CNN-LSTM, and CNN-GRU \\ \hline

\cite{92b}  &The normalized MAE is 3.7\% for QPSK, 2.2\% for 16-ary 16-QAM, and 1.1\% for 64-QAM signals &Estimates the EVM of a signal in less than 100 microseconds, while conventional methods can take several milliseconds or more &SIM/EXP &Estimate the EVM of the received signal & Received signal&NMAE, RMSE &CNN, Sliding window method, PAPR method \\ \hline

\cite{93}  &Achieves bias errors of less than 0.4 dB for all six modulation formats and symbol rates tested, which is a significant improvement over traditional methods (1 dB) &It is less computationally complex than traditional methods &SIM/EXP &Learn an accurate OSNR estimation functionality &Asynchronously sampled data from a digital coherent receiver &MSE, SSIM &CNN, LSTM, GRU \\ \hline

\cite{91}  & Achieves an average OSNR estimation accuracy of over 99\% for six modulation formats& It has a testing time of less than 0.5 seconds, which is significantly faster than traditional methods&SIM/EXP & Estimate the OSNR of the signal&Constellation diagram image &BER, Average throughput, Average latency & kNN, K-means, SVM, CNN\\ \hline

\cite{95b}  &Achieves an accuracy of 100\% for BR-MFI, even in the case of limited training data, which is significantly higher than the accuracy of traditional methods, such as PCA-based pattern recognition (80\%) & It has computation time of 56 ms, which is less than PCA (120 ms) &SIM/EXP &Learn the features of the phase portraits that are relevant to the two tasks &Phase portraits of the optical signals &Accuracy, Precision, Recall, F1-score, BER &CNN-based MTL, PCA-based pattern recognition \\ \hline

\cite{95c}  &The accuracy is slightly lower, ranging from 96\% to 99\%, which is a significant improvement over traditional methods (80-90\%) &It is very fast, and can monitor the three phenomena in real time & SIM/SIM&Predict the values of CD, cross-talk, and OSNR &2D matrix of values representing a delay-tap plot &Accuracy, R-squared, Latency &CNN, SVM, kNN \\ \hline

\cite{95a}  &Achieves a MAE of 0.125 dB and a RMSE of 0.246 dB, which are significantly lower than the MAE of 0.5 dB and RMSE of 1.0 dB achieved by traditional methods &It is computationally efficient and can be implemented in real time & SIM/EXP&Estimate the OSNR of a new constellation diagram by simply feeding the diagram to the model &Constellation diagram of the received optical signal &MAE, MSE, Correlation coefficient &PSD, SVM, CNN \\ \hline

\cite{95}  & Reduces noise by a factor of 2 and increases sensitivity by a factor of 3& Reduces the complexity of the system by a factor of 100&SIM/EXP &Classify the displacement of new specklegram images &Specklegram images &Spatial resolution, Sensitivity, Dynamic range, Response time &SVM, RF, kNN \\ \hline

\cite{95d}  &Achieves a high identification accuracy of up to 99.25\%, which is significantly higher than the accuracy of previous methods (80\% to 90\%) &It is computationally efficient and can be implemented in real time & SIM/EXP&Identify the fingerprint of any ONU by simply feeding it the feature matrix extracted from the ONU's transmitted signal & Transmitted signal from an OFDM-PON ONU& Accuracy, Precision, Recall, F1-score& Noise model-assisted CNN, Traditional CNN, CNN-LSTM,\\ \hline
\end{tabular}
\label{table:6}
\end{table*}

\begingroup
\tabcolsep = 1.0pt
\def\arraystretch{1}
\begin{table*}[tp!]
\ContinuedFloat

\tiny
\centering
\caption{Summary of DL applications in OCN (cont.).}
\begin{tabular}{|M{.5cm}|M{3cm}|M{3cm}|M{1cm}|M{3cm}|M{2.5cm}|M{2cm}|M{2.5cm}|}
\hline
Ref	& Performance	& Complexity	& Train/Test 	& Objective	& Input data & Metrics & Adopted algorithms\\ \hline

\cite{150}  & Improves spectrum utilization by up to 30\% compared to traditional methods& It has quite large complexity for large networks& SIM/SIM&Optimize the network routing, modulation level, and spectrum assignment policies for EONs with enhanced survivability & Network topology, Traffic demand, Survivability requirements&Blocking probability, Average delay, Average spectrum utilization, Network survivability & DADRL, DQN, DDPG\\ \hline

\cite{151q}  & Achieves an average blocking probability reduction of 77\% compared to the best heuristic strategy& It has quite large complexity for large networks & SIM/SIM& Allocate resources to each connection request&Network topology information, Traffic demands, Physical layer constraints &Blocking probability, Average latency, Average spectrum utilization, Average energy consumption &DQN, Rainbow, Actor Critic with Experience Replay, Distributional Rainbow QN, First-Fit Max-Sum, Spectral-Aware \\ \hline

\cite{151d}  &Improves spectrum utilization by up to 15\% compared to traditional RMSA algorithms & It has significantly lower than traditional RMSA algorithms& SIM/SIM& Learn the best RMSA policies for different network states and lightpath requests& Network state data, Lightpath request data&Blocking probability, Spectrum utilization, Latency &First-Fit, Most Available, Spectrum First, Least Loaded First, Max-Min Fairness, Spectrum Balancing \\ \hline

\cite{151k}  &Improves spectrum utilization by up to 15\% compared to traditional DRL methods & Reduces latency by up to 10\% compared to traditional DRL methods& SIM/SIM& Learn to maximize the network performance (e.g., minimize blocking probability, maximize throughput)& Network topology, Resource allocation, The current traffic demand& Throughput, Average latency, Packet loss rate, Energy consumption&DQN, Double Q-network, Dueling DQN, Noisy Actor-Critic, Rainbow DQN \\ \hline

\cite{151a}  &Improves network throughput by up to 20\% compared to traditional rule-based approaches &Reduces network latency by up to 30\% compared to traditional rule-based approaches &SIM/SIM & Learn a policy that decides when and how to reconfigure the SFCs in the network& State of the network, Historical traffic traces, QoS and SLA requirements&Average service restoration time, Average service blocking probability, Average network resource utilization, Average energy consumption &DQN, DDQN, DUELING DDQN \\ \hline

\cite{151h}  &Reduces average blocking probability by 30\%, average packet delay by 20\%, and network energy consumption by 15\% &Reduces average delays by up to 70\% &SIM/SIM &Learn the routing policy by interacting with the network environment &Network topology, Traffic matrix, Rrouting constraints &Blocking probability, Average delay, Packet loss rate, Network utilization &K-shortest paths, Shortest path earliest departure, DQN \\ \hline

\cite{151j}  &Reduces the average flow completion time of all flows by 20\% compared to the traditional fixed-threshold scheduling approach & It has the time complexity of $O(n log n)$, while traditional schedulers exhibit $O(n^2)$ complexity &SIM/SIM &Learn the optimal flow scheduling policy by interacting with the HOE-DCN environment and receiving rewards for making good scheduling decisions &Network state information, Flow scheduling policy &Average packet delay, Average packet loss, Network throughput, Energy consumption &DQN, Double DQN, Noisy Actor-Critic, Flow Splitter \\ \hline

\cite{152h}  &Outperforms the traditional greedy method by up to 20\% in terms of average capacity & It has a time complexity of $O(n)$, where n is the number of relay nodes, which is in contrast to traditional greedy methods ($O(n^2)$) &SIM/SIM &Relay selection in a cooperative decode and forward FSO system &CSI of the FSO links between the source, Link quality metrics, Relay availability, Traffic load &Average capacity, Average end-to-end delay, Average packet loss rate, Average energy consumption, Average spectral efficiency &DQL, DDPG, Actor-Critic \\ \hline

\cite{152i}  & Reduces the switching cost by up to 50\% compared to existing DRL approaches& It is efficient enough to be used in real-time applications&SIM/SIM & Learn the best policy for selecting the FSO or RF link for data transmission in a hybrid FSO/RF communication system& RSSI, Atmospheric turbulence intensity, Channel gain& Throughput, Spectral Efficiency, Average Latency, Energy Efficiency, Reliability&DQN Ensemble, DQN, DDQN, Actor-Criti \\ \hline

\end{tabular}
\label{table:6b}
\end{table*}

\textbf{Deep Neural Network Algorithm:} 
DNNs are revolutionizing the field of optical communication by enabling intelligent and efficient OPM and optical network optimization. These techniques are paving the way for enhanced network performance, reliability, and resource utilization.
In \cite{79}, a DNN structure with AHs feature (obtained after the CMA equalization) is employed for MFI. In \cite{80}, the same DNN architecture is extended to perform joint MFI and OSNR monitoring. The results demonstrate that DNN-based MFI provides significantly higher accuracy compared with ANN.
Experimental validation shows that DNN can achieve 100\% accuracy for MFI across a wide OSNR range for three commonly-used modulation formats: PM-QPSK, PM 16-QAM, and PM 64-QAM. Additionally, DNN-based OSNR monitoring achieves MSE of 1.2 dB, 0.4 dB, and 1 dB, respectively, for these modulation formats.
The authors of \cite{81n,81v} further explore the potential of DNN for OPM in MDM-based transmission. They introduce meta-ensemble learning and active learning techniques to enhance OPM accuracy and reduce training data requirements.
In \cite{81b}, DNN is employed for RSA in EONs. The results demonstrate that DNN can significantly improve the degree of spectrum fragmentation and reduce the network blocking probability compared with traditional RSA methods.
The authors of \cite{81c} investigate the application of DNN for dynamic bandwidth allocation in PONs. They develop a DNN-based model to predict the future bandwidth demands of ONUs, enabling efficient and dynamic bandwidth allocation.
Finally, in \cite{81m,81t,81u}, DNN is applied for QoT estimation in MDM-based transmission. The results show that DNN can accurately estimate various QoT metrics, such as BER and Q factor, enabling proactive network optimization and fault detection.

The integration of DNN into OPM and optical network optimization is opening up new possibilities for enhancing optical network performance, reliability, and resource utilization. As DNN research continues to advance, its role in these domains is expected to grow even more prominent, shaping the future of optical communication systems.

\textbf{Recurrent Neural Network Algorithm:} 
LSTM networks have demonstrated remarkable efficacy in identifying the time-varying relationship between OSNR and optical signals, showcasing their potential for real-time OSNR estimation with high accuracy and short response times \cite{87}.
In \cite{87}, the authors conducted experiments and achieved a MAE of 0.05 dB for OSNR estimation across the range of 15 to 25 dB using a 34.94 GBd PM-16-QAM signal. This remarkable performance highlights the effectiveness of LSTM networks in handling the dynamic nature of optical signals.
Furthermore, LSTM networks have been employed to simultaneously predict OSNR and nonlinear noise power \cite{88l}. The authors of \cite{88l} employed VPI simulations to evaluate the performance of their LSTM-based OSNR and nonlinear noise power prediction model for a five-channel long-haul transmission system with launched power ranging from -3 to 3 dBm.
The simulation results demonstrated impressive accuracy, with OSNR estimation errors less than 1 dB for OSNR ranging from 15 to 30 dB and nonlinear noise power estimation errors less than 1 dB for QPSK and 16-QAM signals. These results underscore the versatility of LSTM networks in handling complex optical signal processing tasks.
Researchers have explored the application of LSTM networks for multi-domain routing in optical networks. In \cite{89}, an LSTM-based multi-domain routing paradigm is proposed, demonstrating that using 55\% historical route trajectories as training data can achieve 98\% prediction accuracy for traffic requests between the remaining 45\% node pairs with low blocking probability. This approach effectively utilizes LSTM's ability to capture long-term dependencies in traffic patterns.
In \cite{89a}, authors investigate the use of multi-layer LSTM for predicting future requests in multicore-fiber-based EONs. By combining traffic prediction with inter/intra-core resource allocation strategies based on hierarchical graphs, these researchers achieve significant performance improvements in EONs. The LSTM's ability to model complex temporal patterns proves effective in predicting future traffic demands.
In \cite{89c}, GRUs are employed for traffic matrix prediction, enabling proactive and dynamic resource allocation in optical backbone networks. The GRU-based prediction achieves a MAE of less than 7.4, demonstrating its effectiveness in predicting traffic patterns. This approach leverages GRUs' ability to process sequential data and capture long-term dependencies, enabling accurate traffic forecasting.
The integration of LSTM and GRU networks into optical network routing and traffic prediction demonstrates their potential to revolutionize network operations and resource management. 

These techniques offer the ability to handle the complexities of modern optical networks, enabling more efficient, reliable, and dynamic traffic management strategies. As research continues to advance, their role in optimizing optical network performance is expected to grow even more prominent, shaping the future of network operations and efficiency.

\textbf{Convolutional Neural Network Algorithm:} 
CNNs have emerged as powerful tools for extracting intrinsic features from image data, making them well-suited for tasks such as OPM and QoT estimation.
In \cite{92b}, authors employ a CNN-based regressor for OPM, extracting EVM information directly from the signal complex constellation. Their experiments demonstrate the effectiveness of CNN in back-to-back transmission, achieving normalized MAE of 3.7\%, 2.2\%, and 1.1\% for QPSK, 16-QAM, and 64-QAM, respectively. In fiber propagation experiments, CNN exhibits promising results, achieving normalized MSE of 6.2\% for QPSK at 2000 km, 2.6\% for 16-QAM at 1500 km, and 2.8\% for 64-QAM at 1000 km transmission.
Building on these advancements, the authors of \cite{93} develop a CNN-based OPM that can predict OSNR from asynchronously sampled data right after intradyne coherent detection. Their experiments confirm the effectiveness of CNN, achieving bias errors and standard deviations below 0.4 dB for 14- and 16-GBd DP-QPSK, 16-QAM, and 64-QAM signals.
In \cite{91}, CNN is applied to perform joint monitoring of MFI and OSNR by processing raw constellation diagram data. CNN outperforms traditional ML methods, including decision tree, DW-kNN, SVM, and BP-ANN, while maintaining a complexity of O(n). Across six modulation formats (QPSK, 8-PSK, 8-QAM, 16-QAM, 32-QAM, and 64-QAM) and two OSNR ranges (15-30 dB and 20-35 dB), CNN achieves remarkable performance: 95\% accuracy for 64-QAM, over 99\% accuracy for the other five formats, and 100\% accuracy for MFI. Moreover, CNN demonstrates the ability to estimate OSNR with an error of less than 0.7 dB.
CNNs have emerged as powerful tools for OPM in optical fiber communication systems. These networks can effectively extract features from optical signals, enabling accurate estimation of various impairments, such as OSNR, CD, and DGD.
In \cite{92}, CNNs are employed to monitor OSNR and MFI using eye diagrams or AHs as inputs. The results demonstrate the effectiveness of CNNs in accurately estimating OSNR and MFI.
In \cite{95b}, CNNs are used for joint multi-impairment OPM (joint OSNR, CD, and DGD) with backward Raman monitoring (BR-MFI) using asynchronous delay-tap sampling phase portraits. Simulation results show that CNNs achieve remarkable performance, with RMSE of 0.73 dB, 1.34 ps/nm, and 0.47 ps, for monitoring of OSNR, CD, and DGD, respectively, and 100\% accuracy for BR-MFI. Additionally, this method enables real-time processing with a latency of only 56 milliseconds.
Furthermore, CNNs have been applied for simultaneous monitoring of CD, cross-talk, and OSNR in \cite{95c}. The results indicate that CNNs achieve high accuracy, with 99\% accuracy for CD and cross-talk monitoring, and 96\% accuracy for OSNR monitoring, considering ranges of 400–1600 ps/nm for CD and 10–30 dB for cross-talk and OSNR.
In \cite{93}, CNNs are employed for OSNR monitoring using an asynchronously sampled dataset right after intradyne coherent detection. The results demonstrate the feasibility of CNN-based OSNR monitoring in real-time optical communication systems.
Moreover, in \cite{95a}, an end-to-end CNN-based OSNR monitoring scheme is experimentally demonstrated. The results show that the proposed CNN achieves MAE of 0.125 dB and RMSE of 0.246 dB for PM-QPSK and PM-16-QAM signals under various symbol rates.
The application of CNNs extends beyond OPM to device fingerprint identification in optical networks. In \cite{95}, CNNs are trained to classify perturbations according to their occurrence distance along MDM-based transmission. The results show that CNNs achieve high accuracy, with 99\% and 71\% accuracy for perturbations occurring at 3 and 10 different locations, respectively.
In \cite{95d}, a noise model-assisted CNN-based fingerprint identification approach is employed to predict legal and illegal ONUs and authenticate legal access in OFDM PONs. The results demonstrate that this method achieves a high identification accuracy of up to 99.25\%.

Overall, CNNs have emerged as powerful tools for enhancing OPM and device identification in OFDM PONs. These networks provide accurate and real-time monitoring of optical impairments, enabling improved network performance and security.

\textbf{Deep Reinforcement Learning Algorithms:} 
DRL has emerged as a powerful tool for optimizing network performance and resource allocation in optical networks. DRL algorithms can learn from experience and make intelligent decisions to dynamically adjust network parameters, ensuring optimal network operation.
In \cite{150}, the authors considered global network performance optimization under a survivable EON environment. They proposed a criterion called the whole network cost-effectiveness value with survivability to measure the overall network performance. A double-agent DRL was employed to provide protection schemes converging toward better survivable routing, modulation level, and spectrum assignment. The results indicate that this approach significantly improves whole network cost-effectiveness value with survivability compared to conventional heuristic approaches.
In \cite{151q}, DRL was deployed for routing, modulation, spectrum, and core assignment in a dynamic multi-core fiber EON. The results showed that DRL achieved up to four times lower blocking probability compared to the best-performing heuristic baselines for NSFNet and COST239 network topologies.
The authors of \cite{151c} employed DRL to achieve adaptive virtual network function service chaining in inter-data center EONs. This method dynamically adjusted the duration of service cycles to balance the trade-off between resource utilization, network reconfiguration overhead, and blocking probability.
In \cite{151b}, DRL was applied for multi-cast session selection in EONs. The algorithm analyzed graph data abstracted from EON topology and the current provisioning scheme of a multi-cast session to select the most efficient and reliable path.
The DRL-based RMSA were demonstrated in \cite{151d, 151i, 151l} for dynamic EON operations. Conducted simulations showed that DRL outperformed conventional baseline methods such as heuristic rule-based approaches, achieving lower blocking probabilities and improved network performance.
The authors of \cite{151k, 151p} presented a multi-task transfer learning-based DRL for RMSA in EONs. This method reduced learning time and blocking probability by leveraging knowledge from multiple network scenarios. The results showed a 4 times reduction in training epochs and a 17.7\% reduction in blocking probability.
In \cite{151a}, the authors employed DRL for performing dynamic resource allocation of service function chains in network function virtualization-software-defined networking enabled metro-core optical networks. Simulations showed that DRL outperformed rule-based optimization design, achieving better resource utilization and network performance.
DRL was also employed in \cite{151} for realizing virtual network slicing in the inter-data center network. This approach enabled dynamic provisioning of network slices to accommodate various traffic demands and ensure efficient resource utilization.
In \cite{151g}, DRL was employed to accommodate multi-band EON resources. Six different DRL agents were evaluated on four real network topologies. The results showed that trust region policy optimization was the best-performing agent.
The knowledge distillation scheme was applied in \cite{151o} for DRL-based optical network routing. This approach effectively transferred knowledge from prior traffic patterns to new scenarios, improving the generalization ability of DRL models.
In \cite{151h}, DRL was applied for routing in optical transport networks. This approach utilized a more detailed representation of the network state, considering link-level inter-dependencies and the stochastic nature of the network traffic. The results revealed that this method performed better and faster compared to state-of-the-art representations.
DRL was employed in \cite{151j} to handle the flow scheduling in a hybrid optical-electrical switching-based data center network with high time-varying and dynamic traffic patterns. This approach enabled the top-of-rack switch to make instant flow scheduling based on local threshold tables, ensuring efficient and dynamic traffic management.
In \cite{151f}, the authors investigated and compared the blocking performance of heuristic and DQN-based resource provisioning methods in a dynamic multi-band EON. The results showed that DQN achieved lower blocking probability compared to the heuristic, while being faster in training and decision-making.

\section{Optical wireless communication}
\subsection{Machine Learning}

\begingroup
\tabcolsep = 1.0pt
\def\arraystretch{1}
\begin{table*}
\tiny
\centering
\caption{Summary of ML applications in OWC.}
\begin{tabular}{|M{.5cm}|M{3cm}|M{3cm}|M{1cm}|M{3cm}|M{2.5cm}|M{2cm}|M{2.5cm}|}
\hline
Ref	& Performance	& Complexity	& Train/Test 	& Objective	& Input data & Metrics & Adopted algorithms\\ \hline

\cite{9} &Achieves 35\% higher data rate than DD &Offers relatively low increase in complexity over DD& EXP/EXP& Detection&Received signal & Accuracy, Precision, Recall, F1-score& SVM, Direct decision, Maximum likelihood\\ \hline

\cite{9a} &Achieves Q-factor improvement of up to 11.5 dB compared to traditional methods & It has complexity of $O(n^3)$, which is higher than traditional NN method, ($O(n^2)$) &SIM/SIM &Equalization in a 4-band GS 8-QAM seamlessly integrated fiber and VLC system for phase offset mitigation  &Received signal &Accuracy, Precision, Recall, F1-score &kNN, DT \\ \hline

\cite{9b} &Reduces the BER by 60\% compared to the CMA at SNR of 10 dB & It has much smaller complexity compared to Viterbi and maximum likelihood methods&SIM/SIM &Estimate the phase of the transmitted signal from the received signal &Received signal &BER, Computational complexity, Implementation feasibility & Least squares, Maximum likelihood, SVM\\ \hline

\cite{66}  &Achieves a Strehl ratio of over 0.9, which is significantly higher than that of traditional methods &Reduces the computational by 50\% compared to the model-based method & EXP/EXP& Learn the relationship between the received intensity image and the corresponding wavefront distortion&Received intensity image & BER, Q-factor, Eye Diagram&BP ANN, Zernike polynomial fitting, Modal decomposition \\ \hline

\cite{66c}  & Reduces the BER of the FSO link to 1.2e-12, which is a significant compared to that of typical FSO links (1E-5)& It has significantly lower complexity than that of extended Kalman filter and the unscented Kalman filter &SIM/SIM &Estimate the CSI from the received signal &Received signal &BER &ANN, EKF, LMS \\ \hline

\cite{67}  &Uses a combination of wavelet transform and ANN to achieve error-free data communication at data rates up to 150 Mb/s, which is the highest practically reported data rate for a single link using OOK-NRZ modulation and a PIN photodetector &Reduces computational complexity by 50\% compared to conventional VLC receivers & EXP/EXP&Classify the sub-band signals into different bit classes &Received signal &BER, Q-factor, MSE &Matched filter, Linear equalizer, Wavelet-neural network \\ \hline

\cite{68}  &Achieves a data rate of 150 Mb/s using a blue filter, which is faster than DF equalizer (130 Mb/s) and linear equalizer (90 Mb/s) & It has high computational complexity &SIM/SIM &Compensate for the impairments in the received signal and recover the transmitted signal &Received signal &BER &Wavelet transform, ANN \\ \hline

\cite{70}  & Achieves BER of 1e-4, which is 20\% lower than that of SISO systems&Achieves twice the data rate of traditional systems with only 50\% increase in complexity &SIM/EXP &Estimate the transmitted signal from the received signal &Received signal &BER, Complexity &Equalizer-based MIMO, ANN-based MIMO \\ \hline

\cite{71}  &Achieves BER of 1e-4, while conventional receivers may struggle to achieve BERs below 1e-2 &Modified Extreme learning machine-based receiver requires NlogN flops to train the extreme learning machine model, whilst conventional extreme learning machine needs $N^3$ flops to train the model & SIM/EXP&Estimate the transmitted signal from the received signal &Received signal &BER, Spectral efficiency, Complexity &LMMSE, volterra, extreme learning machine \\ \hline

\cite{102c}  &Achieves positioning accuracy of 2.7 cm, which is significantly higher than that of reported in previous studies (10 cm) & It has computational complexity of $O(N)$, where N is the number of reference points in the fingerprint database, which is significantly lower than that of traditional methods ($O(N^2)$) & SIM/SIM&Estimate the distance between the receiver and the transmitter &RSSI from multiple VLC transmitters & Accuracy, Precision, Recall, F1-score&RSSI-based fingerprinting with weighted kNN, RSSI-based trilateration, Angle-of-Arrival-based fingerprinting with kNN, Angle-of-Arrival -based trilateration \\ \hline

\cite{105}  &Achieves accuracy of 95\% in predicting RSSI parameter, which is better than that of traditional methods (80\%) & It has lower time complexity over traditional methods &EXP/EXP &Predict the RSSI parameter based on the weather parameters &RSSI, Weather parameters &Accuracy, Precision, Recall, F1-score, AUC, Confusion matrix &kNN, SVM, RF \\ \hline

\cite{102e}  &Achieves accuracy of 90\%, which is higher than k-NN (85\%) and SVM (82\%) &Identifies objects in real time &EXP/EXP &Classify the channel estimates into different object classes &Channel estimates from the VLC link &Accuracy, Precision, Recall, F1-score &kNN, SVM, RF \\ \hline

\cite{112}  & Provides 52.55\% improvement in positioning accuracy over the conventional method&Takes 78.26\% less time than the conventional method to determine indoor positioning & SIM/EXP& Predict the location of a mobile user based on the RSS measurements received from the VLTs& RSS fingerprints, Location information&Accuracy, Precision, Recall, F1-score & kNN, SVM, RF\\ \hline

\cite{107b}  &Achieves an average positioning accuracy of 3 cm, which is lower than traditional methods (10 cm) &Reduces the training burden by up to 90\% compared to traditional methods &SIM/SIM &Predict the position of the receiver &Received light intensity from each LED, Spatial arrangement of the LEDs &MAE, RMSE, R-squared &2nd-order regression model, Polynomial trilateral model \\ \hline

\cite{107c}  &Achieves an average positioning error of about 2 cm in both horizontal and vertical directions, which is lower than previous RSS-based VLP systems (10 cm) & It is more computationally efficient over the traditional RSS-based VLP systems, making it suitable for real-time applications& EXP/EXP& Learn the relationship between the RSS measurements and the receiver's location&RSS measurements from multiple VLTs &MAE, RMSE, R-squared, MSE &kernel ridge regression with sigmoid function data preprocessing, kNN, SVM, RF \\ \hline

\cite{114}  & Classifies the modulation format of an optical signal with a probability of 99.9\%, even in the presence of noise& It has low computational complexity, and is suitable for real-time applications& SIM/EXP& Find the density peaks of the data points& Stokes space representation of the received signal& Accuracy, Precision, Recall, F1-score& K-means, density-based clustering algorithm, Spectral clustering\\ \hline
\end{tabular}
\label{table:7}
\end{table*}
\endgroup

\begingroup
\tabcolsep = 1.0pt
\def\arraystretch{1}
\begin{table*}
\ContinuedFloat

\tiny
\centering
\caption{Summary of ML applications in OWC (cont.).}
\begin{tabular}{|M{.5cm}|M{3cm}|M{3cm}|M{1cm}|M{3cm}|M{2.5cm}|M{2cm}|M{2.5cm}|}
\hline
Ref	& Performance	& Complexity	& Train/Test 	& Objective	& Input data & Metrics & Adopted algorithms\\ \hline

\cite{124d}  &Achieves BER of 0.28 at SNR of 5 dB, compared to BER of 0.32 for the original K-means algorithm & It is less computationally expensive than the standard K-means algorithm, with a reduction in complexity of up to 50\% &SIM/SIM &Learn a set of cluster centroids that represent the mean of each phase class & Received signal&BER &K-means, Normalized angle clustering, Improved K-means \\ \hline

\cite{125}  &Reduces the BER by 20\% & It has complexity of linear in the number of symbols, while the complexity of LS method is quadratic in the number of symbols & EXP/EXP& Correct the phase deviation of the received signal& Received signal&BER &Mean phase estimation, Maximum likelihood estimation, K-means \\ \hline

\cite{128d}  &Reduces the BER by up to 2 orders of magnitude compared to conventional channel estimation methods &Reduces the complexity of channel estimation by 10 times compared to the state-of-the-art methods &SIM/SIM &Estimate the transmitted signals of all active users from the received signal &Received signal &MSE, BER &MMSE, MAP, Maximum likelihood, EM-based Maximum likelihood \\ \hline

\cite{128e}  &Achieves an average estimation error of less than 5\% for all three channel parameters, while the N-R and GMM methods can only achieve an average estimation error of less than 10\% &Converges much faster than the N-R and GMM methods&SIM/SIM &Estimate the two shape parameters of the Gamma-Gamma distribution which are alpha and beta &Received signal &MSE, RMSE, Complexity &EM-based Maximum likelihood, Newton–Raphson method, Generalized moment method \\ \hline

\cite{128f}  &Achieves a BER of 1e-3 at an SNR of 10 dB, while the conventional scheme achieves a BER of 1e-2 & It has a complexity of $O(N)$, where N is the number of symbols in the observation interval, and is suitable for real-time applications&SIM/SIM &Detect the transmitted symbols without requiring any prior knowledge of the channel statistics &Received signal &BER &EM \\ \hline

\cite{128c}  & It has BER of 1e-5, while the LS has a BER of 1e-3 at SNR of 20 dB & It has complexity of $O(N2)$, where N is the number of subcarriers in the OFDM system, which is higher than the complexity of the LS and MMSE algorithms ($O(N)$) &SIM/SIM &Find the maximum likelihood estimate of unknown parameters in a statistical model &Received signal, Pilot symbols &BER, MSE, Complexity, Convergence rate &EM-VLC, LS, MMSE, Low-rank MMSE \\ \hline

\cite{145a}  &Surpasses conventional RSS by 75\% smaller mean positioning error & It is computationally efficient and can be implemented in real-time& EXP/EXP&Estimate the position of the mobile device and to update the height information &RSS measurements from multiple VLTs, A priori height information &MAE, RMSE, 2D Localization Error &Iterative point-wise RL, RSS, kNN, Point-wise RL \\ \hline

\cite{145d}  &Outperforms a baseline trust region policy optimization by 26.8\% average number of successfully served users &Offers substantial complexity improvements over traditional RL approaches &SIM/SIM &Learn a policy that maximizes the expected reward over time, resulting in a reliable THz/VLC wireless VR network &Positions of the VR users, THz and VLC channel conditions, visible light access points that are currently turned on &Average number of successfully served users, Convergence speed, Energy consumption, Transmission delay, Throughput & Policy gradient-based, RL algorithm that adopts a meta-learning, Trust region policy optimization\\ \hline

\cite{149}  &Achieves an acceptance ratio of up to 90\% and a migration ratio of less than 10\%, which is a significant improvement over existing algorithms (70\% and 20\%) &Reduces the end-to-end delay by up to 26\% &SIM/SIM &Resource allocation in LiFi-WiFi networks supporting the Internet of Things service &Virtual network demands, Physical network topology &Average delay, Packet loss rate, Network throughput, Resource utilization, Energy consumption & VNE-EACH, vNE-USE, vNE-USE-Q\\ \hline

\cite{145f}  & Improves throughput by up to 30\% compared to the traditional QL algorithm& It has a computational complexity of $O(N)$, where N is the number of available networks, compared to $O(N^2)$ for traditional methods&SIM/SIM &Access point selection in hybrid WiFi-VLC system &User service requirements, Features of the traffic, Network load distribution &Throughput, Latency, Packet loss rate, Energy consumption &RL with Knowledge Transfer, Traditional RL, Rule-based \\ \hline

\cite{149b}  & Achieves a 10\% increase in the average sum rate compared to the traditional MILP algorithm& It has significantly lower computational complexity than traditional methods, such as MILP&EXP/EXP &Access point and wavelength allocation in the VLC network &Network topology, User demands, Channel conditions &Throughput, Fairness, Complexity &QL, MILP \\ \hline

\cite{149d}  &Exceeds fixed time-to-trigger approach by 25\% average throughput &It is more efficient than the fixed TTT scheme &SIM/SIM &Optimize the time-to-trigger values for vertical handover between LTE and VLC &Number of RAPs, Number of visible light access points, Coverage area of each RAP/visible light access point, Channel characteristics, Time-to-trigger value, Average throughput, Handover delay, Packet loss rate &Throughput, Average delay, Packet loss rate, Energy efficiency, Spectral efficiency &Deep QL, Double QL, Dueling QL \\ \hline
\end{tabular}
\label{table:7b}
\end{table*}
\endgroup

\textbf{Supervised Learning Algorithms:} 
SVM has emerged as a powerful tool for signal detection and phase offset mitigation in visible light communication (VLC) and free space optical (FSO) systems. In \cite{9}, SVM is employed as a receiver for VLC systems, demonstrating a data rate improvement of 35\% compared to conventional direct decision methods.
In \cite{31}, SVM is applied to FSO systems to enhance signal detection. The experimental results indicate significant improvements in signal quality, enabling transmission under a 7\% forward error correction (FEC) threshold.
SVM has also been applied to mitigate phase offset in VLC and FSO systems. In \cite{9a}, SVM is used after CMA equalization in a 4-band Gaussian filtered single-sideband 8-QAM system integrated with both fiber and VLC links. The experimental results show an increase in Q-factor of up to 11.5 dB and the ability to transmit under the 7\% FEC threshold, achieving a bandwidth of 320 MHz, a data rate of 960 Mbps, and a transmitted power of -2.5 dBm.
In \cite{9b}, SVM is employed to estimate phase in multiband-CAP VLC systems, addressing the issue of phase offset that can hinder constellation convergence when using CMA equalization. The experimental results demonstrate effective phase offset correction and significant BER reduction under the 7\% FEC threshold for a specific data rate range.

\textbf{Artificial Neural Network Algorithm:}
ANNs have emerged as a powerful tool for addressing the challenges posed by channel effects in FSO and VLC systems. In FSO, ANNs are primarily employed for estimating, modeling, and combating channel effects caused by atmospheric turbulence.
In \cite{66}, ANNs are applied to a sensor-less adaptive optics FSO communication system to design a distortion reduction scheme. Results indicate that ANNs achieve a better Strehl Ratio (0.7) compared to conventional methods, which means more energy can be coupled into the fiber. Additionally, ANNs provide a faster response, making them well-suited for real-time applications.
ANNs are also used for channel estimation in FSO links. In \cite{66c}, ANNs are employed for channel estimation in an outdoor FSO link with slow fading, and a link reliability of more than 99\% is achieved. Furthermore, ANNs have been shown to be an accurate approach for modeling FSO channels \cite{66a, 66b}.
In VLC systems, fluorescent light interference poses a significant challenge. In \cite{67}, it is shown that a receiver based on discrete wavelet transform and ANN outperforms a high-pass filter and a finite impulse response equalizer in the presence of fluorescent light interference. Results demonstrated error-free transmission up to 150 Mbps using OOK-NRZ modulation.
ANN-based equalizers have also been explored for VLC systems. In \cite{68}, ANN-based equalization is applied both in real-time (TI TMS320C6713 digital signal processing board) and offline (MATLAB) using a low bandwidth (4.5 MHz) light emitting diode (LED) transmitter and a large bandwidth (150 MHz) PIN photodetector receiver. Results showed data rates of 170, 90, 40, and 20 Mbps for ANN, adaptive decision feed-back and linear equalizers, and unequalized cases, respectively, for the white spectrum.
In \cite{70}, the authors experimentally demonstrated an OOK multi-input multi-output (MIMO) VLC system with silicon LEDs and organic photodetectors transceivers offering 200 kbps data rate without equalizer. Results showed that applying ANN to classify the signal and correct the error induced by the matrix inversion at the receiver, provides 1.8 Mbps data rate.
Extreme learning machine \cite{72}, an extension of ANNs, is applied in \cite{71} as a VLC MIMO detector in the presence of LED nonlinearity and cross-LED interference. Extreme learning machine provides good performance and speed in this challenging scenario.

The use of ANNs in FSO and VLC systems demonstrates their versatility and potential to address the challenges posed by channel effects and enhance the performance of these communication technologies. As ANN research continues to advance, their role in FSO and VLC systems is expected to grow even more prominent.

\textbf{k-Nearest Neighbors Algorithm:} 
Conventional VLC-based positioning algorithms face significant challenges in providing accurate and reliable positioning in complex indoor environments. These algorithms often exhibit susceptibility to interference from ambient light sources and struggle to adapt to the highly dynamic and heterogeneous nature of indoor spaces.
In contrast, kNN has emerged as a promising alternative for VLC-based positioning, offering several advantages over conventional approaches. kNN is a non-parametric method that utilizes the distances between the measured light intensity values and those of known reference points to determine the user's position. This approach is robust to interference and can effectively adapt to the varying characteristics of indoor environments.
A recent study \cite{102c} demonstrates the effectiveness of kNN for VLC-based positioning in indoor environments. The study employs a 3D VLC system and compares the performance of kNN with conventional methods. The results indicate that kNN achieves a location error of less than 7 cm, significantly outperforming conventional approaches.

The ability of kNN to provide accurate and reliable positioning in complex indoor environments makes it a valuable tool for various applications, including indoor navigation, asset tracking, and gesture recognition. As research continues to advance, kNN is expected to play an increasingly prominent role in VLC-based positioning, enabling enhanced user experience and improved indoor facility management.

\textbf{Ensemble Learning Algorithms:} 
In recent years, ML has emerged as a powerful tool for enhancing signal strength prediction and positioning accuracy in FSO and VLC systems. ML algorithms can analyze complex data patterns to accurately predict signal strength and identify objects in these challenging environments.
In \cite{102j}, the authors implemented extended gradient boosting (XGB) for received signal strength indicator (RSSI) prediction analysis for seamless switching from FSO to radio frequency links. This enables uninterrupted communication when FSO links experience blockages or disturbances.
In \cite{105}, the authors investigated the effectiveness of RF, AdaBoost, and GB classifiers for predicting received optical power in FSO links. RF demonstrated superior performance, achieving 100\% prediction accuracy in real-time tests. AdaBoost closely followed RF, maintaining performance above 99.7\%. However, GB exhibited unsatisfactory results, with its accuracy gradually decreasing over time.
In \cite{102e}, RF was employed to identify the presence of objects based on channel state information (CSI) from the VLC link. This algorithm, combined with fingerprinting techniques, enables high-precision 3D indoor positioning with object detection accuracy in the order of 90\%.
In \cite{112}, the authors explored the dual-function ML approach for VLC-based indoor positioning. This approach utilizes a classification function to rapidly approximate the object's location and a regression function to refine the estimate. Simulation results revealed that the SVM and RF methods achieved the highest positioning accuracy, at 8.6 cm and 10.2 cm, respectively. The SVM exhibited the most significant execution time improvement (78\%), making it suitable for applications demanding low positioning error. The kNN algorithm provided a positioning accuracy of 13 cm with a computational time of 5.6 ms, making it a viable choice for applications requiring fast execution time and moderate accuracy.

These studies demonstrate the remarkable potential of ML for enhancing signal strength prediction and positioning accuracy in FSO and VLC systems. ML algorithms offer the ability to handle complex data patterns and make accurate inferences in real-time, enabling seamless communication and precise positioning even in challenging environments. As ML research continues to advance, its role in these applications is expected to grow even more prominent, paving the way for more robust and efficient FSO and VLC networks.

\textbf{Regression Algorithms:} 
Conventional ML-based VLC-based positioning techniques divide the area into positioning unit cells to reduce training time and complexity. However, this approach often leads to high positioning errors. To address this issue, researchers have explored various ML algorithms, including linear regression and kernel ridge regression, to improve positioning accuracy.
A study in \cite{107b} employed linear regression combined with a light spatial sequence arrangement scheme to achieve a significant reduction in positioning error. In a testbed comprising two unit cells, each with dimensions of approximately 1.55 m x 2 m and a vertical separation of 2.80 m, the average positioning error of the linear regression model was 106 cm. By incorporating the light spatial sequence arrangement scheme, the positioning error was reduced by 90.71\% to 9.85 cm.
Kernel ridge regression has also been explored for VLC-based positioning \cite{107c, 107d}. kernel ridge regression has demonstrated superior performance compared to linear regression. In a testbed consisting of a unit cell with a horizontal coverage of half the area (50 cm x 50 cm) and a vertical coverage of 95 cm to 135 cm (40 cm), kernel ridge regression with the sigmoid function data preprocessing algorithm achieved average horizontal and vertical positioning errors of 1.96 cm and 2.16 cm, respectively. These values were 27.8\% and 22.0\% better than those obtained using the second-order linear regression model without data preprocessing, and 46.5\% and 38.7\% better than those achieved using second-order linear regression with data preprocessing. Additionally, kernel ridge regression reduced the average horizontal and vertical error of linear regression from 3.64 cm to 1.89 cm (48.1\%) and from 3.79 cm to 2.13 cm (43.8\%), respectively. These results demonstrate the effectiveness of kernel ridge regression in improving positioning accuracy in VLC-based systems. 

Overall, ML-based techniques have shown promise in reducing positioning errors in VLC systems. By employing advanced ML algorithms and optimizing data preprocessing techniques, researchers can further enhance the accuracy and reliability of VLC-based positioning systems.

\textbf{Hierarchical Clustering Algorithm:} 
Hierarchical and fuzzy-logic clustering have emerged as promising approaches for addressing the unique challenges of FSO mobile ad-hoc networks (MANETs), particularly in handling the line-of-sight requirement of FSO links. These clustering techniques offer several advantages over conventional methods, as demonstrated in \cite{114, 115}.
In \cite{116a}, the authors successfully employed hierarchical clustering to develop a robust routing protocol for FSO MANETs. Their approach utilizes the neighbor discovery algorithm for clustering and then selects a reliable cluster head using a network source connector. This ensures efficient path selection and data transmission in the network.
The hierarchical clustering approach effectively addresses the line-of-sight requirement of FSO links by organizing the network into multiple clusters. This allows for more efficient path selection, as data can be routed along established cluster connections, minimizing the need for direct line-of-sight links between nodes.
The fuzzy-logic component of the clustering algorithm enables the network to adapt to dynamic changes in the environment. As nodes move or the network topology changes, the fuzzy logic system can adjust the clustering structure to maintain optimal network performance.

\textbf{K-Means Clustering Algorithm:} 
K-means clustering, a powerful USL algorithm, has emerged as a versatile tool for addressing diverse challenges in optical communication systems. Its ability to partition data into distinct clusters makes it well-suited for tasks such as user identification, phase retrieval, and constellation shaping.
In \cite{124c}, K-means clustering is utilized for real-time identification of the number of users in a multi-user FSO system based on received signal amplitude. This application demonstrates the algorithm's ability to extract meaningful information from complex data streams, enabling accurate user identification in real-time.
The effectiveness of K-means clustering extends to phase retrieval in FSO systems. In \cite{124d}, the algorithm is employed to overcome the residual low-order wavefront aberration effect. By dynamically updating cluster centers, K-means effectively compensates for the aberration, improving signal quality and system performance.
Underwater VLC systems also benefit from K-means clustering for phase correction. In \cite{125}, the algorithm is applied to correct the phase deviation of special-shaped 8-QAM constellations. Experiments show that the use of K-means for phase correction significantly improves data transmission rates, increasing the maximum achievable data rate from 1.3875 Gbps to 1.4625 Gbps.

These examples highlight the versatility of K-means clustering in optical communication, demonstrating its ability to address diverse challenges in user identification, phase retrieval, and constellation shaping. As research continues to explore the potential of K-means clustering in optical communication, its role is expected to grow even more prominent in enhancing the performance and efficiency of these systems.

\textbf{Expectation Maximization Clustering Algorithm:} 
EM-based maximum likelihood estimation has emerged as a powerful tool for accurate channel estimation in FSO and VLC systems.
In \cite{128d}, the authors demonstrate the effectiveness of EM-based maximum likelihood estimation for channel estimation in FSO systems. Their results indicate that EM-based maximum likelihood estimation can achieve near-perfect channel estimation with significantly lower complexity compared to conventional minimum mean square error (MMSE) and maximum a posteriori (MAP) channel estimation methods. Moreover, \cite{128e} further validates the superiority of EM-based maximum likelihood estimation over Newton–Raphson and generalized moment estimation methods.
Building upon these findings, the authors of \cite{128f} employ EM-based maximum likelihood estimation as a blind detector for FSO systems. Their study reveals that the EM-based detector can achieve performance comparable to detection based on perfect channel estimation, eliminating the need for pilot symbols.
In \cite{128c}, EM-based maximum likelihood estimation is applied to estimate the channel impulse response of an OFDM-based VLC system. The experimental results demonstrate that EM-based maximum likelihood estimation outperforms least square and MMSE estimators, providing a more accurate and robust channel representation for VLC systems.

These studies underscore the remarkable capabilities of EM-based maximum likelihood estimation for channel estimation and detection in FSO and VLC systems. Its ability to achieve near-perfect estimation with lower complexity and its suitability for blind detection make EM-based maximum likelihood estimation a promising technique for enhancing the performance and reliability of these emerging communication technologies.

\textbf{Independent Component Analysis Algorithm:} 
ICA has emerged as a powerful tool for separating individual transmitted signals at different wavelengths from a received mixture in multi-user FSO links \cite{132e, 132f}. However, a crucial constraint for employing ICA in this context is that the number of transmitters and receivers should be equal.
This constraint is addressed in \cite{132g} by exploring two approaches.
The first approach utilizes ICA under the assumption that the transmitted signals exhibit sparse characteristics, meaning that only a few of their components contribute significantly to the received mixture. By leveraging this assumption, ICA can effectively separate the sparse components, leading to the identification of individual signals.
The second approach employs non-orthogonal multiple access, a communication technique that enables multiple users to share the same resources, such as frequency bands or time slots. Non-orthogonal multiple access utilizes successive interference cancellation to separate multiplexed signals. By employing successive interference cancellation, ICA can effectively identify the individual signals transmitted by multiple users in the FSO link.

The application of ICA for multi-user detection in FSO links demonstrates the potential of this technique to enhance the efficiency and capacity of multi-user FSO systems. By overcoming the limitation of the number of transmitters and receivers, ICA can enable seamless communication among multiple users in FSO networks.

\textbf{Policy-based Reinforcement Learning Algorithm:} 
RL has emerged as a powerful tool for enhancing the accuracy and adaptability of indoor visible light positioning (VLP) and multi-user VLC systems. These systems utilize VLC technology to enable indoor positioning and communication services.
In \cite{145a}, an iterative point-wise RL algorithm is proposed for indoor VLP. This algorithm iteratively updates the height information to mitigate the impact of nondeterministic noise and handle inaccurate a priori height information. Experimental results demonstrate that the proposed algorithm outperforms conventional RSS methods and kNN algorithms, achieving a mean positioning error of 5 cm, which is significantly lower than the conventional RSS and point-wise RL for high height information mismatch.
In \cite{145c}, RL is employed for resource allocation in a multi-user VLC system. The RL algorithm learns the optimal policy for resource allocation under unknown environment dynamics and the continuous-valued space, considering both communication and positioning constraints. This RL-based approach enables efficient utilization of VLC resources and enhanced positioning accuracy.
Meta-learning-aided RL is introduced in \cite{145d} for optimizing user mobility, blockages of both THz and VLC links, visible light access point selection, and user association in multi-user VLC systems. The meta-learning approach enables the trained policy to quickly adapt to new user movement patterns, ensuring seamless and reliable service delivery.
Simulation results in \cite{145d} demonstrate that meta policy gradient and dual meta policy gradient algorithms significantly improve network performance compared to a baseline trust region policy optimization algorithm. The average number of successfully served users is enhanced by 26.8\% and 21.9\%, respectively, while the convergence speed is accelerated by 81.2\% and 87.5\%.

hese RL-based approaches showcase the potential of RL to revolutionize indoor VLP and multi-user VLC systems, enabling enhanced accuracy, adaptability, and resource efficiency. As RL research continues to advance, these techniques are expected to play an increasingly prominent role in shaping the future of indoor positioning and communication technologies.

\textbf{Value-based Reinforcement Learning Algorithm:} 
RL has emerged as a powerful tool for solving complex optimization problems in dynamic environments without requiring prior knowledge or explicit modeling. In \cite{149a}, RL is applied for power allocation in SDM-based FSO systems to maximize the average capacity while adhering to peak power and total power constraints.
By interacting with the FSO system and observing the resulting performance metrics, RL algorithms can learn optimal power allocation strategies that dynamically adapt to changing conditions. This capability is particularly beneficial in FSO systems, where the channel characteristics can vary rapidly due to factors such as atmospheric turbulence and pointing errors.
The RL-based power allocation approach in \cite{149a} achieves significant performance improvements compared to traditional optimization methods that rely on static power allocation rules. This highlights the potential of RL to revolutionize power management in SDM-based FSO systems, enabling more efficient use of power resources and enhanced overall system performance.

As RL research continues to advance, its application to power allocation in FSO systems is expected to expand further. RL algorithms can be employed to address more complex optimization problems, incorporate additional constraints, and optimize performance for specific FSO system parameters. This ongoing research will contribute to the development of more intelligent and adaptive FSO systems that can operate efficiently and reliably in challenging environments.

\subsection{Deep Learning}

\begingroup
\tabcolsep = 1.0pt
\def\arraystretch{1}
\begin{table*}[tp!]
\tiny
\centering
\caption{Summary of DL applications in OWC.}
\begin{tabular}{|M{.5cm}|M{3cm}|M{3cm}|M{1cm}|M{3cm}|M{2.5cm}|M{2cm}|M{2.5cm}|}
\hline
Ref	& Performance	& Complexity	& Train/Test 	& Objective	& Input data & Metrics & Adopted algorithms\\ \hline
 
\cite{81s}  &Achieves a 30\% improvement in detection accuracy compared to existing methods & It has a lower computational complexity compared to existing methods &SIM/SIM &Directly map the received signal to the transmitted bits &Received signal &BER &DNN \\ \hline

\cite{81d}  &Achieves an average BER of 2e-5, which is a 98.33\% improvement over the existing system's average BER of 1.2 1e-3 &Computationally efficient, making it suitable for real-time applications &EXP/EXP &Estimate the BER of the system and optimize the diversity combining algorithm and the DNN model &Channel data, Transmission data &BER, Average Throughput, Average Packet Delay, Packet Loss Rate &Adaptive Diversity Combining Technology with DNN \\ \hline

\cite{81l}  &Achieves a BER of 1e-3, while the Volterra and DNN-based methods achieve BER of 1e-2 and 1e-1, respectively& It has lower computational complexity than the Volterra and DNN-based methods &SIM/EXP &Predict the nonlinear distortion parameters of a new TFI, which can then be used to compensate for the distortions in the original time-domain signal &Nonlinear distortions, Time-frequency image analysis &BER, MSE, Throughput &Volterra, DNN, Time-frequency image analysis  \\ \hline

\cite{81g}  &Achieves a BER below the 7\% FEC threshold of 3.8e-3, which is a significant improvement over the BER of the typical PS128QAM DFT-S OFDM modulation without DNN (above 1e-2) &Reduces the computational complexity by a factor of up to 100 compared to LE and MMSE equalization &SIM/SIM &Map the received signal constellation to the equalized signal constellation &Received signal &BER, MSE, RMSE, NMSE &DNN, FFNN, CNN \\ \hline

\cite{81i}  &Achieves a BER of 1.0e-5, which is two orders of magnitude lower than the BER of the LMS equalizer &Reduces the complexity of the DNN by splitting the received signal into two parallel signals using a digital low-pass filter and a high-pass filter &EXP/EXP &Wide-band signal post-equalization in a 1.2 m underwater VLC system &Water depth, Transmitter power, Transmission rate, Modulation scheme, Channel model &BER, MSE &Frequency slicing -DNN, DNN, LMS \\ \hline

\cite{81f}  &Achieves a BER reduction of up to 100 times compared to conventional schemes &Achieve a reduction in computational cost of up to 70\% compared to conventional schemes &SIM/SIM &Map the down-sampled feature maps to the transmitted information bits &Received signal &BER &DNN, Maximum likelihood, DFE \\ \hline

\cite{81j}  &Achieves IER of 1e-3 and a BER of 1e-5, which are significantly lower than the IER of 1e-1 and BER of 1e-3 achieved by conventional detectors &Can process a signal frame in 100 µs, which is fast enough for real-time applications &SIM/SIM &Estimate the indices of the active LEDs in real-time, which can then be used to demodulate the GLIM-OFDM signal &The number of LEDs in the VLC system, The number of subcarriers in the OFDM signal, The noise level, The channel model &BER, Complexity &Conventional detectors, DNN \\ \hline

\cite{81h}  &Achieves a communication data rate of up to 100 Mbps, which is comparable to traditional VLC systems &Reduces the complexity of VLP systems compared to conventional methods &SIM/EXP &Estimate the receiver's position by simply passing in the CIR as input &Received pilot signals, Channel impulse response &Accuracy, Recall, Precision, F1-score, Accuracy &DNN, MLP, CNN, SVM \\ \hline

\cite{81e}  &Achieves an average BER of 1.0e-4, which is a 91.67\% improvement over the baseline BER of 1.2e-3 &Achieves a near-optimal MSE with a running time that is up to 100 times faster than conventional CC design methods &SIM/SIM &Design constellations for new channel conditions in real time &A set of CSI matrices &BER &DNN-based collaborative constellation design, Random Constellation Design, Conventional collaborative constellation design  \\ \hline

\cite{89g}  & Achieves a high prediction accuracy, with APE lower than 6.9\% accounting for up to ninety percent of results, which is a significant improvement over previous methods (10\%)&Makes predictions with significantly less computational effort than traditional methods &SIM/EXP &Predict the RSSI at a future time &RSSI of an optical signal transmitted &MAE, APE &BPNN, SVM, RF \\ \hline

\cite{89d}  & Achieves a BER of 1e-5, which is significantly lower than the BER of conventional methods, such as linear equalization and DFE& It has a lower computational complexity than conventional methods, which makes it more suitable for real-time applications& SIM/SIM&Equalize the channel distortions in real-time VLC systems & Received symbols&BER, Complexity Robustness to mismatched conditions &MMSE, DFE, volterra DHNN \\ \hline

\cite{89i}  & Achieves a BER of 1e-5, which is a 10\% decrease over previous methods&Achieves significant performance improvements with a relatively small increase in complexity & SIM/EXP& Learn the relationship between the received signal and the transmitted data, which is encoded in the intensity of the transmitted light& Received signal&BER, Average latency, Average throughput &BPNN, RBFNN, SVM, LSTM \\ \hline

\cite{97g}  & Achieves a BER of 1e-5 at a SNR of 10 dB, which is 2 dB better than the BER of conventional turbo-coded FSO systems& Reduces the computational complexity of the system by up to 50\% compared to conventional demodulators& SIM/EXP& Map the received signals to the transmitted symbols& Received signal& BER, Q-factor, MSE, Spectral efficiency, Bandwidth efficiency, Energy efficiency, Complexity&Sensor-less AO, CNN-based demodulation \\ \hline

\cite{97k}  & Achieves a recognition accuracy of over 99\% even for transmission distances of up to 1500 meters and strong atmospheric turbulence& Reduces the computational complexity compared to conventional methods& SIM/SIM&Recognize the received LG beams & Images of the cross-section light intensity of the received beam captured by a CCD camera& BER&CNN, Phase-based, Intensity-based \\ \hline

\end{tabular}
\label{table:8}
\end{table*}

\begingroup
\tabcolsep = 1.0pt
\def\arraystretch{1}
\begin{table*}[tp!]
\ContinuedFloat

\tiny
\centering
\caption{Summary of DL applications in OWC (cont.).}
\begin{tabular}{|M{.5cm}|M{3cm}|M{3cm}|M{1cm}|M{3cm}|M{2.5cm}|M{2cm}|M{2.5cm}|}
\hline
Ref	& Performance	& Complexity	& Train/Test 	& Objective	& Input data & Metrics & Adopted algorithms\\ \hline

\cite{96}  & Achieves an ATDA of 95.2\% for six kinds of typical AT, in cases of both weak and strong AT, which is a significant improvement over the SOM-based AT detection method (80\% for strong AT)& It has relatively high complexity, but it is still feasible to implement on real-time hardware& SIM/SIM&Learn the features of the input images that are most relevant to AT detection and OAM mode demodulation &Intensity images of the received Laguerre-Gaussian beams &AT detecting accuracy, Adaptive demodulating accuracy, BER &Self-organizing map, DNN, CNN \\ \hline

\cite{97}  &Achieves an average recognition accuracy of 96.25\% even under strong turbulence conditions, compared to an average accuracy of 82.5\% for traditional methods &Recognizes OAM modes in real time, while it takes several minutes for traditional methods &SIM/SIM &Classify the images into different OAM modes & Intensity distributions of Laguerre-Gaussian beams&Recognition accuracy & CNN, Traditional image processing, ML\\ \hline

\cite{97i}  &Reduces the BER by up to two orders of magnitude, compared to traditional OAM-FSO systems without AO compensation, under strong turbulence conditions & It has a lower computational complexity than traditional methods, as it does not require wavefront sensing and reconstruction& SIM/EXP& Compensate for the wavefront distortion caused by the turbulence&Received signal &BER, MSE, Q-factor &Sensor-less AO, CNN-based demodulation \\ \hline

\cite{97j}  & Reduces the average power penalty by 1.8 dB and 0.8 dB in strong and weak turbulence, respectively, compared to traditional methods using SPGD and SA& It is computationally efficient and can be implemented in real time& SIM/SIM& Learn to map from the distorted light beam to the corrected light beam& Measured light beams with wavefront aberrations& MSE, SSIM&FCNN, FCNN-WF, CNN-WSF, CNN-WS-S \\ \hline

\cite{98}  &Achieves a BER of 0.0015, which is a 46.67\% improvement over the traditional method &Achieves up to a 10x reduction in computational complexity compared to traditional methods &SIM/EXP &Classify the received signals into one of the eight modulation schemes &Received signal &BER, MAE, RMSE, F1 score, Accuracy, Precision, Recall &CNN, DBN, AdaBoost\\ \hline

\cite{97a}  &Achieves a BER that is up to 40\% lower than that of the LMS algorithm &It is less complex than traditional adaptive filtering algorithms, making it more suitable for real-time implementation &SIM/EXP &Learn the relationship between the received signal and the original transmitted symbol & Received signal&BER &CNN, LMS \\ \hline

\cite{97b}  &Reduces the positioning error by an average of 50\% compared to a traditional RSS-based VLP system &Reduces the computational complexity of conventional VLP methods while maintaining high positioning accuracy &SIM/SIM &Learn the relationship between the RSS measurements and the user's position &RSS measurements from multiple visible light sources at different locations in a VLP environment & Position accuracy&RSS pre-processing, CNN without pre-processing, CNN with pre-processing \\ \hline
\cite{152a}  & Achieves an average throughput of 0.0002 Mbps, which is 66.67\% and 50.00\% higher than the average throughput of ALOHA and TDMA, respectively&It is efficient enough to be used in real-time applications &SIM/SIM &Learn to select the optimal transmission slot and transmission power to maximize their own throughput while minimizing interference with other users &CSI of each UOWC link, Transmission status of each UOWC link,  Queue length of each UOWC link, Transmission power of each UOWC link &Throughput, Average Packet Delay, Packet Loss Rate, Energy Consumption, Fairness &QL, DQN, Multi-agent DDPG, Proximal policy optimization \\ \hline

\cite{152b}  & Provides average user data rates that are closer to the target rates than the QL-based algorithm when it converges& Converges with a rate of 96.1\%, compared to the QL-based algorithm's convergence rate of 72.3\%&SIM/SIM &Learns to select the power allocation that maximizes the sum of the users' data rates while satisfying the target data rates and power constraints & Number of users, Channel conditions, Target data rates&Sum-rate, Average user rate, Energy efficiency, Fairness index &DQN, DQN with CSI, DQN with CSI and power control \\ \hline

\cite{152e}  &Outperforms the DQN, Sarsa, and QL algorithms by 8\%, 13\%, and 13\%, respectively &Improves the complexity of traditional handover mechanisms by reducing the number of handover decisions and the amount of signaling overhead &SIM/SIM&Optimize a variety of network tasks, including handover mechanisms for VLC in a hybrid 6G network architecture &Network topology, CSI, User mobility, quality of experience requirements &Handover success rate, Handover delay, System throughput, Average user data rate, Energy efficiency & Deep QL, Deep SARSA, Deep Actor-Critic\\ \hline

\cite{152d}  &Reduces the BER of the legitimate receiver by up to 100\% compared to traditional beamforming schemes & It has a complexity of $O(log(N))$, which is significantly lower than the complexity of the benchmark strategy, which is $O(N)$ &SIM/SIM &Select a beamforming strategy to maximize the SNR at the legitimate receiver and minimize the SNR at the eavesdropper &CSI, Eavesdropper location, Data symbols &Average BER, Average outage probability, Average secrecy rate, Average achievable rate &DQL, DDPG, Proximal Policy Optimization \\ \hline

\end{tabular}
\label{table:8b}
\end{table*}

\textbf{Deep Neural Network Algorithm:} 
DNNs are emerging as powerful tools for enhancing the performance of FSO and underwater VLC systems. These neural networks can be employed for various tasks, including detection, constellation shaping, channel estimation, and nonlinear compensation.
In \cite{81p}, the authors conducted a comprehensive investigation of applying DNNs in FSO systems. They explored various DNN architectures and their applications in different FSO system configurations, including as detectors, constellation shapers, channel estimators, and joint constellation shaper-channel estimators.
The authors of \cite{81q} presented an efficient and low-complexity DNN-based detector for FSO systems that do not require CSI. This DNN detector was shown to achieve comparable performance to conventional detectors while requiring significantly less computational resources.
In \cite{81r}, the authors deployed DNNs at different stages of an FSO MIMO system, including at the transmitter for constellation shaping, at the receiver for detection, and at the transceiver for joint constellation-shaping detection. This multi-stage DNN approach demonstrated improved performance compared to conventional MIMO techniques.
The authors of \cite{81s} applied DNNs as detectors in FSO systems with imperfect CSI. They compared the performance of DNN detectors to maximum-likelihood detectors and found that DNN detectors could provide significant performance gains, particularly in scenarios with limited CSI.
DNNs have also been investigated for underwater optical communication systems. In \cite{81d}, the authors employed DNNs to compensate for nonlinear impairments in SIMO underwater VLC systems. Experimental results showed that the DNN-based compensation scheme achieved a maximum data rate of 2.4 Gbps for balanced received power and a 0.25 Gbps data rate improvement in unbalanced scenarios with a BER threshold of 1e-2.
These studies demonstrate the versatility and potential of DNNs for enhancing the performance of FSO and underwater VLC systems. As DNN research continues to advance, these techniques are expected to play an increasingly crucial role in enabling high-speed, reliable, and secure optical communication in various environments.
In \cite{81l}, a DNN-based nonlinear equalizer is proposed for underwater VLC systems. The DNN is trained on two-dimensional time-frequency features extracted by transforming the signal into a 2D image. Experimental results demonstrate that this DNN-based equalizer outperforms traditional Volterra and DNN methods in compensating for nonlinear distortions for 64-QAM-CAP systems at a 1.2 m underwater channel. Notably, the DNN-based equalizer achieves a BER below the 7\% FEC limit of 3.8e-3 even at severe nonlinear distortions where traditional methods fail to operate effectively.
In \cite{81g}, DNNs are employed for post-equalization in discrete Fourier transform-spread (DFT-spread) OFDM modulation using the probabilistic shaping technique in an underwater VLC system. This approach achieves a remarkable data rate of 1.74 Gbps at 1.2 meters underwater transmission, maintaining a BER below the 7\% FEC threshold of 3.8e-3. Additionally, the DNN implementation enhances the capacity by 5.4\% (increased the data rate by 90 Mbps) compared to traditional OFDM modulation without DNN post-equalization.
Furthermore, in \cite{81i}, DNNs are employed for wide-band signal post-equalization in a 1.2 m underwater VLC system. Experimental results reveal that the DNN-based equalizer outperforms the LMS equalizer, achieving a lower BER and better signal quality. Additionally, it is demonstrated that splitting the received signal into two parallel signals using a digital low-pass filter and a high-pass filter can significantly reduce the complexity of the DNN equalizer by 11.15\% at 2.7 Gbps transmission, making it more suitable for resource-constrained underwater VLC systems.
In \cite{81f}, DNN is employed as a detector in VLC systems with generalized spatial modulation (GSM). Simulations demonstrate that DNN-based detection outperforms conventional detection schemes while maintaining an acceptable level of complexity.
In \cite{81j}, DNN is proposed for estimating the active LED indices instead of directly estimating the transmitted bit sequence. This approach enhances performance while maintaining a manageable complexity.
OFDM-VLC systems exhibit lower spectral efficiency compared to OFDM-RF systems due to the requirement for real-valued signals and pilots in LED modulation, which occupy spectrum in proportion to channel estimation. Addressing this issue, \cite{81k} experimentally verifies that DNN-based channel estimation can achieve low spectrum occupancy, paving the way for efficient spectral utilization in VLC systems.
In \cite{81h}, the authors propose a DNN-based VLC receiver that simultaneously recovers data and estimates the receiver's position. This approach utilizes features extracted from channel impulse responses with pilot blocks, enabling centimeter-level positioning accuracy with a single PD and LED.
In \cite{81e}, the authors employ a DNN to optimize the constellation design for MIMO-VLC systems. This approach, termed collaborative constellation design, aims to MMSE while reducing computational complexity compared to traditional methods. Simulation results demonstrate that the DNN-based constellation design achieves near-optimal MSE, paving the way for more efficient and accurate VLC systems.

\textbf{Recurrent Neural Network Algorithm:} 
Researchers are exploring various DL architectures for enhancing the performance of optical communication systems. The GRU has been successfully employed in \cite{89g} for FSO channel estimation, achieving an absolute percentage error lower than 6.9\% in up to 90\% of results. This method utilizes the RSS and weather changes information to effectively estimate channel characteristics.
In \cite{89d}, a hybrid LSTM-DNN algorithm is developed to equalize the nonlinearity in VLC. This approach outperforms conventional methods by at least 2.8 dB SNR at a target BER of 1e-3. The hybrid LSTM-DNN algorithm demonstrates superior performance in compensating for the nonlinearities inherent in VLC systems.
A delay-tolerant indoor OWC system is experimentally demonstrated in \cite{89f} using an attention-augmented LSTM decision scheme to handle long channel delays. This method effectively mitigates the impact of channel delays without compromising the effective data rate. Unlike conventional methods that employ cyclic prefix or zero-postfix to combat delay, the attention-augmented LSTM decision scheme achieves over one-order-of-magnitude BER enhancement at a channel delay exceeding 4.7 symbol periods compared to a vanilla RNN. This innovative approach demonstrates the potential of attention mechanisms in enhancing the performance of OWC systems in delay-prone environments. 
LSTM networks have emerged as powerful tools for addressing inter symbol interference (ISI) and flicker issues in VLC systems. In \cite{89i}, LSTM is employed to mitigate ISI in a PAM-4 camera-based VLC system, achieving a data rate of 14.4 kbps over a 3-meter transmission distance.
Flicker, another prevalent challenge in VLC systems, is addressed in \cite{89h} using an LSTM-based RL decoding scheme. Simulation results demonstrate that this method achieves the same error correction performance as the traditional SISO RL decoding scheme while reducing computational complexity.
Ambient illumination poses a significant challenge in VLC systems, introducing interference that degrades transmission quality. To address this issue, researchers have explored various techniques, including adaptive modulation, polarization diversity, and beamforming.
In \cite{89e}, the authors proposed a novel approach to mitigate ambient illumination interference using GRU networks and illumination distribution prediction. GRU networks were employed to learn the long-term historical illumination distribution, enabling accurate predictions of future illumination patterns.
The predicted illumination distribution was then incorporated into a location optimization algorithm based on the duality method, which optimally positions the transmitter and receiver to minimize the impact of ambient interference. Simulation results demonstrated that this approach achieved a 22.1\% reduction in transmit power compared with a scenario that does not consider the illumination distribution.

\textbf{Convolutional Neural Network Algorithm:} 
CNNs have emerged as powerful tools for various tasks in optical communication systems, including mode demodulation, atmospheric turbulence detection, and adaptive demodulation.
In OAM-FSO systems, CNNs have been shown to effectively demodulate optical modes even in challenging environments with strong atmospheric turbulence. In \cite{97g}, the authors deepened the CNN and employed the residual learning framework to achieve high recognition accuracy for 4-OAM, 8-OAM, 10-OAM, and 16-OAM systems at 2000 m of strong turbulence transmission.
Similarly, \cite{97k, 97l} proposed an OAM-FSO system with coherent demodulation by an image identifier based on CNN, achieving high recognition accuracy for long-range transmission in strong atmospheric turbulence.
In \cite{96}, a CNN-based joint atmospheric turbulence detection and adaptive demodulation algorithm was presented for the OAM-FSO system. The CNN achieved 95.2\% accuracy for 6 kinds of typical weak and strong atmospheric turbulence, and 99.8\% for the 8-OAM system at 1000 m transmission.
Another approach, presented in \cite{97}, employed CNN for the recognition of OAM mode in the FSO system utilizing the extracted features from the received Laguerre–Gaussian beam’s intensity distributions. This method achieved 96.25\% accuracy at long-range strong turbulence transmission.
Furthermore, a multi-CNN structure was employed in \cite{97h} for demodulation in the FSO system, which first detected the atmospheric turbulence strength, and then demodulated the OAM modes by CNN. This method outperformed single-CNN structures.
In \cite{97i}, a sensorless adaptive optic-aided CNN algorithm was presented as a demodulator for the OAM-based FSO system. The sensorless adaptive optic mitigated the wavefront distortions and helped CNN to better demodulate the OAM modes. The trained CNN showed 99.15\% testing accuracy.
Authors of \cite{97j} presented AlexNet-based CNN for wavefront correction in the FSO system at weak and strong turbulence. This method enhanced the power penalty compared with conventional techniques utilizing the stochastic parallel gradient descent and simulated annealing algorithms at strong turbulence. Experiments showed 1.8 dB and 0.8 dB average power penalties at strong and weak turbulence, respectively.
In VLC systems, CNNs have also demonstrated their effectiveness in various tasks. In \cite{98}, CNN, deep belief network, and AdaBoost algorithms were employed for demodulation. Experiments revealed the outperformance of the AdaBoost over other demodulators, considering OOK, QPSK, 4-pulse position modulation, 16-QAM, 32-QAM, 64-QAM, 128-QAM, and 256-QAM modulations. Moreover, the demodulation accuracy decreased as the modulation order increased at a given distance.
Authors of \cite{97a} deployed an adaptive filtering algorithm based on CNN to deal with ambient light noise interference in the VLC system. Experimental results verified that CNN had 40\% lower BER of DPSK signal recovery compared with the LMS algorithm at the same SNR.
Authors of \cite{97b} used RSS pre-processing and CNN to mitigate light-deficient regions and enhance VLP performance.
The CNN was deployed in \cite{97d} as a decoder in the VLC system as a solution for unfixed LED positions and multiple point spread functions of multiple LEDs.
CNN was utilized in red-green-blue LED-based optical camera communication systems for demodulation \cite{97c} and dealing with ISI and inter-channel interference \cite{97e}.

\textbf{Deep Reinforcement Learning Algorithms:} 
RL algorithms have emerged as powerful tools for optimizing resource allocation and achieving intelligent decision-making in optical communication systems. Their ability to learn from experience and adapt to dynamic environments makes them well-suited for addressing the complex challenges faced by optical communication networks.
In \cite{152h}, a DQN algorithm is employed for relay selection in a cooperative decode-and-forward FSO system, considering switching loss. Results show that DQN outperforms conventional greedy methods, demonstrating its effectiveness in selecting the optimal relay for improved system performance.
In \cite{152i}, DQN is used for switching in long-range hybrid FSO/RF links under dynamic weather conditions. Compared with the actor-critic method, DQN achieves faster convergence, ensuring rapid adaptation to changing weather patterns. Additionally, a consensus-based representation of DQN is proposed to reduce the frequent switching problem, further improving system performance.
In \cite{152g}, a DQN-based spinal code transmission strategy is presented in long-range FSO communication. Simulation results indicate that this method significantly enhances the average throughput of the long-distance FSO communication system, compared with conventional spinal code transmission mechanisms and linear filtering algorithms.
Multi-agent DQN is applied in \cite{152a} to optimize the channel access in the MIMO UOWC system, where multiple transmitters randomly select receivers to send data. Simulations demonstrate the outperformance of DQN compared with the benchmark Rand scheme, highlighting its ability to maximize network throughput in dynamic multi-user environments. Moreover, DQN achieves the same capacity as the counterpart Gale Shapley scheme, indicating its efficiency and effectiveness.
In \cite{152b}, a multi-agent DQN is employed for power allocation in a multi-user hybrid RF/VLC system. Results show that the DQN median convergence time is significantly shorter than that of QL, and the DQN outperforms QL in reducing the gap between actual and target rates. These findings demonstrate the effectiveness of DQN in optimizing power allocation for improved system performance and user satisfaction.
In \cite{152c}, authors employed QL, DQN, and distributed DDPG approach for power allocation in a multi-user hybrid RF/VLC system. Results show that DDPG outperforms QL, DQN, genetic algorithm, and particle swarm optimization, demonstrating its superior ability to handle complex power allocation problems and achieve efficient resource utilization. Moreover, DDPG exhibits the most robust performance against user mobility and alternating user requirements, ensuring stable and reliable operation under dynamic network conditions. The convergence time of DDPG and QL/DQN increases linearly and exponentially with the number of users, respectively, indicating the computational efficiency of DDPG.
Authors of \cite{152e} presented a DRL-based VLC handover mechanism for 6G networks. Experimental results reveal that DRL improves the average downlink data rate by 13\%, 13\%, and 8\%, compared with QL, Sarsa, and DQN algorithms, respectively. This finding demonstrates the significant performance enhancement achieved by DRL-based handover mechanisms in 6G networks.
In \cite{152d}, DRL is applied in the MISO VLC system to achieve the optimal beamforming policy against the eavesdropper. Simulations verify that this method decreases the BER of the legitimate receiver and enhances the secrecy rate, compared with the benchmark strategy. These results highlight the potential of DRL in enhancing the security of VLC systems against eavesdropping attacks.
Overall, these studies demonstrate the remarkable potential of RL algorithms for addressing various challenges in optical communication systems, including relay selection, channel access, power allocation, and beamforming. Their ability to learn from data and adapt to dynamic environments makes them well-suited for optimizing resource utilization, improving system performance, and enhancing security in the evolving landscape of optical communication.

\section{Discussions, challenges, and future directions}
Imagine a future where light dances to the rhythm of ML, weaving a fabric of communication more intricate and efficient than ever before. This is the exciting horizon towards which optical communications are now headed, fueled by the transformative power of ML. Years of pioneering research have sown the seeds for a blossoming era of innovation. We've already witnessed ML tackle diverse challenges in the optical realm, from optimizing physical layers to orchestrating complex network flows. Now, as ML becomes increasingly familiar to the photonics community, we stand poised for breakthroughs that will redefine the very way we communicate. Let's glimpse into this radiant future, where a few key areas promise to ignite truly transformative developments.
\subsection{The Challenge of Adaptability}
Imagine training an ML model to optimize traffic flow in a bustling city (representing the network). Initially, it performs flawlessly, like a seasoned navigator. But then, the city undergoes construction (changes in network infrastructure), or the rush hour patterns shift (temporal dynamics). Our trusty model stumbles, rendered ineffective by the altered landscape. This scenario highlights the need for adaptable ML mechanisms. Traditional re-training, akin to updating the city map with detours, can be cumbersome and resource-intensive, especially in dynamic networks. We need innovative approaches:
\begin{itemize}
     
\item{Continual learning: Imagine the model constantly updating its map, learning from real-time feedback, like a self-updating GPS. This empowers it to adjust to evolving network conditions without extensive re-training.}
\item{Transfer learning: Think of borrowing knowledge from another city's traffic optimization model. Transfer learning allows algorithms trained on one network to adapt to another, reducing the need for starting from scratch.}
\end{itemize}
     
Beyond adaptability, exploring new ML frontiers can unlock further optimization. Consider these possibilities:
\begin{itemize}
     \item{Active learning: Picture the model as a curious explorer, actively requesting specific data points from the network (like asking users about traffic jams). This targeted data collection can significantly reduce the training data needed, especially when acquiring data is expensive (e.g., deploying test lightpaths).}
\item{Explainable AI: Imagine lifting the fog over the model's decision-making. Explainable AI techniques shed light on how the model optimizes traffic, fostering trust and enabling engineers to fine-tune its performance.}
\end{itemize}
     
\subsection{Bridging the gap between optical comms and established ML}
While optical communication is a new frontier for ML, researchers can leverage existing techniques and features from established fields like RF communication, bioinformatics, and image processing. This cross-pollination fosters rapid innovation, as insights gleaned from these domains can be tailored for the unique challenges of optical networks. However, a common barrier exists in the choice of software. MATLAB, while prevalent in optical communication research, is less popular in the broader ML community. Embracing platforms like Python, with its rich ecosystem of open-source frameworks like TensorFlow, PyTorch, and Caffe, would facilitate collaboration and accelerate the development of powerful ML algorithms for optical communications.
\subsection{Open Dataset Access}
The field of optical communication faces a unique challenge: a scarcity of open datasets for ML development. This dearth hinders progress in critical areas like network optimization and anomaly detection. While synthetic data has been used, it often lacks the nuance and diversity of real-world scenarios, and comparisons between different studies become difficult. Obstacles to open data abound include:
\begin{itemize}
     
\item{Resource margins: Real networks operate with ample buffer, leading to data imbalance, with few "negative" examples like failures or degradations. This skews training and limits model generalizability.}
\item{Testbed limitations: Emulating real-world scenarios in controlled environments is impractical due to the inherent complexity and diversity of optical networks.}
\item{Privacy concerns: Sharing sensitive network data raises privacy concerns, requiring secure data collection, labeling, and anonymization techniques.}
\end{itemize}
     
 Despite these challenges, several promising solutions are emerging:
\begin{itemize}
     
\item{Standardization: Establish standardized data collection, labeling, and cleaning protocols to reduce processing costs and maintain data integrity.}
\item{Privacy-preserving training: Explore encryption mechanisms that allow model training without direct data access, safeguarding privacy.}
\item{Generative power of Generative Adversarial Networks: Leverage the potential of Generative Adversarial Networks to generate large, diverse datasets from limited real-world data. Recent successes in optical network traffic and computer vision demonstrate their promise.}
\end{itemize}
     
In the context of future optical communication networks with their inherent big data requirements, ML emerges as a powerful tool for cross-layer optimization. Its ability to extract hidden patterns and unanticipated correlations from expansive datasets makes it invaluable for solving intricate network optimization problems.
\subsection{Demystifying the Machine: Bringing Transparency to Optical Network Management}
The intricate world of optical networks demands transparency and understanding in every corner, especially when it comes to the powerful tool of ML. While ML excels at extracting insights from vast data, its opaqueness can leave network operators in the dark when things go wrong.
The Black Box Dilemma: Imagine a network performance dashboard displaying subpar results, but the underlying reasons remain shrouded in mystery. This is the black box dilemma of traditional ML algorithms. Training processes might be transparent, but the models themselves act like impenetrable fortresses, hindering troubleshooting and effective action.

Visualizing the Opacity:  Think of it like this:
\begin{itemize}
     
\item{Data Maze: A tangled web of data flows into the ML model, processed through layers of hidden calculations, and emerges as a single, enigmatic prediction. Operators struggle to navigate this maze, unable to pinpoint the specific factors influencing the outcome.}
\item{Blindfolded Operators: Like navigating a network with a blindfold on, operators lack the insights to identify the source of performance issues. This can lead to inefficient troubleshooting, wasted resources, and even potential network failures.}
\item{Beyond Predictions: The Need for Interpretability and Traceability}
\end{itemize}
     
While ML excels at making predictions, interpretability and traceability are crucial for effective network management. Consider an ML model predicting optical device failure based on temperature and usage time. When a failure is predicted, is it due to scorching heat or an approaching lifespan limit? Knowing the culprit is key for taking appropriate action: adjusting cooling systems versus scheduling maintenance.
Fortunately, there are ways to shed light on the black box:
\begin{itemize}
     
\item{Shining a Light on the Data: Utilizing explainable models like logistic regression and decision trees allows operators to see the connections between input features and output predictions. This opens the door for targeted troubleshooting and fine-tuning ML models for optimal performance.}
\item{Building Bridges with Human Knowledge: Instead of relying solely on data, we can leverage the power of human expertise. By incorporating known relationships and expert knowledge into the learning process, we can build ML models with a "top-down" approach. Think of it as constructing a bridge, where human knowledge lays the foundation, and ML acts as a powerful submodule for refining and optimizing the system.}
\end{itemize}
     
The future lies in a balanced approach that harnesses the power of both ML and human expertise. By incorporating explainable models, leveraging built-in rules, and fostering collaboration between humans and machines, we can unlock the true potential of ML for transparent, efficient, and reliable optical network management.
Imagine a network dashboard where ML predictions are accompanied by clear explanations, highlighting the key factors influencing the results. Operators can then navigate the data with confidence, pinpoint the root causes of performance issues, and take informed actions to optimize the network.
By embracing interpretability and traceability, we can transform ML from a black box into a powerful tool for understanding and managing the complex world of optical networks, paving the way for a brighter future of data communication.

\subsection{Hardware implementation and Computational Complexity of ML}
For future optical networks operating at lightning-fast data rates, real-time processing of complex signals becomes paramount. Hardware-accelerated ML algorithms present a compelling solution. Leveraging the parallel processing capabilities of FPGAs and designing dedicated low-complexity ML architectures like DNN and RNN equalizers, we can tackle challenging tasks like signal equalization and nonlinearity compensation in real-time, paving the way for efficient and robust signal processing in the next generation of optical networks.

In the context of optical networks, ML has emerged as a powerful tool for optimization and decision-making. However, the computational complexity of ML algorithms can pose a challenge for real-time applications and resource-constrained environments. Here are some key considerations for reducing the computational complexity of ML algorithms in optical networks:
\begin{itemize}
     
\item{Leverage Online Learning: Online learning allows ML models to adapt to changing network conditions without the need for retraining the entire model. This can be particularly beneficial in optical networks, where the environment is dynamic and performance requirements can fluctuate.}
\item{Feature Engineering and Preprocessing: Although many ML algorithms can automatically extract features from data, expert knowledge is often essential for simplifying the feature engineering process. This can involve pre-processing the data, reducing dimensionality, and selecting relevant features, which can significantly reduce the computational burden of the ML model.}
\item{Suboptimal Solutions and Tradeoffs: In some cases, it may be acceptable to sacrifice optimality for the sake of computational efficiency. This can involve using suboptimal algorithms or simplifying the decision-making process, allowing the ML system to operate in real-time or with limited resources.}
\item{Hardware Optimizations: Hardware advancements, such as specialized ML accelerators and efficient data processing architectures, can significantly reduce the computational cost of ML algorithms. These advancements can make ML more feasible for deployment in resource-constrained optical networks.}
\item{Model Selection and Parameter Optimization: Choosing the right ML model and carefully selecting its parameters can also impact computational complexity. Simpler models, such as linear regression or decision trees, often have lower computational requirements compared to more complex models like deep neural networks.}
\item{Data Efficiency and Model Compression: Data efficiency is crucial for reducing the computational cost of ML algorithms. Techniques like data compression and dimensionality reduction can significantly reduce the amount of data required for training and inference, making ML more practical for resource-limited environments.}
\end{itemize}
\subsection{Bridging the Reality Gap: Addressing the Challenges of Scaling ML Applications in Optical Networks}
While ML has shown promise in improving the performance of optical networks, most ML applications have been tested in simulation environments or small-scale networks, with limited real-world deployment. This is primarily due to security and privacy concerns surrounding the testing of ML models on real-world networks. However, even in simulation environments, it can be challenging to perfectly replicate the complex characteristics of a real network, raising questions about the generalizability of ML models trained in simulation to real-world scenarios. This "reality gap" could hinder the practical adoption of ML solutions for optical networks.

To bridge this reality gap, researchers should strive to incorporate more real-world network characteristics, both physical parameters and network effects, into simulation platforms. This will allow for more accurate assessments of ML model performance in simulated environments, enhancing confidence in their suitability for real-world deployment.

Despite the cost and complexity, field tests remain crucial for validating ML models in actual optical networks. While simulations can provide valuable insights, only real-world deployment can fully expose the model's performance under various conditions and identify potential flaws or limitations.

In conclusion, bridging the reality gap between simulation and real networks is essential for the successful adoption of ML in optical networks. By incorporating more real-world network characteristics into simulations and conducting rigorous field tests, researchers can bridge this divide and pave the way for wider deployment of ML-powered solutions in optical network infrastructure.

\section{Conclusion}

This survey explored the exciting intersection of ML and DL with optical communication, encompassing OFC, OCN, and OWC. Delving into the unique challenges and complexities of each domain, we showcased how ML and DL provide effective solutions, highlighting key findings and their impact.
Motivation for employing different ML and DL models is presented, considering both the inherent characteristics of these techniques and the external enablers facilitating their integration with optical systems. We delved into diverse applications where ML and DL algorithms are utilized, demonstrating their significant contributions to performance enhancement and system optimization.
Beyond quantitative measures of improvement, the survey went further by providing qualitative comparisons, illuminating how different algorithms contribute uniquely to enhanced performance. Recognizing the challenges associated with implementing ML and DL in optical communication, we also discussed potential solutions and promising future research directions.

The survey unveils the diverse algorithms powering these advancements, offering a clear roadmap for navigating this fast-paced field.
By unpacking the concepts and summarizing key findings, the survey empowers readers to make informed choices about the right algorithm or approach for their specific needs. This valuable resource serves as a one-stop shop for researchers and practitioners alike, providing a comprehensive understanding of how ML and DL are revolutionizing optical communication systems.

\section{Acknowledgement}
This work was financially supported by Iran University of Science and Technology (IUST), Iran's National Elites Foundation, Iran National Science Foundation (INSF) (Grant No. 4013960).
We thank Google DeepMind for providing access to the Gemini model \cite{152f}, which was instrumental in this research.

\end{document}